\documentclass[aps,prb,twocolumn,groupedaddress,longbibliography]{revtex4-2}%

\usepackage{amsmath}
\usepackage{amsfonts}
\usepackage{amssymb}
\usepackage{graphicx}
\usepackage{xcolor}
\usepackage{lineno}  %
\usepackage{subcaption}
\usepackage[normalem]{ulem}

\begin{document}
%\linenumbers  %
\title{Modeling of a twisted-Kagome HoAgGe spin ice using Reduced-Configuration-Space Search and Density Functional Theory}

\author{Gunnar F. Schwertfeger}
\affiliation{Department of Physics and Astronomy, George Mason University, Fairfax, VA
22030, USA}
\affiliation{Quantum Science and Engineering Center, George Mason University, Fairfax, VA
22030, USA}
\author{Po-Hao Chang}
\affiliation{Department of Physics and Astronomy, George Mason University, Fairfax, VA
22030, USA}
\affiliation{Quantum Science and Engineering Center, George Mason University, Fairfax, VA
22030, USA}
\author{Predrag Nikoli\'{c}}
\affiliation{Department of Physics and Astronomy, George Mason University, Fairfax, VA
22030, USA}
\affiliation{Quantum Science and Engineering Center, George Mason University, Fairfax, VA
22030, USA}
\author{Igor I. Mazin}
\affiliation{Department of Physics and Astronomy, George Mason University, Fairfax, VA
22030, USA}
\affiliation{Quantum Science and Engineering Center, George Mason University, Fairfax, VA
22030, USA}
\begin{abstract}
The Kagome lattice is a 2D network of corner sharing triangles found in several rare earth materials resulting in a complicated and often frustrated magnetic system. In the last decades, modifications of the motif, such as breathing Kagome, asymmetric Kagome, and twisted Kagome were brought into the limelight. In particular, the latter has lower symmetry than the original Kagome and thus allows implementations of an ``Ising-local'' Hamiltonian, leading to a 2D spin ice. One such material implementation, HoAgGe, was recently reported to have an exceptionally rich  phase diagram and is a strongly frustrated 2D spin-ice material with a twisted-Kagome geometry. In the presence of an  external magnetic field the compound exhibits step-like magnetization plateaus at simple fractions of the saturation magnetization. It is believed that this phenomenon results from strong single-site anisotropy, which in HoAgGe was found to be in-plane and along a high-symmetry direction. Previous  Monte Carlo simulations with empirical exchange parameters explain some, but not all experimental observations. In this work we present (a) first-principle calculations of the crucial model parameters and (b) direct energy minimization via a Reduced-Configuration-Space search, as well as Monte-Carlo simulations of the field-dependent phase diagram. We find that for HoAgGe the calculated exchange parameters are very different from the earlier suggested empirical ones, and describe the phase diagram much more accurately. This is likely because the first-principles parameters are, in addition to geometrically, also parametrically frustrated.
%\IM{ likely because the first-principles parameters are, in addition to geometrically, also parametrically frustrated.} 
%than empirically found parameters, largely by the virtue of being more \PN{geometrically \sout{parametrically}} frustrated.
\end{abstract}
\maketitle
\section{Introduction}

Magnetic frustration has been the source of many exotic phenomena in matter. In 3D rare-earth pyrochlore compounds, for instance, one finds a rather unique phenomenon dubbed ``spin ice'' \cite{spinice1999}. 
In a new twist, a {\it 2D} material, HoAgGe, was recently reported\cite{doi:10.1126/science.aaw1666} to be an example of 2D spin ice. Spin ice is, typically, a highly frustrated magnetic system where each magnetic site has a strong uniaxial anisotropy, with the axis varying from site to site according to the crystal symmetry.

\begin{figure}[h!]
\centering
\includegraphics[scale=0.18]{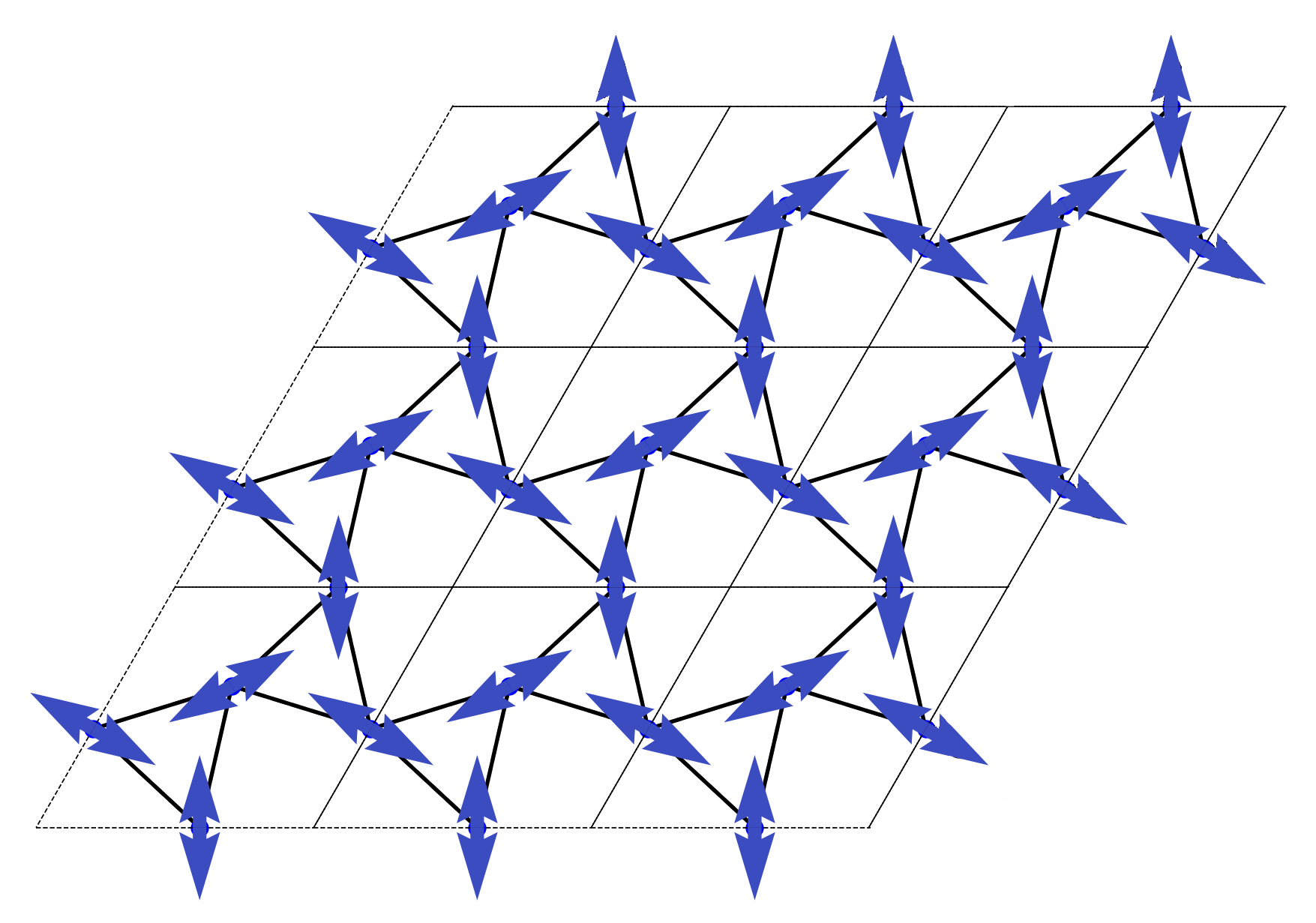}
\caption{ 
Twisted Kagome Lattice with 
$J_1$ 
bonds visualized (a 
$3 \times 3$ supercell is shown). The arrows indicate the easy axis in HoAgGe.
}
\label{fig:3x3_lattice}
\end{figure}

The kagome lattice is formed from corner sharing equilateral triangles (Fig. \ref{fig:3x3_lattice}). As the kagome lattice is the 2D analogue of the 3D pyrochlore lattice, it is natural to look for spin ice behavior there. In the last decades, various interesting modifications of the kagome lattice have also been discussed in the literature, such as breathing kagome\cite{PhysRevLett.120.026801}, distorted kagome\cite{Hering2022}, square kagome\cite{DOI:10.5488/CMP.12.3.507} etc. 
%\PN{[cite examples?]} 
One of the newest developments is the so-called\cite{Huang2023} twisted Kagome lattice, where the alternating triangles in the lattice are twisted around their centers in the opposite directions. This operation lowers the symmetry from  P6/mmm  to P$\bar{6}$/2m. In particular, the site symmetry is lowered from mmm to m2m, as one can see from Fig. \ref{fig:3x3_lattice}, and the global inversion symmetry is also lost.

The recently reported magnetic compound, HoAgGe\cite{doi:10.1126/science.aaw1666} has this twisted kagome structure, with the twisting angle  $\sim 15.6^\circ$ (again as depicted in Fig. \ref{fig:3x3_lattice}).
Ho$^{3+}$ atoms in this compound have 10 f-electrons, a large orbital moment and strong spin-orbit coupling. Therefore, the singe-site anisotropy is large and, generally speaking, can have any easy axis, in or out of plane\cite{PhysRevB.108.045132}. Experimental evidence suggests that the easy axis for Ho is along a high-symmetry direction as indicated by the arrows in Fig. \ref{fig:3x3_lattice}. Our first-principle calculations described below confirm this.

In this case, the magnetic Hamiltonian, to a good approximation, is local-Ising, in the sense that each spin can assume only two directions. This results in a "two-in-one-out" spin-ice rule in plane for the triangular nearest neighbor interactions between Ho atoms in the twisted kagome lattice. This is analogous to the "two-in-two-out" rule observed in the tetrahedra in pyrochlore spin-ice compounds \cite{spinice1999}.
%\PN{[The "spin-ice rule" refers to a finite ground-state degeneracy of spin configurations on a single lattice unit, e.g. "two-in-two-out" on pyrochlore tetrahedra; "spin-ice" requires an Ising spin anisotropy, but it's not analogous to it...]}
Existing experimental data are reasonably well consistent with this picture\cite{doi:10.1126/science.aaw1666,nirmal2024} (albeit small, but distinct deviations from the infinite-anisotropy model were reported in Ref. \cite{nirmal2024}). As the dominant interaction is antiferromagnetic, the ground state in zero field has zero magnetization, and the spins form a $\sqrt3\times\sqrt{3}$ supercell, as depicted in Fig. \ref{fig:ground}, with the magnetic space group retaining the hexagonal symmetry, P$\bar{6}'$m2$'$. In the in-plane saturation field, whose value depends on the field direction, all spins assume the direction most parallel to the applied field,

Experimentally, the ``iceness'' of the system is reflected in the multiple magnetization steps in an in-plane magnetic field $\mathbf{h}$. Indeed, HoAgGe exhibits, in an external magnetic field, several of these unique magnetic steps. These steps are simple fractions of the total magnetization. While some of the larger steps (such as the 1/3 and 2/3 magnetization plateaus observed for $\mathbf{h}||y$) have been explained in earlier works \cite{doi:10.1126/science.aaw1666}, many smaller magnetization steps, particularly those observed for $\mathbf{h}||x$, have not been explained.

To this end, we have, first, performed first-principles DFT+U calculations, making sure that the Hund's rule on Ho is respected\cite{PhysRevB.108.045132,lee2024}, as discussed below, and extracted six Heisenberg exchange parameters, up to the 5th nearest neighbors (as opposed to three parameters used in the empirical model of Ref. \cite{doi:10.1126/science.aaw1666}). The resulting parameters are completely different from latter, yet provide a better description of experimentally observed plateaus not only for $\mathbf{h}||y$, but also for $\mathbf{h}||x$. Using these parameters, we constructed full exchange Hamiltonians (as discussed below, the Dzyaloshinskii-Moriya and dipole-dipole interactions appear to be not important), and directly minimized the energy by performing a search over a Reduced-Configuration-Space for all possible supercells up to 18 formula units and constructed the zero-temperature phase diagram for both field directions. We found that all major phases are very stable, and secondary phases, corresponding to smaller magnetization plateaus, are on the borderline of stability, but still present. Thus, we identified all 9 experimentally observed phases.
To further validate the experimentally determined local easy-axis orientation, we also performed total energy calculations comparing several in-plane spin configurations (see Supplementary Materials\cite{supplementary}). These calculations confirm that the local easy axis for Ho lies along a high-symmetry direction, as indicated in Fig. \ref{fig:3x3_lattice}.

In addition, we attempted Monte-Carlo simulations similar to those in Ref. \cite{doi:10.1126/science.aaw1666}, but, as opposed to their model, our calculated Hamiltonian is much more frustrated, so finding the global energy minimum in MC calculations is challenging.

%\begin{figure}[h!]
%\centering
%\includegraphics[width=0.32\linewidth]{Figures/ising0.png}
%\includegraphics[width=0.32\linewidth]{Figures/potts2.png}
%\includegraphics[width=0.32\linewidth]{Figures/potts4.png}
%\caption{ Possible spin models on a 
%twisted kagome lattice with a strong easy axis. Left:
%Ising-ice model (same as Fig. \ref{fig:3x3_lattice}). Center:
%4-Potts-ice model. Right: 8-Potts-ice model.
%The red rods indicate the hypothetical easy axes.
%}
%\label{fig:spins}
%\end{figure}

\section{general model}
\subsection{Crystal Lattice Hamiltonian}
\begin{figure}[!h]
\includegraphics[scale=0.36]{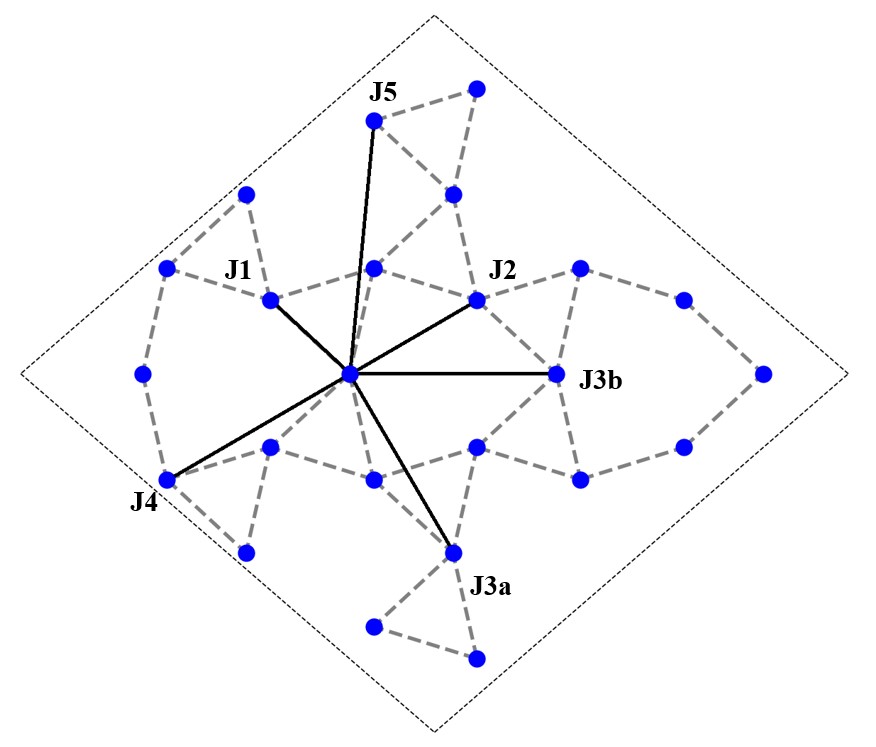}
\centering
\caption{Nearest neighbor exchange interactions, $J_1$ to $J_5$, depicted for a single atom. }
\label{fig:ref_bonds}
\end{figure}

The Hamiltonian for the Spin-Ice crystal HoAgGe is:
\\
\begin{equation}
H[M;J_\alpha,h] = -\sum_\alpha J_{\alpha} \sum_{\left< i, j\right>_\alpha} \hat{\bf S}_i\cdot\hat{\bf S}_j - h \sum_i \hat{\bf S}_i \cdot \hat{\bf h} \label{eq:hamiltonian} \ ,
\end{equation}
\\
where $M = \{ \hat{\bf S}_0, \hat{\bf S}_1, ... \hat{\bf S}_N \}$ defines a particular state of the crystal lattice, $J_\alpha$ are Heisenberg exchange parameters and $h$ the applied magnetic field along $\hat{\bf h}$. All spins, $\hat{\bf S}_i$ and $\hat{\bf h}$ are unit length vectors in $\mathbb{R}^3$. Here, $i,j$ label lattice sites, and $\alpha$ enumerates the Ho-Ho bonds of different lengths on which two spins interact (shown in Fig. \ref{fig:ref_bonds}). Again, as the Ho atoms have large magnetic moments, we can treat these spins classically.

Using a method described in a previous paper by some of us \cite{pohao2024}, we considered periodic magnetic orders as ground-state candidates, with tessellating supercells of various sizes and shapes. A supercell effectively has periodic boundary conditions; if a supercell is very small, then a long-range exchange parameter doubles back on the same supercell atom, which results in a constant positive addition to the lattice energy per antiferromagnetic pair or a constant negative addition per ferromagnetic pair. However as we search supercells containing up to octodecuple ($\times18$) Ho atoms, all six exchange parameters play nontrivial roles.

Exchange interactions are defined by their relative distances and crystallographic symmetry. For $J_1$, $J_2$, $J_4$, and $J_5$ it is enough to use only the corresponding bond lengths to distinguish interaction types. However, for $J_3$ we have elected to divide these interactions further into crystallographically inequivalent $J_{3a}$ and $J_{3b}$ (in Ref. \cite{doi:10.1126/science.aaw1666} they were assumed to be equal, however there is no microscopic justification for that).

Atoms joined by $J_2$, $J_4$, or $J_5$ couplings also satisfy the two-in-one-out spin-ice rule, further compounding the system's frustration. This is particularly true in the case of $J_5$ interactions as, like the nearest neighbor $J_1$, neighboring $J_5$ pairs also form corner sharing triangles between atoms. These interactions form decoupled lattices, each still obeying the spin-ice rule between triangles of interacting atoms.

By finding the lattice configurations which minimize Eq. \ref{eq:hamiltonian} for particular values of $J_\alpha$ and $h$, the magnetic plateaus observed experimentally should be reproduced given a sufficiently large maximum magnetic unit cell.

For a particular super cell of $N$ atoms, if each atom is in one of $q$ possible states, the total number of configurations of the super cell is simply $q^N$, meaning the total number of elements in the configuration space, $\mathbb{M}$, is finite. On sufficiently small lattices with up to $N=18$ sites, one formally can generate every possible spin configuration and check its energy, but the number of configurations grows exponentially with system size. Using Eq. \ref{eq:hamiltonian} however, we can greatly reduce the number of states in $\mathbb{M}$ that need to be considered by eliminating symmetrically equivalent states.

For any particular state, $M\in\mathbb{M}$, we define the constants; $C_\alpha\lbrack M\rbrack$ and $\left< M \right>_i$ using Eq. \ref{eq:ca} and Eq. \ref{eq:Ma}.

\begin{equation}
    C_\alpha[M] = \sum_{\left< i, j\right>_\alpha} \hat{\bf S}_i \cdot \hat{\bf S}_j \\
\label{eq:ca}
\end{equation}
\begin{equation}
    \left< M \right>_i = \sum_{j} \hat{\bf S}_j \cdot \hat{\bf e}_i \\
\label{eq:Ma}
\end{equation}
Using Eq. \ref{eq:hamiltonian}, the Hamiltonian for any given state may be written as Eq. \ref{eq:simplified_ham}. 
\begin{equation}
H_M[J_\alpha, h] = -\sum_\alpha J_\alpha C_\alpha -\sum_i h_i \left< M \right>_i\\
\label{eq:simplified_ham}
\end{equation}
meaning each Hamiltonian for a state in $\mathbb{M}$ can be characterized by $\alpha + 3$ parameters. By collecting only the states which exhibit unique values for $C_\alpha$ and $\left< M \right>_i$ and therefore by Eq. \ref{eq:simplified_ham} constitute a unique Hamiltonian, the total number of unique states of the super cell is greatly diminished. Meaning the search over configuration space for the lowest energy state for any combination of $J_\alpha$ and $h$ is significantly reduced. For this reason, we refer to this method as a Reduced-Configuration-Space Search (RCS Search).

\subsection{Magnetic Dipole Interaction}

In principle, in rare-earth systems the magnetic dipole-dipole interactions may be relevant; they can be introduced to the Hamiltonian as follows: 
\begin{align}
&E_{dd} = - \frac{g_{dd}}{ \left| \vec{\bf r}_{ij} \right|^3} \left( 3 \left( \hat{\bf S}_i \cdot \hat{\bf r}_{ij} \right)   \left( \hat{S}_j \cdot \hat{\bf r}_{ij} \right) - \left( \hat{\bf S}_i \cdot \hat{\bf S}_{j}\right) \right)\nonumber\\
&g_{dd} = \frac{\mu_0 \mu_B^2 g^2}{4 \pi}\nonumber\\
 &   \left< E_{dd} \right> = \sum_{\alpha,\left< i, j\right>_\alpha} \frac{\left( 3 \left( \hat{\bf S}_i \cdot \hat{\bf r}_{ij} \right)   \left( \hat{\bf S}_j \cdot \hat{\bf r}_{ij} \right) - \left( \hat{\bf S}_i \cdot \hat{\bf S}_{j}\right) \right)}{ \left| \vec{\bf r}_{ij} \right|^3}\nonumber \\
&H_M[J_\alpha,h,g_{dd}] = -\sum_\alpha J_\alpha C_\alpha -\sum_i h_i \left< M \right>_i - g_{dd} \left< E_{dd} \right>\label{eq:edda}
\end{align}
An additional term to the Hamiltonian means that each Hamiltonian is now characterized by $\alpha + 4$ parameters and when searching through all forms of $M$ in $\mathbb{M}$ some degeneracies may be lifted allowing for more unique Hamiltonians to be observed. Using a g-factor of 10 (as was reported in Ref. \cite{doi:10.1126/science.aaw1666}), $g_{dd}$ is calculated to be $ 5.3681$ $meV {\r{A}^3}$. With the distance between nearest neighbor Ho-Ho atoms of $3.67$ $\r{A}$, the dipole-dipole interaction is estimated to have an average energy of $-0.1085$ $m eV$, far below the $J_1$ interaction energy of $3.154$ $m eV$. While this may initially appear negligible, especially when an external magnetic field is applied to the crystal, the minute effect of the dipole interaction is critical in distinguishing between potential states that are extremely close in energy. 
%This is particularly true in the case of degenerate potential ground states and states for $h||x$ as (assuming infinite anisotropy) a third of the Ho atoms will still be free to align themselves as their magnetic moments would be perpendicular to the applied field.

\subsection{Dzyaloshinskii-Moriya Interaction}

As the twisted Kagome lattice of the HoAgGe crystal is two dimensional with the XY plane being a mirror plane for the lattice, it is easy to determine the unit DMI vector, $\hat{\bf D}_{ij}$,  for the nearest neighbor interactions to be aligned perpendicular to the plane along the $\pm \bf{\hat{z}}$ direction. Through further symmetry considerations, it is then possible to determine the orientation of the $\hat{\bf D}_{ij}$ for each interaction.

In a similar fashion as was done to include the contribution of the dipole interaction, to account for the energy contribution to the Hamiltonian from the DM interaction we must introduce an additional term, $B_\alpha$, to the Hamiltonian of Eq. \ref{eq:simplified_ham}. We can then define the DM interaction scalar, $D_\alpha$, similar to the Heisenberg exchange parameters.

\begin{align}
B_\alpha \left[ M \right] &=  \sum_{\left< i, j\right>_\alpha} \left( \hat{\bf S}_i \times \hat{\bf S}_j \right) \cdot \hat{\bf D}_{ij} \nonumber \\
& \nonumber \\
H_M[J_\alpha,D_\alpha,h,g_{dd}] &= -\sum_\alpha \left( J_\alpha C_\alpha + D_\alpha B_\alpha \right) \nonumber\\
&-\sum_i h_i \left< M \right>_i - g_{dd} \left< E_{dd} \right> \label{eq:DMI_ham}
\end{align}

With the introduction of additional parameters to the Hamiltonian, even more degeneracy is expected to be lifted. However, in this particular case, as the magnetic moments of each Ho atom are either at ${2 \pi}/{3}$ rad or parallel to all other neighboring moments, the Dzyaloshinskii-Moriya interactions simplify into a scalar of the corresponding Heisenberg interaction and may safely be absorbed into the exchange parameters, $J_\alpha$. This is confirmed through direct calculation as for all allowed super cell Hamiltonians for the Ho lattice, $B_\alpha = \sqrt{3} C_\alpha$.

% I can throw a phase diagram in here to demonstrate this, but its not very interesting.

\subsection{Calculation of exchange parameters}
\begin{table}[h]
\begin{center}
\caption{
    The comparison between the effective exchange coupling constants up to 4th nearest neighbor assumed by Zhao {\it et al}\cite{doi:10.1126/science.aaw1666} and up to 5th nearest neighbor calculated using DFT and GF. The values of the former are normalized to $J_1=2$ meV and the latter are normalized to $J_1=3.154$ meV. NN here is the coordination number of the corresponding bond.
   Note that in Ref.  \cite{doi:10.1126/science.aaw1666} $J_{3a}$, $J_{3b}$ and $J_4$ were assumed to be equal.}
    \renewcommand*{\arraystretch}{1.4}
    \setlength{\tabcolsep}{8.5pt}
    \begin{tabular}{c|c|c|c}
    \hline 
     && {Ref. \protect\cite{doi:10.1126/science.aaw1666}} &  {DFT-GF}\tabularnewline
    \hline 
     & NN & \multicolumn{2}{c}{$J_{\alpha}/|J_{1}|$ }\tabularnewline
    \hline 
    $J_{1}$ & 4 & 1 & 1\tabularnewline
    $J_{2}$ & 2 & -0.115 &  -0.442\tabularnewline
    {$J_{3a}$} & {4} & {-0.064} & -0.131\tabularnewline
    $J_{3b}$ &2& {-0.064} & -0.175\tabularnewline
    $J_{4}$ & 2 & -0.064 & -0.023\tabularnewline
    $J_{5}$ & 4 && -0.067\tabularnewline
    \hline 
    \end{tabular}
   \end{center}
   \label{table:GW_params}
\end{table}
The calculations described below have been, to a large part, described in an earlier paper\cite{nirmal2024}, but we briefly outline them below for completeness.
A numerical-orbital-based \cite{ozakiVariationallyOptimizedAtomic2003}
DFT code OpenMX \cite{OpenMX} was first used to obtain the ground
state Hamiltonian. The Perdew-Burke-Enzerhof (PBE) \cite{perdewGeneralizedGradientApproximation1996}
generalized gradient approximation was employed to describe exchange-correlation effects. To ensure the proper f-shell occupation, a large Hubbard
U correction is used to account for the strongly correlated Ho $4f$
states \cite{liechtensteinDensityfunctionalTheoryStrong1995,dudarevElectronenergylossSpectraStructural1998} (since these states are well removed from the Fermi level, the exact value of $U$ in this case is not important, as compared to $d$-metals.
The exchange coupling constants are then calculated perturbatively
using Green's function method \cite{antropovExchangeInteractionsMagnets1997,katsnelsonFirstprinciplesCalculationsMagnetic2000} implemented in OpenMX 3.9 \cite{terasawaEfficientAlgorithmBased2019,omxgf2004}. Calculated exchange parameters are compared with exchange parameters assumed by Zhao {\it et al} \cite{doi:10.1126/science.aaw1666} in Table .1.

\begin{figure*}[t]
\includegraphics[scale=0.65]{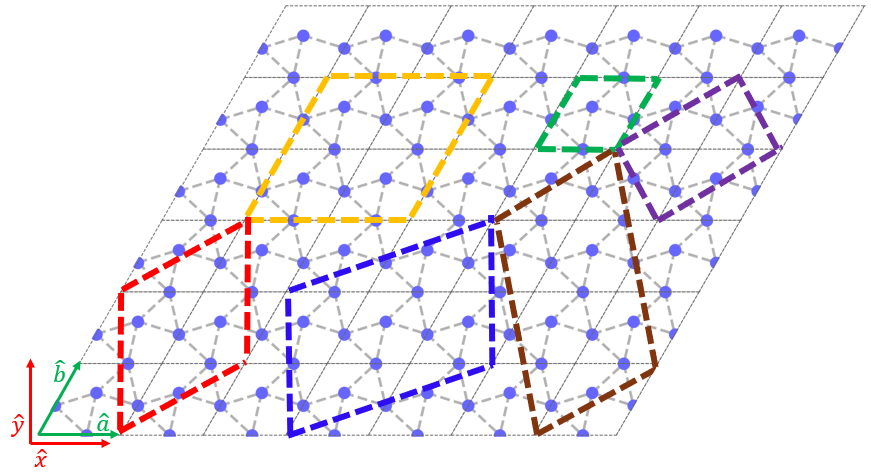}
\centering
\caption{
Magnetic supercells exhibiting the lowest energy for different in-plane fields.
The $1 \times 1$ Super cell (\textcolor{green}{Green}).
The $\sqrt{3} \times 1$ Super cell (\textcolor{purple}{purple}).
The $\sqrt{3} \times \sqrt{3}$ Super cell (\textcolor{red}{Red}).
The $2 \times 2$ Super cell (\textcolor{yellow}{Yellow}).    
%The $\sqrt{3} \times 2$ Super cell (\textcolor{pink}{Pink}). 
%The $\sqrt{7} \times \sqrt{3}$ Super cell (\textcolor{black}{Black}).
The $\sqrt{3} \times \sqrt{7}$ Super cell (\textcolor{brown}{Brown}).
The $\sqrt{7} \times \sqrt{3}$ Super cell (\textcolor{blue}{Blue}).
Unit vectors $\hat{x}$, $\hat{y}$, $\hat{a}$, and $\hat{b}$ included for reference clarity.
}
\label{fig:PH_states}
\end{figure*}

\section{Analysis of ${\text{HoAgGe}}$ model}

\subsection{Average Magnetization}

The minimum antiferromagnetic supercell is the $\sqrt{3} \times \sqrt{3}$ (Fig. \ref{fig:PH_states} Red). Using this supercell and imposing periodic boundary conditions gives a system of nine atoms with each atom in one of the two states. There are then $2^9 = 512$ total elements in the configuration space meaning 512 possible states can be represented on the $\sqrt{3} \times \sqrt{3}$ super cell. However, the number of states with unique Hamiltonians is only 88. These states are then compared to find the lowest energy configuration as the applied magnetic field is increased. 

This is done for all possible guesses of magnetic unit cells with the volume less than or equal to six times the volume of a single unit cell. This gives 3540 acceptable supercells. The states that exhibit unique Hamiltonians are then compared against each other to create a list of 220170 states with unique Hamiltonians covering 33 possible supercell shapes. Of those, only 6 shapes are appear to be relevant to minimum-energy states for any in-plane field. These shapes are shown in Fig. \ref{fig:PH_states} .

\begin{figure}[!h]
\includegraphics[scale=0.5]{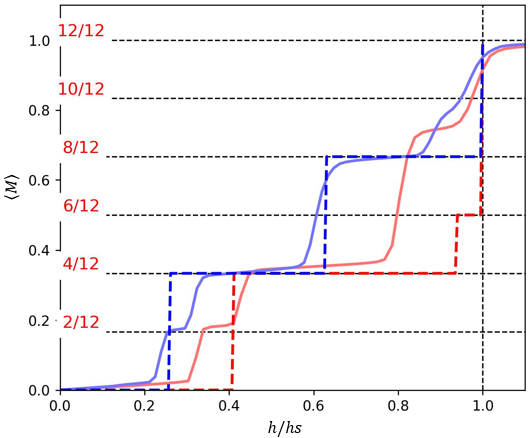}
\centering
\caption{Average Ho$^{3+}$ magnetization as an external magnetic field is applied perpendicular (X, \textcolor{red}{Red}) and parallel (Y, \textcolor{blue}{Blue}) to the easy axis as  found by direct minimization and empirical exchange parameters of Ref. \cite{doi:10.1126/science.aaw1666} (dashed), compared to experiment\cite{nirmal2024}.}
\label{fig:GW_mag}
\end{figure}

Using the exchange parameters conjectured by Zhao et al \cite{doi:10.1126/science.aaw1666} and applying the above protocol, the resulting average magnetization of the minimum energy states can then be directly compared with the experiment in Fig. \ref{fig:GW_mag}.
Note again that there is no distinction between bonds of type $J_{3a}$ and $J_{3b}$ for their set of parameters.

While the resulting magnetization curves (dashed) do represent, to some extent, experimental measurements (specifically the two most prominent magnetization plateaus at the $1/3$ and $2/3$ magnetizations), this Hamiltonian does not fully describe the experiment.

\begin{figure}[!h]
\includegraphics[scale=0.5]{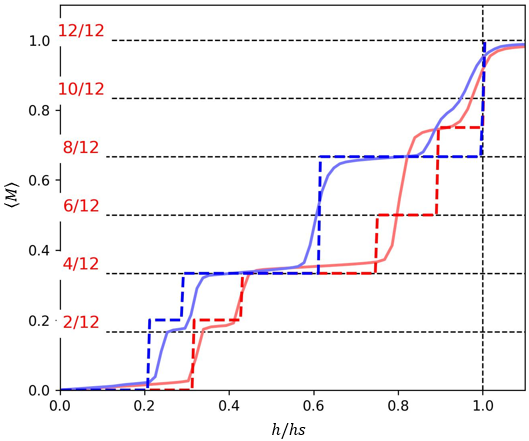}
\centering
\caption{Average Ho$^{3+}$ magnetization as an external magnetic field is applied perpendicular (X, \textcolor{red}{Red}) and parallel (Y, \textcolor{blue}{Blue}) to the easy axis found using direct minimization and our calculated DFT-GF exchange parameters (dashed) compared to experiment\cite{nirmal2024}.}
\label{fig:PH_mag}
\end{figure}

Using the other set of exchange parameters, calculated, as described, using DFT-GF method, the same states can again be analyzed with new parameters and the total magnetization is plotted in Fig. \ref{fig:PH_mag}.

Importantly, this set of parameters distinguishes between $J_{3a}$ and $J_{3b}$ while including the further neighbor coupling parameter, $J_5$. For this reason, the calculated values of $J_\alpha$ appear significantly different than those found empirically but still describe the system as sufficiently. In fact, these inclusions significantly improve the accuracy of the model, particularly in the X direction, perpendicular to the easy axis of the Ho atoms.

\begin{figure}[!h]
\includegraphics[scale=0.45]{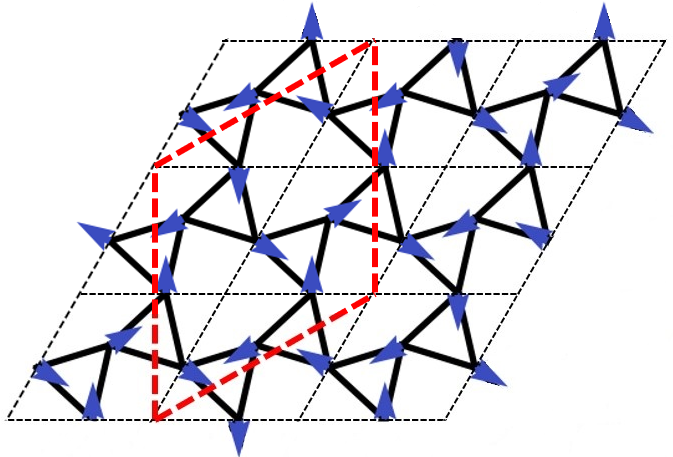}
\centering
\caption{The ground state $\sqrt{3} \times \sqrt{3}$ supercell (\textcolor{red}{Red}) with Magnetic Space Group P$\bar{6}$'m2' 
}
\label{fig:ground}
\end{figure}

Surprisingly, what was initially assumed to be a small $1/6$ magnetic plateau is not directly observed using this method. However, a $1/5$ magnetization state is observed. These new parameters exhibit several small steps between the larger $1/3$, $2/3$ and the fully saturated state plateaus. These are the $1/5$, $1/2$, and $3/4$ states. These states were not observed using the empirical parameters, but are observed experimentally. In total, the inclusion of the further nearest neighbor, $J_5$, and distinguishing between $J_{3a}$ and $J_{3b}$ type interactions resulted in ten distinct magnetization steps\footnotemark[3] again defined by the six shaped cells shown in Fig. \ref{fig:PH_states}. The inclusions of further neighbor interactions, $J_5$, does not generate significant change for magnetic fields applied in the Y direction, but still describe the magnetic behavior as accurately as the empirical parameters, even finding the same ground state (Fig. \ref{fig:ground}).

\footnotetext[3]{These are the $0/1$, $1/5 (x)$, $1/5 (y)$, $1/3 (x)$, $1/3 (y)$, $1/2 (x)$, $2/3 (y)$, $3/4 (x)$, and $1/1 (x)$, $1/1 (y)$ magnetization plateaus observed.}

It should be noted that for the ground state in particular, there are several states\footnotemark[5] we find to be exceptionally close in energy. Among the potential minimum energy states found using this method was the previously found HoAgGe magnetic ground state. So while we find several other states with similar energy, as the previously reported state was the only configuration of the correct magnetic space group, this is in agreement within the accuracy of our model. Any further degeneracy between these states may potentially be lifted through the employment of further neighbor interactions.

\footnotetext[5]{at least three.}

%Similarly, we also observe a $1/2$ state for fields applied in the X direction. While this step appears quite prominent, by inspecting the average energy per Ho atom of the competing states, it is clear that this $1/2$ state, while broad, is again extremely close in energy to the $1/3$ and $2/3$ states, indicating low stability. As the competing $\sqrt{3} \times \sqrt{3}$ and $2 \times 2$ cells have a difference in energy less than $40$ $\mu eV$ (i.e., $0.46$ $K$ in temperature units) for any value of $h$ between the $1/2$ and $2/3$ phase transitions. This energy difference is well within the range of thermal fluctuations to trigger these phase transitions, As well as being likely beyond the accuracy of some of the model assumptions, such as infinite anisotropy.

In total, using a RCS Search method, we found the ground state and all major magnetization steps observed in experiment ($0$, $1/5 (x)$, $1/5 (y)$, $1/3 (x)$, $1/3 (y)$, $2/3 (y)$, $3/4 (x)$, $1 (x)$, and $1 (y)$. Shown in supplemental\cite{supplementary}).
%Thus finding all experimentally observed magnetic states of the HoAgGe crystal. 
Among them, all states previously found empirically in Ref. \cite{doi:10.1126/science.aaw1666} were again seen and confirmed using this method.

%It should be pointed out however that this is also a place where the model may diverge from reality due to the infinite anisotropy assumed in the orientation of the Ho atoms. In a real crystal, due to the applied field we would expect some canting of the magnetic moments in the direction of the field. This will slightly change the energy of the system and due to the extremely closely competing states around several phase boundaries, and may result in the observation of more magnetic states.

\subsection{$J_\alpha-h$ phase diagrams and minimum energy curves}

\begin{figure}[!h]
\includegraphics[scale=0.6]{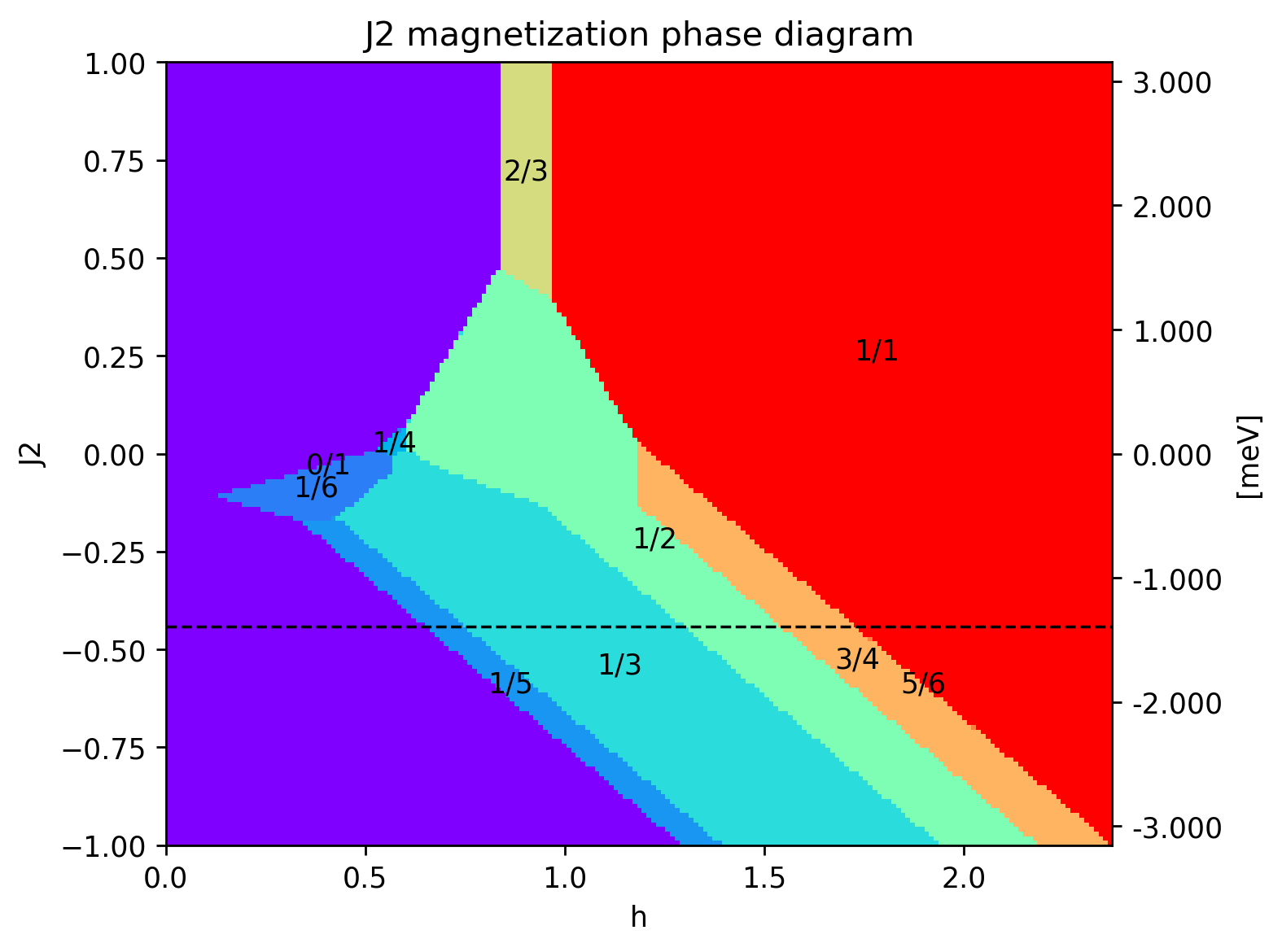}
\centering
\caption{$J_{2}$ Magnetic phase diagram for a magnetic field applied in the X direction. Colors represent average magnetization in the X direction. The dotted line is the theorized value for $J_2$. Again, all interactions normalized to $J_1=3.154$ meV}
\label{fig:phase_J2x}
\end{figure}

Using this procedure, it is relatively simple to vary the parameters, $J_\alpha$ and $h$, and search for which Hamiltonian of which magnetic unit cell has the lowest energy. By repeating this process and varying any one $J_\alpha$, phase diagrams of the system's behavior can be produced. As all exchange  parameters are normalized to the nearest neighbor, $J_1$, we calculate the phase diagrams for all interactions other than the first. This allows a visual inspection for the stability against any of these parameters and the behavior of the model. The resulting $J_2$ phase diagram for a field applied in the X direction is shown in Fig. \ref{fig:phase_J2x}.
From inspecting the remaining $J_{3a}$, $J_{3b}$, $J_4$, and $J_5$ magnetic phase diagrams\footnotemark[1], for fields applied in the X direction in particular, it is clear that almost no adjustments can easily be made to the internal coupling parameters without seriously altering the behavior of the model, in some cases introducing new magnetic plateaus, in others removing steps. 

\begin{figure}[!h]
\includegraphics[scale=0.6]{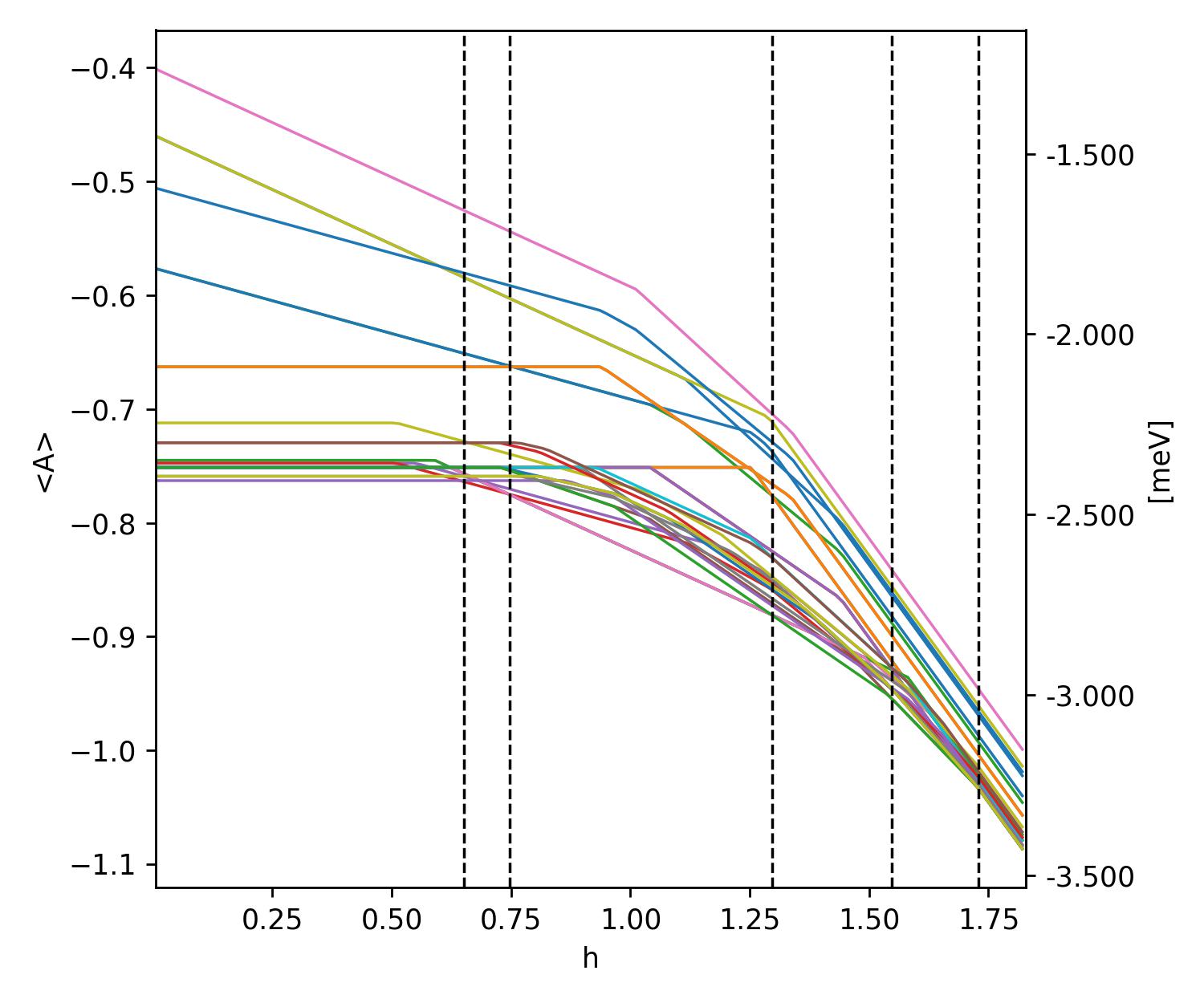}
\centering
\caption{Minimum average energy per Ho atom ($\left< A \right>$) for all 33 acceptable magnetic unit cells for a magnetic field applied in the X direction. Dotted lines indicating phase transitions.}
\label{fig:energy}
\end{figure}

The close competition between states can be seen directly in Fig \ref{fig:energy} as many states of different shapes and magnetization are found to be close in energy, especially near phase transitions. This indicates that any further adjustment to the model would likely require either the introduction of further neighbor interactions, new interactions, or a search over a larger RCS. 

Notably, this direct comparison of competing energy states cannot be done easily using a more standard Monte Carlo approach and further demonstrates the utility of this method.

\footnotetext[1]{$J_{3a}$, $J_{3b}$, $J_4$, $J_5$ phase diagrams are shown in the supplemental materials\cite{supplementary} and follow similar trends to the $J_2$ phase diagram}

\subsection{Monte Carlo}

Despite being systematic, a direct search of the Reduced-Configuration-Space only applies to zero temperature and  is limited in terms of the size of the magnetic unit cell. With this in mind, we have investigated the phase diagrams using Monte Carlo (MC) simulations. The classical dynamics of our model is adequately addressed with the Metropolis-Hastings algorithm and simulated annealing. The main challenge is that due to high frustration of our DFT-GF Hamiltonian the free energy landscape is crowded with metastable states;  we mitigate this by a combination of annealing, aggressive averaging and database bookkeeping of lowest energy states.

These MC simulations were performed on a periodic $18 \times 18$ lattice using the same annealing procedure at zero temperature for both the empirically found and the calculated parameters.

%Due to the extreme frustration of the system, modeling the behavior of the HoAgGe crystal using Monte Carlo simulations is difficult. This issue is only further compounded with the inclusion of further neighbor interactions. As was shown directly in Fig. \ref{fig:energy}, for any applied magnetic field, there may exist many states with extremely similar energies, especially for a large lattice, and it is extremely likely for a Monte Carlo simulation to find a metastable state instead of the true ground state.

\begin{figure}[!h]
\includegraphics[scale=0.5]{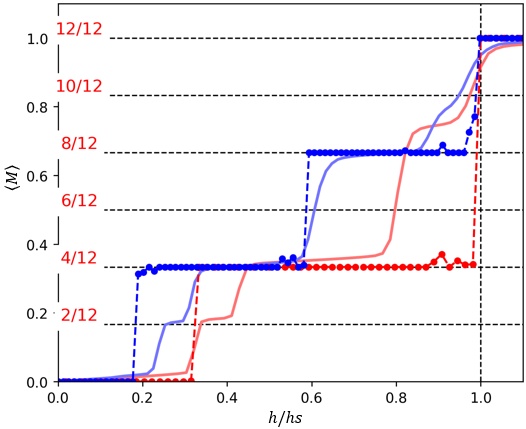}
\centering
\caption{Monte Carlo simulation of average Ho$^{3+}$ magnetization as an external magnetic field is applied perpendicular (X, \textcolor{red}{Red}) and parallel (Y, \textcolor{blue}{Blue}) to the easy axis calculated using empirical exchange parameters from Ref. \cite{doi:10.1126/science.aaw1666} (dashed), compared to experiment\cite{nirmal2024}.}
\label{fig:GW_MC_mag}
\end{figure}

\begin{figure}[!h]
\includegraphics[scale=0.5]{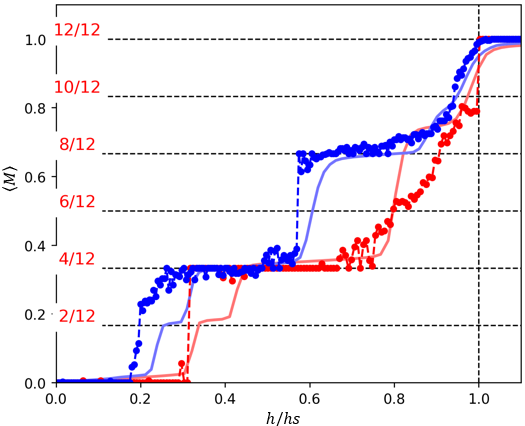}
\centering
\caption{Monte Carlo simulation of average Ho$^{3+}$ magnetization as an external magnetic field is applied perpendicular (X, \textcolor{red}{Red}) and parallel (Y, \textcolor{blue}{Blue}) to the easy axis calculated using DFT-GF first-principles exchange parameters (dashed), compared to experiment\cite{nirmal2024}.}
\label{fig:PH_MC_mag}
\end{figure}

MC simulations using the empirical exchange parameters result in clear magnetization steps (Fig. \ref{fig:GW_MC_mag}) which again match some, but not all of the steps observed experimentally, and are in agreement with the MC simulation reported in Ref. \cite{doi:10.1126/science.aaw1666}\footnotemark[6]. When using our first-principles parameters, which include $J_5$ bonds and distinguishing between $J_{3a}$ and $J_{3b}$ bonds, MC results in a less clear image of the magnetization plateaus (Fig. \ref{fig:PH_MC_mag}), but matches experimental observations far more closely for fields applied in both the X and Y directions\footnotemark[2]. 
Thus, the first principles parameters pass the test of being relevant in the thermodynamic limit better than the empirical ones.

\footnotetext[6]{The referenced work does use a slightly different Monte Carlo loop-algorithm simulation technique, however the Metropolis Hastings algorithm employed here also yielded results in sufficient agreement, meaning these technique differences are likely negligible in this system.}

The noise present in the Monte Carlo simulations performed with our calculated parameters demonstrates how much more frustrated the system becomes with the inclusion of further neighbor interactions. To 'smooth out' this noise would require exceedingly long simulations.

\footnotetext[2]{All Monte Carlo simulations across the different parameter sets were performed using the same simulation protocol.}

\section{conclusions}

By searching over a Reduced-Configuration-Space for the minimum energy of differently shaped magnetic unit cells of the twisted Kagome lattice and through the use of first principle DFT-GF calculations to determine exchange parameters, we have been able to reproduce the magnetization plateaus observed in experimentation. We have also determined that the employment of larger magnetic unit cells and further neighbor interactions are crucial in reproducing the behavior observed in experiment. 

In addition, we have shown the utility of a unique combination of approaches (DFT-GF and RCS Search method) to a  complicated system like HoAgGe twisted Kagome, with advantages over more standard methods. We note that it is possible that the full phase diagram could be better captured using magnetic unit cells larger than six times the single unit cell volume\footnotemark[4]. Other limitations include even longer-range exchange interactions (HoAgGe being a metal, we expect the long range exchange to be of RKKY type, i.e., only cubically decaying) and deviations from the ideal Ising model.

\footnotetext[4]{This was later confirmed using the GMU Cluster computer, HOPPER, for magnetic unit cells of eight times the unit cell volume.}

\begin{acknowledgments}
We are grateful to Nirmal Ghimire for attracting our attention to this compound, sharing their experimental data, and for many useful discussions.
This work was supported by the Office of Naval Research through grant \#N00014-23-1-2480.

\end{acknowledgments}

\bibliography{bib.bib}

\end{document}

\title{Modeling of a twisted-Kagome HoAgGe spin ice using Reduced-Configuration-Space Search and Density Functional Theory \\ 
    Supplemental materials}

\author{Gunnar F. Schwertfeger}
\affiliation{Department of Physics and Astronomy, George Mason University, Fairfax, VA
22030, USA}
\affiliation{Quantum Science and Engineering Center, George Mason University, Fairfax, VA
22030, USA}
\author{Po-Hao Chang}
\affiliation{Department of Physics and Astronomy, George Mason University, Fairfax, VA
22030, USA}
\affiliation{Quantum Science and Engineering Center, George Mason University, Fairfax, VA
22030, USA}
\author{Predrag Nikoli\'{c}}
\affiliation{Department of Physics and Astronomy, George Mason University, Fairfax, VA
22030, USA}
\affiliation{Quantum Science and Engineering Center, George Mason University, Fairfax, VA
22030, USA}
\author{Igor I. Mazin}
\affiliation{Department of Physics and Astronomy, George Mason University, Fairfax, VA
22030, USA}
\affiliation{Quantum Science and Engineering Center, George Mason University, Fairfax, VA
22030, USA}

\maketitle

%\clearpage
\section{Validation of local easy axis}
\begin{figure}
    \includegraphics[scale=0.9]{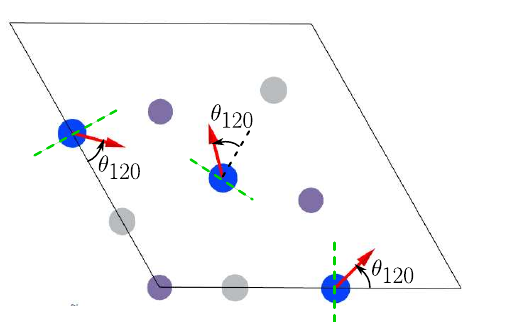}
    \caption{\label{fig:local_axis}
    The high-field $120$ noncollinear configuration considered in the analysis. The angle $\theta_{120}$ indicates the angle measured from the starting local axes, while green dashed lines indicate the local easy axes.}
\end{figure}

To further validate the experimentally determined local easy-axis orientation, we performed total energy calculations comparing several in-plane spin configurations. Based on the in-plane symmetry, We considered the high-field $120$ noncollinear configuration characterized by a $1\times1$ unit cell, as shown in Fig. \ref{fig:local_axis}. The magnetic state was then rotated rigidly in-plane, so that the exchange energy does not change, with angle $\theta$ defined in Fig. \ref{fig:local_axis}. The green dashed lines indicate the experimentally suggested local easy axes defined in Fig. 1 of the main text, which corresponds to $\theta_{120}=90^{\circ}$. Indeed, the configurations exhibit their lowest energy at $\theta=90^{\circ}$.

\begin{figure}
    \includegraphics[scale=0.9]{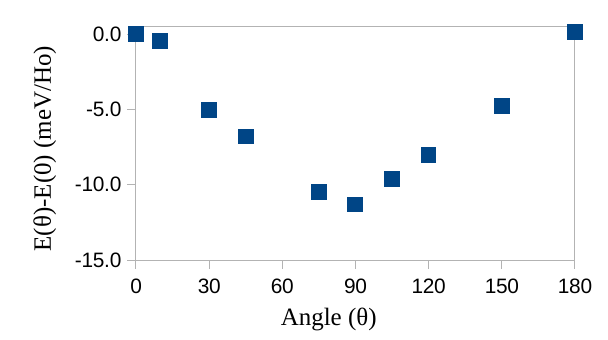}
    \caption{\label{fig:evstheta_in}
    The angle dependence of in-plane magnetic anisotropy of the high-field $120$ noncollinear configuration. 
    }
\end{figure}

%\begin{table}
%    \fontsize{8}{8}\selectfont
%    \def\arraystretch{2.0}% 
%    \begin{tabular}{r@{\extracolsep{0pt}.}lr@{\extracolsep{0pt}.}l|r@{\extracolsep{0pt}.}lr@{\extracolsep{0pt}.}l}
%    \multicolumn{4}{l|}{120-AF-config} & \multicolumn{4}{l}{In-plane FM}\tabularnewline
%    \hline 
%    \multicolumn{2}{c}{$\theta_{120}$} & \multicolumn{2}{c|}{$E(\theta_{120})$} & \multicolumn{2}{c}{$\theta_{FM}$} & \multicolumn{2}{c}{$E(\theta_{FM})$}\tabularnewline
%    \hline 
%    0&0 & 11&31 & 0&0 & 2&23\tabularnewline
%    45&0 & 5&28 & 45&0 & 1&44\tabularnewline
%    90&0 & 0&00 & 90&0 & 0&00\tabularnewline
%    \hline 
%    \end{tabular}
%    \caption{\label{tab:e_vs_q}Total energy versus rigid rotation angles. The
%    angles are defined in Fig. \ref{fig:local_axis}.}
%\end{table}
% ===========================================================================

\section{Magnetic States}

The specific magnetic states found using the RCS Search method for the HoAgGe crystal are shown below in Figs. \ref{fig:1X} - \ref{fig:4Y}. The average magnetization of the Ho atoms (normalized to the saturation field for ${\bf h }|| x$ and ${\bf h }|| y$ respectively) and the Magnetic Space Group (MSG) are listed as well. It should be noted that several of these states are not unique and characterize a family of symmetrically equivalent states. For example, there are 6 equivalent ground states representable on the $\sqrt{3} \times \sqrt{3}$ Magnetic Unit Cell. This number of equivalent states only increases with the consideration of differently shaped cells. However, as described in the main text, only one state actually needs to be considered.

\subsection{States found for magnetic fields applied in the X direction}

\begin{table}[h]
\begin{center}
\caption{
    Hamiltonian equations for minimum energy states for fields applied in the X direction}
    \renewcommand*{\arraystretch}{1.4}
    \setlength{\tabcolsep}{8.5pt}
    \begin{tabular}{c | c}
    \hline  
     Figure & H/N\tabularnewline
    \hline 
    Fig. \ref{fig:1X} & H/N = -1/3 $J_1$ + 1/2 $J_2$ + 2/3 $J_{3a}$ + 1/3 $J_{3b}$ + 1/2 $J_4$ + -1/3 $J_5$ + 0/1 $h_x$ + 0/1 $h_y$ + 0/1 $h_z$ + -12/365 $g_{dd}$\tabularnewline
    \hline
    Fig. \ref{fig:2X} & H/N = -1/3 $J_1$ + 11/30 $J_2$ + 2/5 $J_{3a}$ + 3/5 $J_{3b}$ + 11/30 $J_4$ + -1/3 $J_5$ + -$\sqrt{3}/15$ $h_x$ + -1/15 $h_y$ + 0/1 $h_z$ + -13/444 $g_{dd}$\tabularnewline
    \hline
    Fig. \ref{fig:3X} & H/N = -1/3 $J_1$ + 5/18 $J_2$ + 2/3 $J_{3a}$ + 1/3 $J_{3b}$ + 5/18 $J_4$ + -1/3 $J_5$ + -$\sqrt{3}/9$ $h_x$ + -1/9 $h_y$ + 0/1 $h_z$ + -5/186 $g_{dd}$\tabularnewline
    \hline
    Fig. \ref{fig:4X} & H/N = -1/3 $J_1$ + 1/6 $J_2$ + 0/1 $J_{3a}$ + 1/3 $J_{3b}$ + 0/1 $J_4$ + 0/1 $J_5$ + -$\sqrt{3}/6$ $h_x$ + 0/1 $h_y$ + 0/1 $h_z$ + -1/41 $g_{dd}$\tabularnewline
    \hline
    Fig. \ref{fig:5X} & H/N = -1/3 $J_1$ + 0/1 $J_2$ + -2/3 $J_{3a}$ + 0/1 $J_{3b}$ + 0/1 $J_4$ + 0/1 $J_5$ + -$\sqrt{3}/4$ $h_x$ + -1/12 $h_y$ + 0/1 $h_z$ + -19/845 $g_{dd}$\tabularnewline
    \hline
    Fig. \ref{fig:6X} & H/N = -1/3 $J_1$ + -1/6 $J_2$ + -4/3 $J_{3a}$ + -1/3 $J_{3b}$ + -1/6 $J_4$ + -1/3 $J_5$ + -$\sqrt{3}/3$ $h_x$ + 0/1 $h_y$ + 0/1 $h_z$ + -13/643 $g_{dd}$\tabularnewline
    \hline 
    \end{tabular}
   \end{center}
   \label{table:GW_params}
\end{table}

\begin{figure}[h!]
\centering
\includegraphics[scale=0.5]{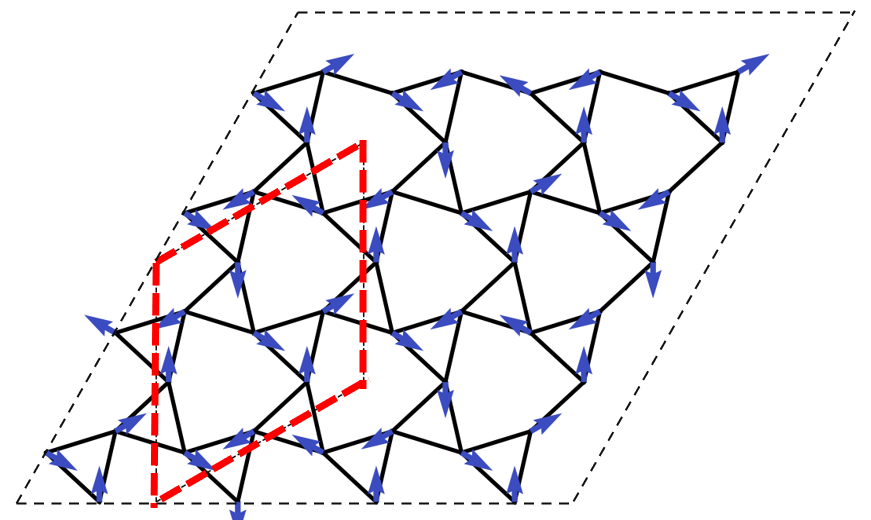}
\includegraphics[scale=0.5]{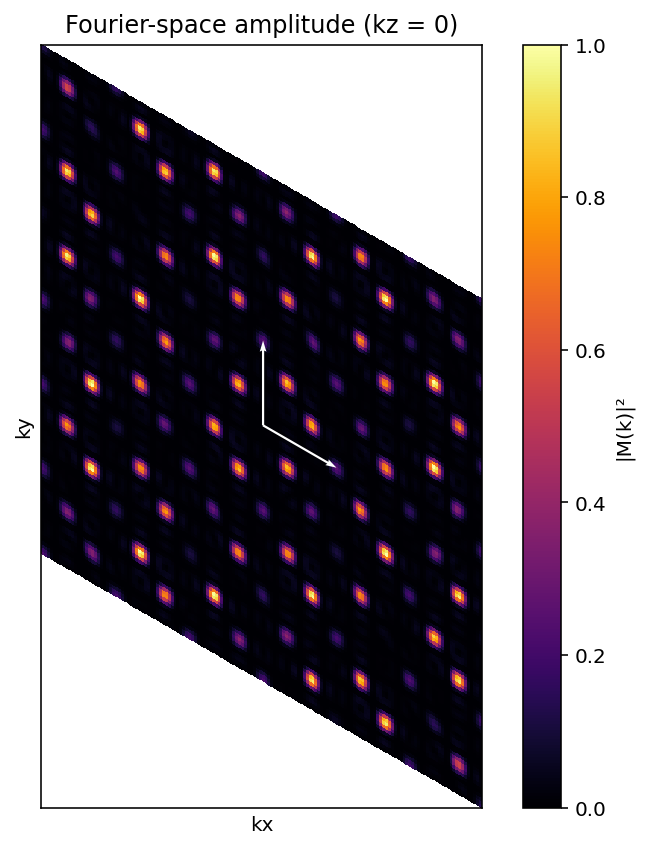}
\caption{The ground ($\left< M \right>_x = 0$) magnetic state for the Ho crystal. Its magnetic fourier space amplitude is also shown. (MSG: P$\bar{6}$'m2')}
\label{fig:1X}
\end{figure}

%\begin{figure}[h!]
%\centering
%\includegraphics[scale=0.5]{Figures_supplemental/ground_state_MFA}
%\caption{The magnetic Fourier amplitude of the ground state}
%\label{fig:MFA}
%\end{figure}

\begin{figure}[h!]
\centering
\includegraphics[scale=0.5]{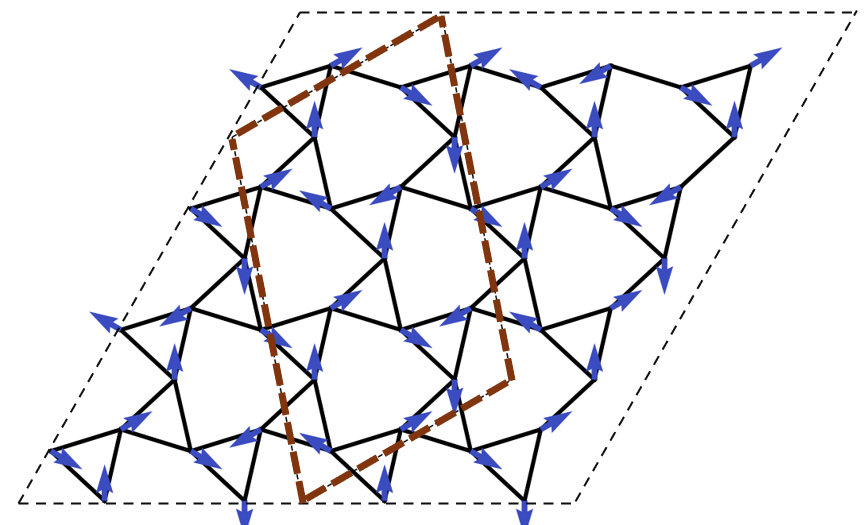}
\caption{The $\left< M \right>_x = 1/5$ magnetic state for the Ho crystal. (MSG: Am'm2')}
\label{fig:2X}
\end{figure}

\begin{figure}[h!]
\centering
\includegraphics[scale=0.5]{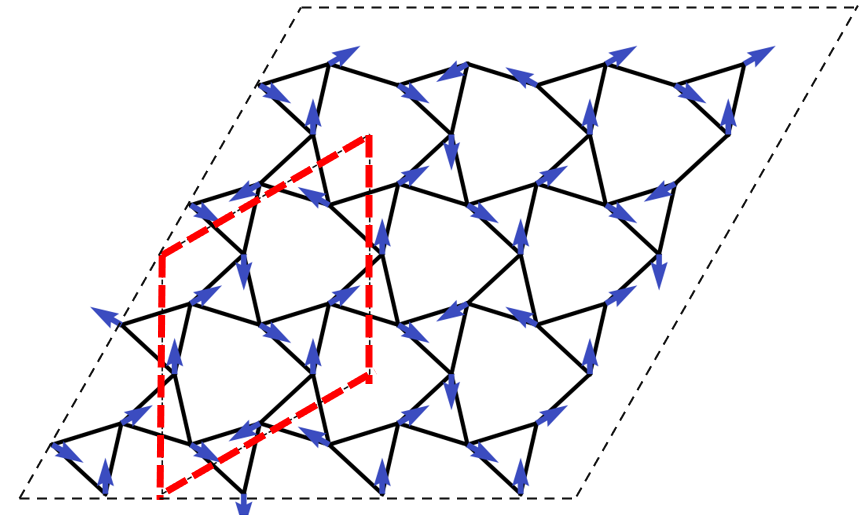}
\caption{The $\left< M \right>_x = 1/3$ magnetic state for the Ho crystal. (MSG: Am'm2')}
\label{fig:3X}
\end{figure}

\begin{figure}[h!]
\centering
\includegraphics[scale=0.5]{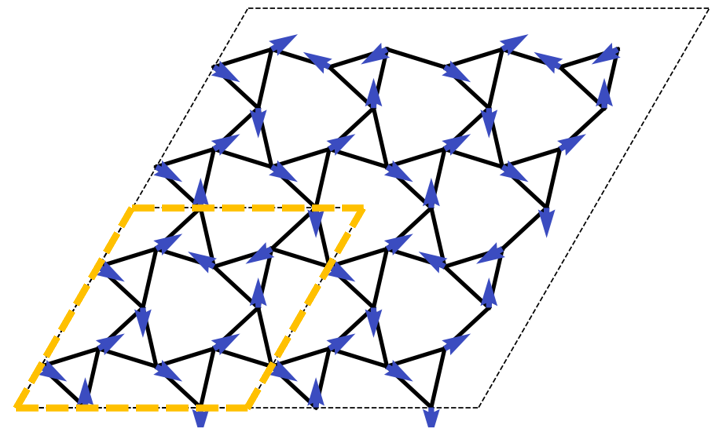}
\caption{The $\left< M \right>_x = 1/2$ magnetic state for the Ho crystal. (MSG: Pm')}
\label{fig:4X}
\end{figure}

\begin{figure}[h!]
\centering
\includegraphics[scale=0.5]{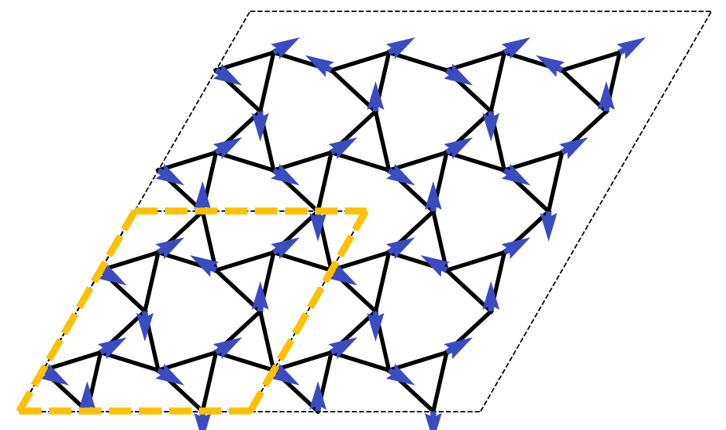}
\caption{The $\left< M \right>_x = 3/4$ magnetic state for the Ho crystal. (MSG: Pm')}
\label{fig:5X}
\end{figure}

\begin{figure}[h!]
\centering
\includegraphics[scale=0.5]{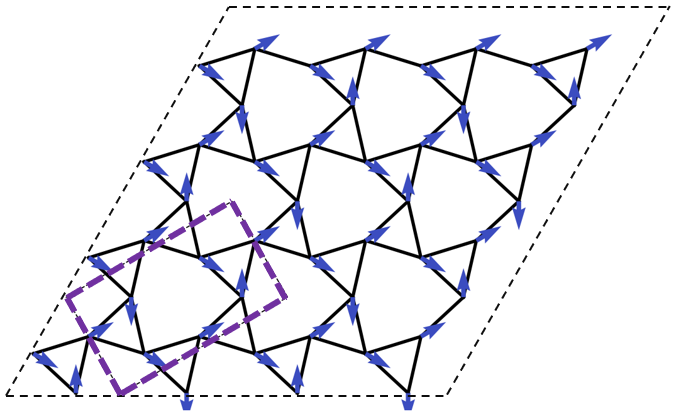}
\caption{The saturated ($\left< M \right>_x = 1/1$) magnetic state for the Ho crystal. (MSG: Pm')}
\label{fig:6X}
\end{figure}

\clearpage
\subsection{States found for magnetic fields applied in the Y direction}

\begin{table}[h]
\begin{center}
\caption{
    Hamiltonian equations for minimum energy states for fields applied in the Y direction}
    \renewcommand*{\arraystretch}{1.4}
    \setlength{\tabcolsep}{8.5pt}
    \begin{tabular}{c | c}
    \hline  
     Figure & H/N\tabularnewline
    \hline 
    Fig. \ref{fig:1Y} & H/N = -1/3 $J_1$ + 1/2 $J_2$ + 2/3 $J_{3a}$ + 1/3 $J_{3b}$ + 1/2 $J_4$ + -1/3 $J_5$ + 0/1 $h_x$ + 0/1 $h_y$ + 0/1 $h_z$ + -12/365 $g_{dd}$\tabularnewline
    \hline
    Fig. \ref{fig:2Y} & H/N = -1/3 $J_1$ + 11/30 $J_2$ + 2/5 $J_{3a}$ + 3/5 $J_{3b}$ + 11/30 $J_4$ + -1/3 $J_5$ + 0/1 $h_x$ + -2/15 $h_y$ + 0/1 $h_z$ + -13/444 $g_{dd}$\tabularnewline
    \hline
    Fig. \ref{fig:3Y} & H/N = -1/3 $J_1$ + 5/18 $J_2$ + 2/3 $J_{3a}$ + 1/3 $J_{3b}$ + 5/18 $J_4$ + -1/3 $J_5$ + 0/1 $h_x$ + -2/9 $h_y$ + 0/1 $h_z$ + -5/186 $g_{dd}$\tabularnewline
    \hline
    Fig. \ref{fig:4Y} & H/N = -1/3 $J_1$ + 1/18 $J_2$ + -2/9 $J_{3a}$ + -1/9 $J_{3b}$ + 1/18 $J_4$ + -1/3 $J_5$ + 0/1 $h_x$ + -4/9 $h_y$ + 0/1 $h_z$ + -16/701 $g_{dd}$\tabularnewline
    \hline
    Fig. \ref{fig:5Y} & H/N = -1/3 $J_1$ + -1/6 $J_2$ + -2/1 $J_{3a}$ + -1/1 $J_{3b}$ + -1/6 $J_4$ + -1/3 $J_5$ + 0/1 $h_x$ + -2/3 $h_y$ + 0/1 $h_z$ + -13/628 $g_{dd}$\tabularnewline
    \hline
    \end{tabular}
   \end{center}
   \label{table:GW_params}
\end{table}

\begin{figure}[h!]
\centering
\includegraphics[scale=0.5]{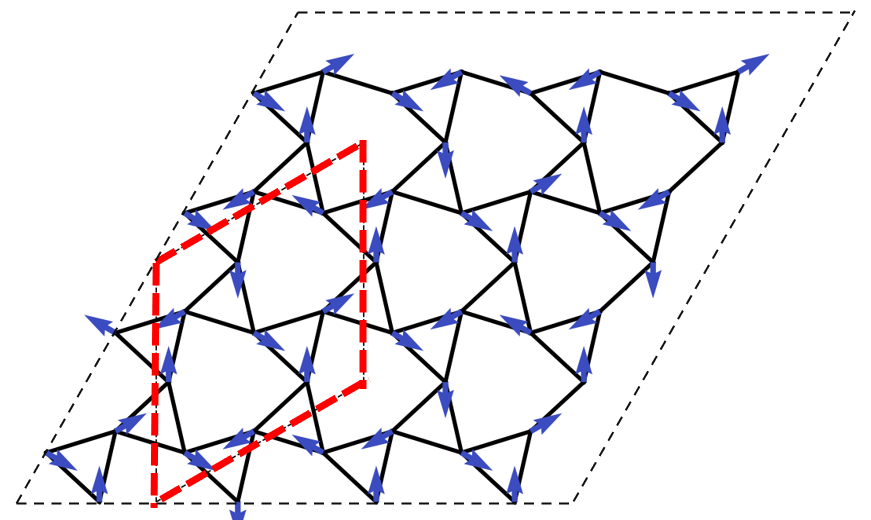}
\caption{The ground ($\left< M \right>_y = 0$) magnetic state for the Ho crystal. (MSG: P$\bar{6}$'m2')}
\label{fig:1Y}
\end{figure}

\begin{figure}[h!]
\centering
\includegraphics[scale=0.5]{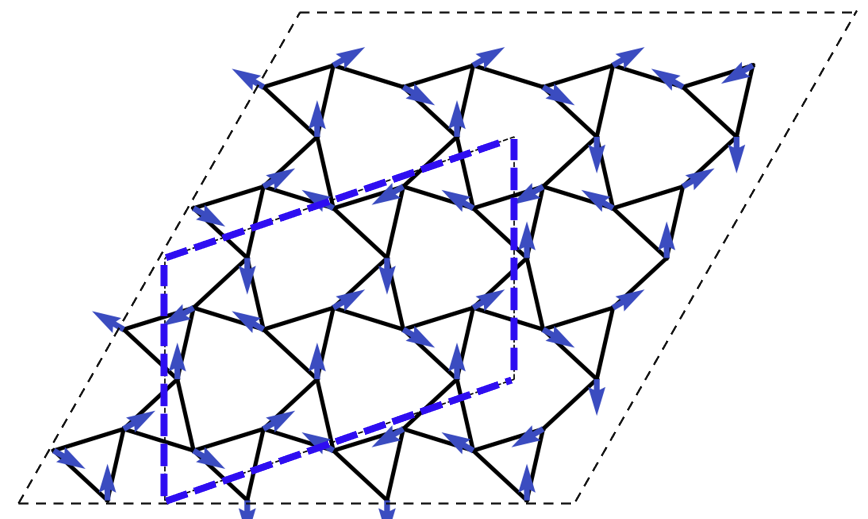}
\caption{The $\left< M \right>_y = 1/5$ magnetic state for the Ho crystal. (MSG: Am'm2')}
\label{fig:2Y}
\end{figure}

\begin{figure}[h!]
\centering
\includegraphics[scale=0.5]{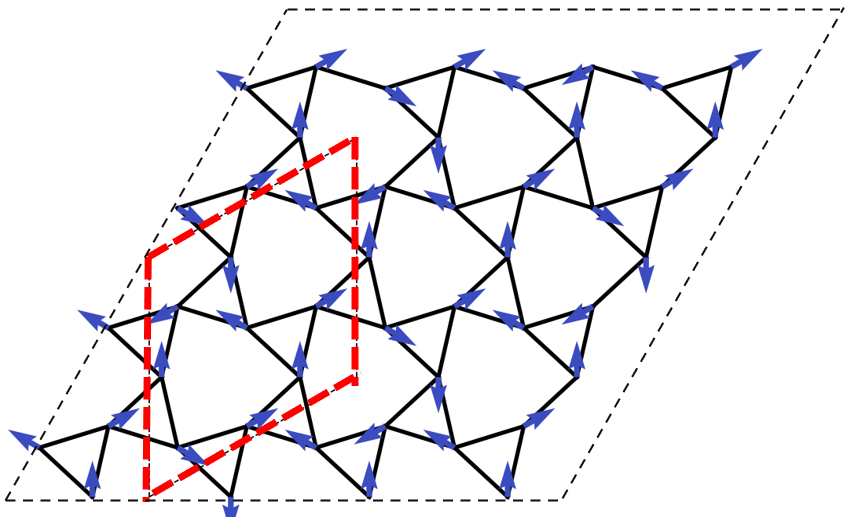}
\caption{The $\left< M \right>_y = 1/3$ magnetic state for the Ho crystal. (MSG: Am'm2')}
\label{fig:3Y}
\end{figure}

\begin{figure}[h!]
\centering
\includegraphics[scale=0.5]{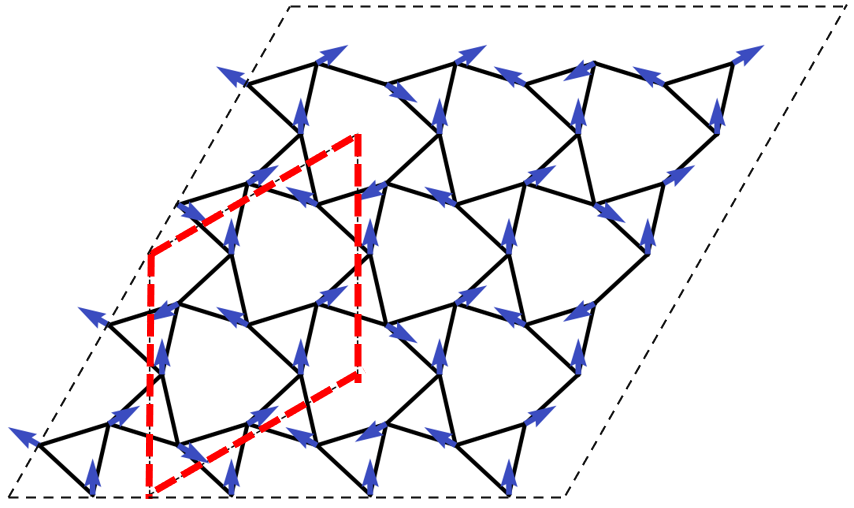}
\caption{The $\left< M \right>_y = 2/3$ magnetic state for the Ho crystal. (MSG: Am'm2')}
\label{fig:4Y}
\end{figure}

\begin{figure}[h!]
\centering
\includegraphics[scale=0.5]{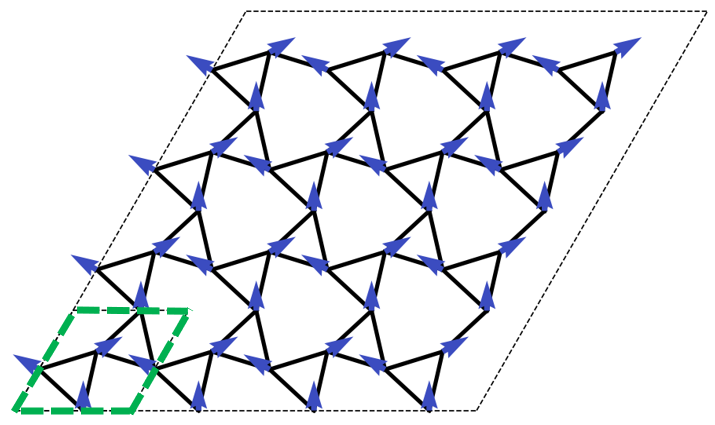}
\caption{The saturated ($\left< M \right>_y = 1/1$) magnetic state for the Ho crystal. (MSG: Am'm2')}
\label{fig:5Y}
\end{figure}

\clearpage
\section{Phase Diagrams}

The complete $J_\alpha - h$ magnetic and state phase diagrams are shown below. These were again found using an RCS Search method.

\subsection{Magnetic Phase Diagrams}

\begin{figure*}[!h]
\includegraphics[width=0.195\linewidth]{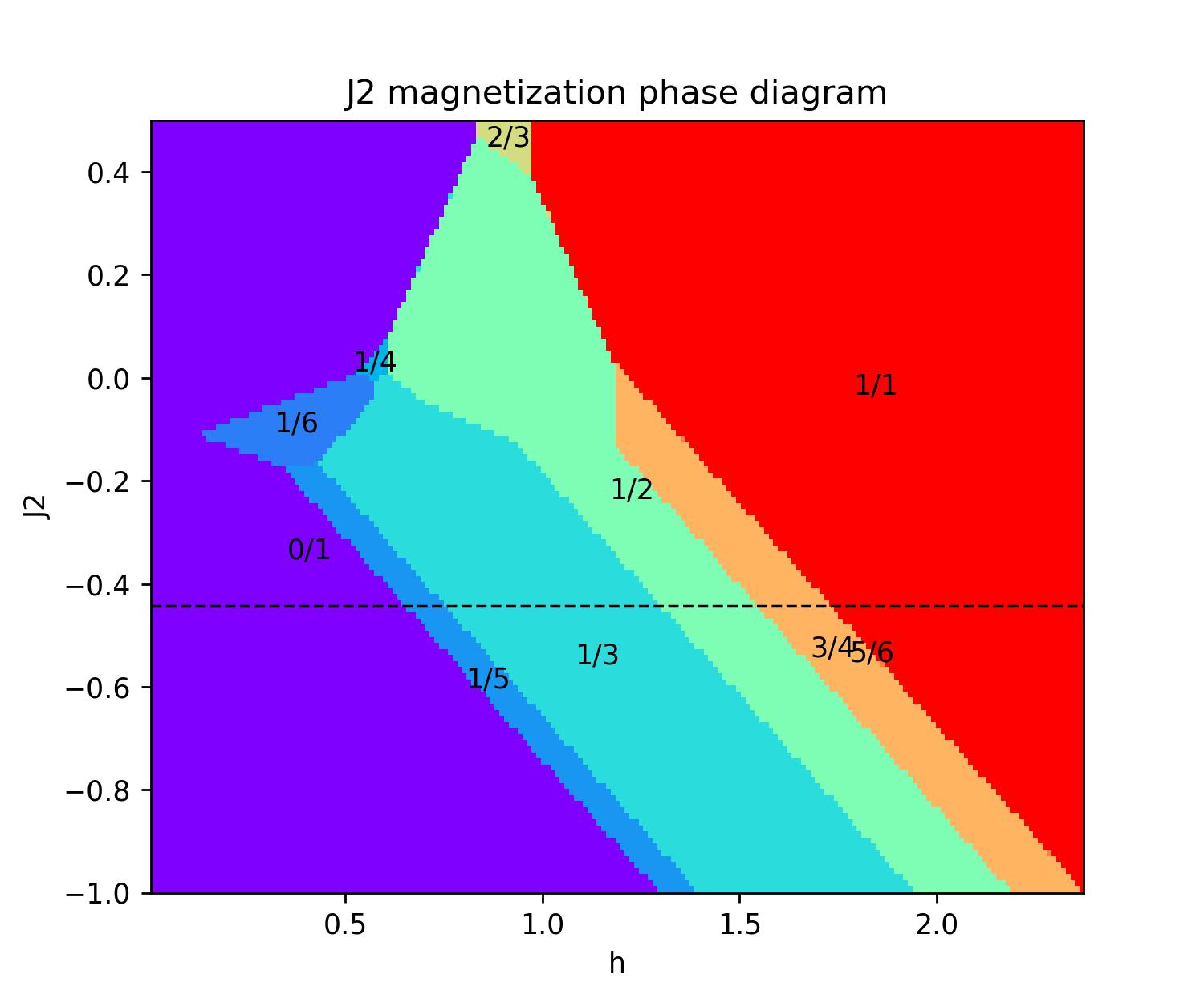}
\includegraphics[width=0.195\linewidth]{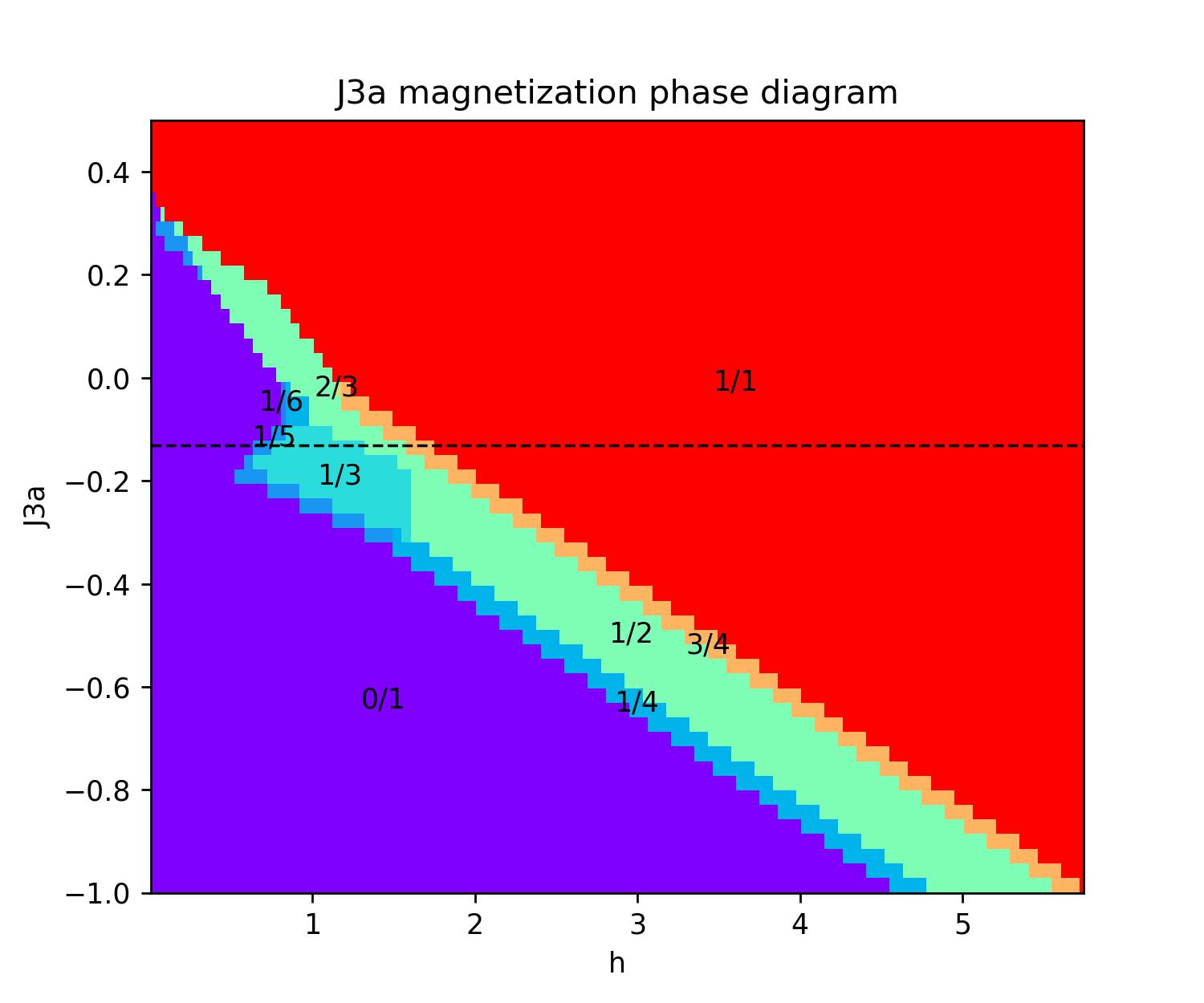}
\includegraphics[width=0.195\linewidth]{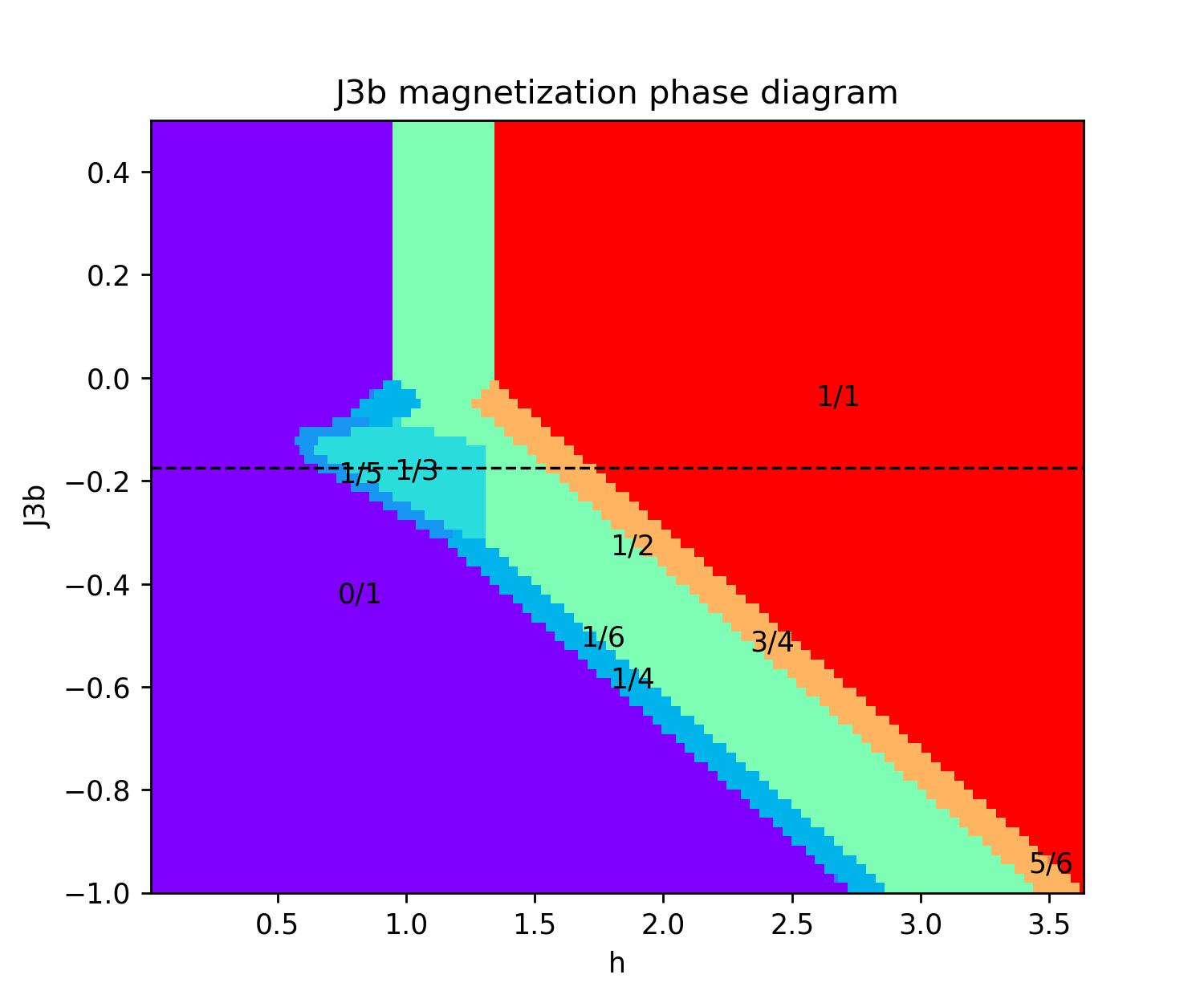}
\includegraphics[width=0.195\linewidth]{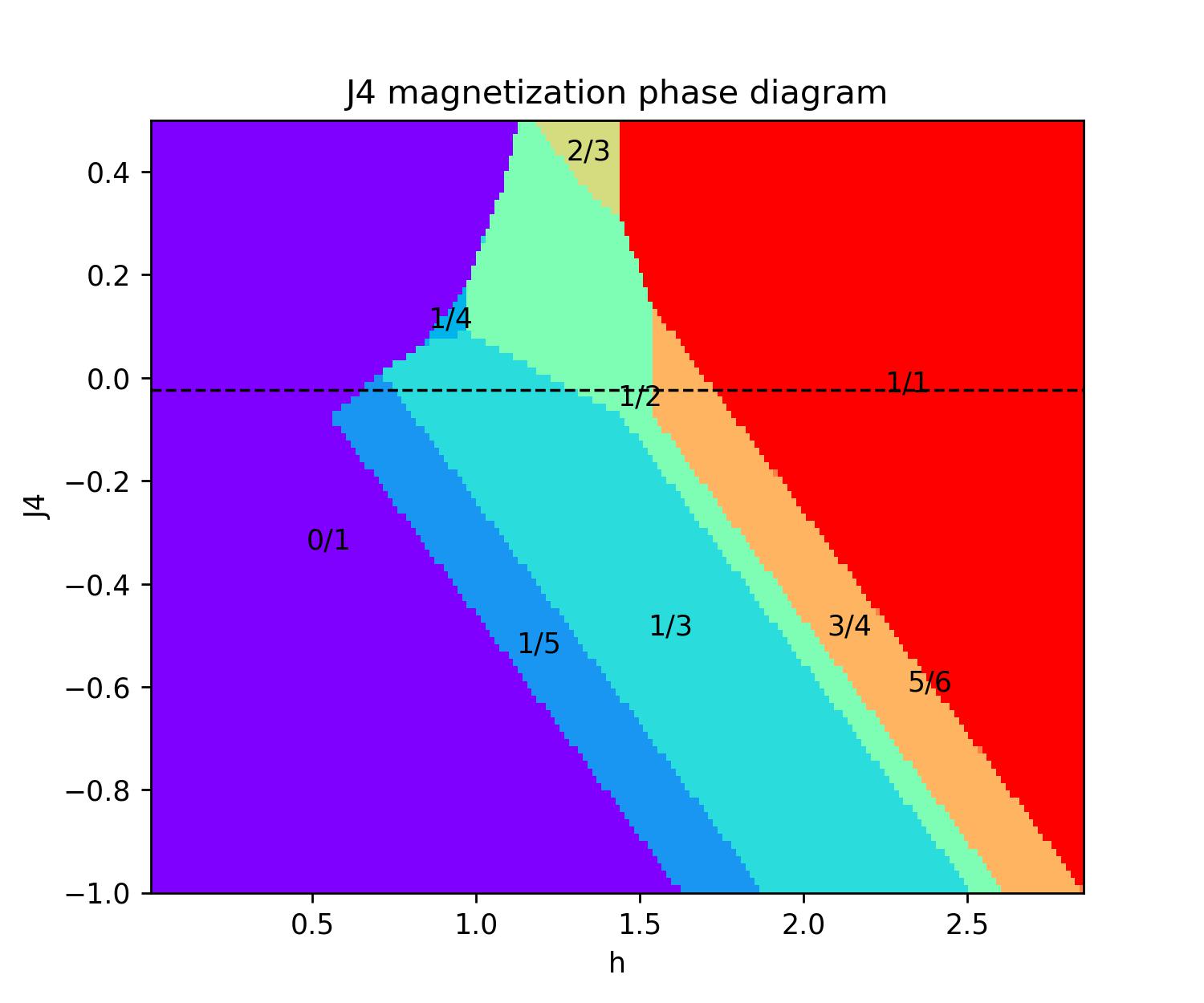}
\includegraphics[width=0.195\linewidth]{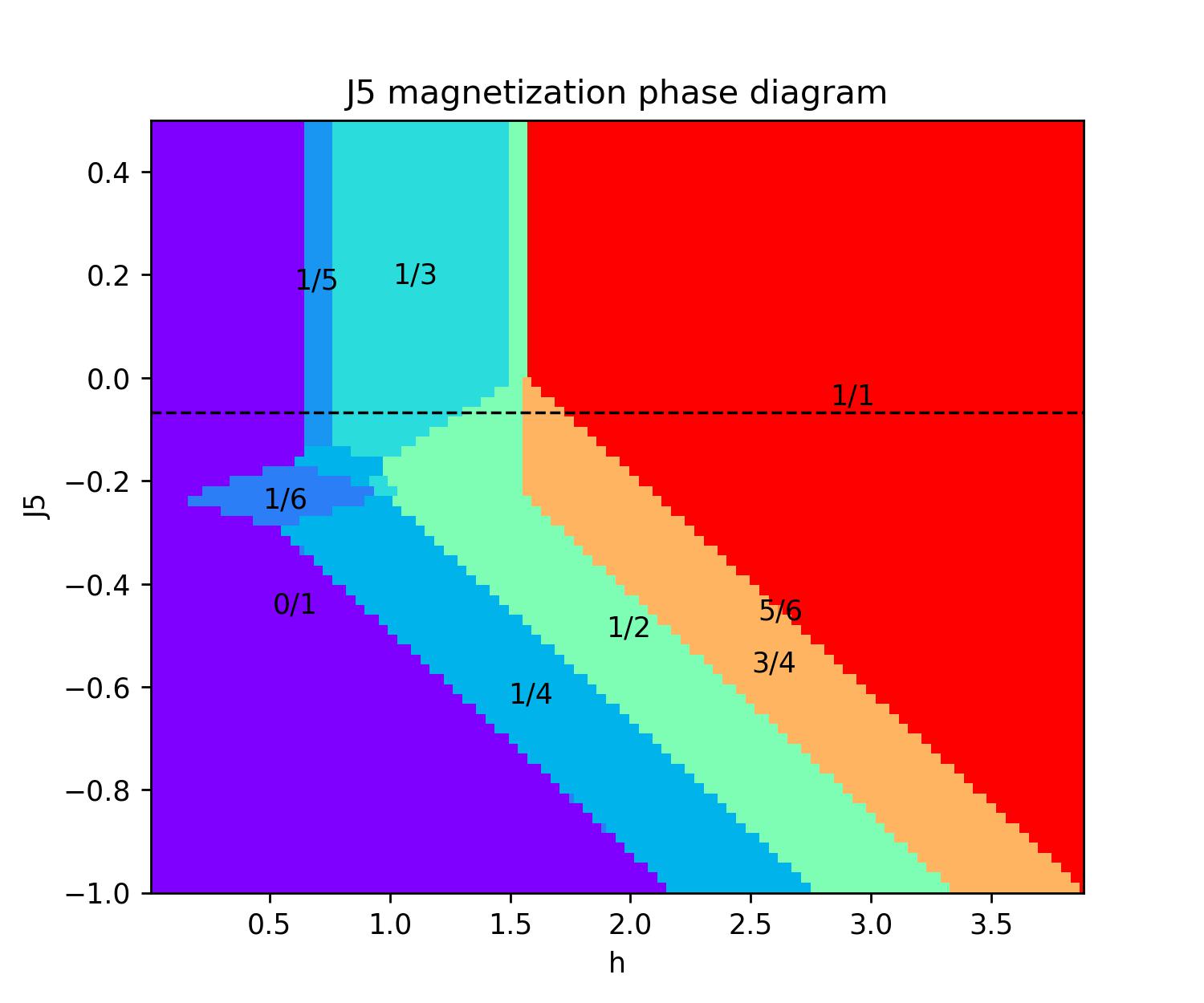}
\includegraphics[width=0.195\linewidth]{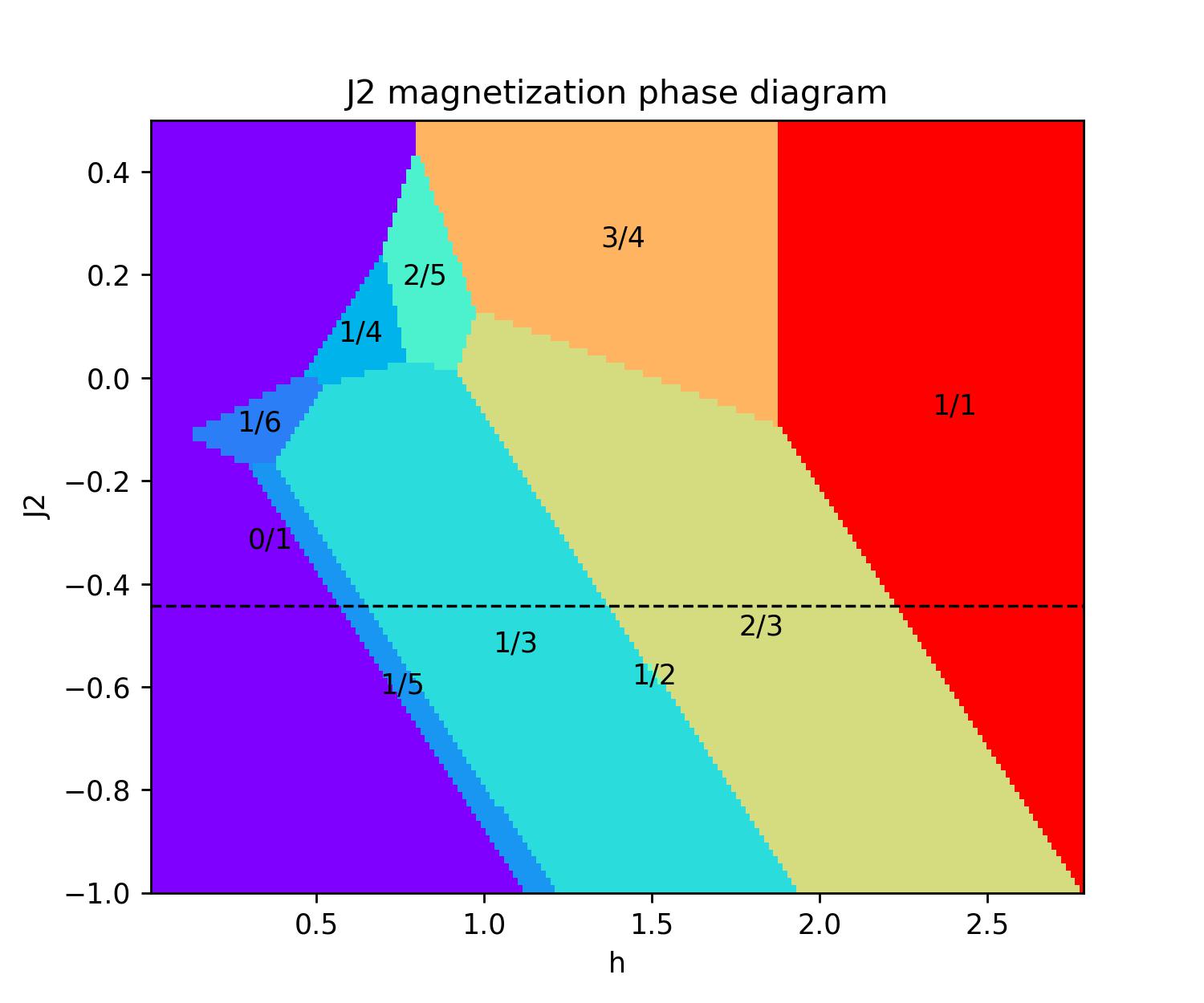}
\includegraphics[width=0.195\linewidth]{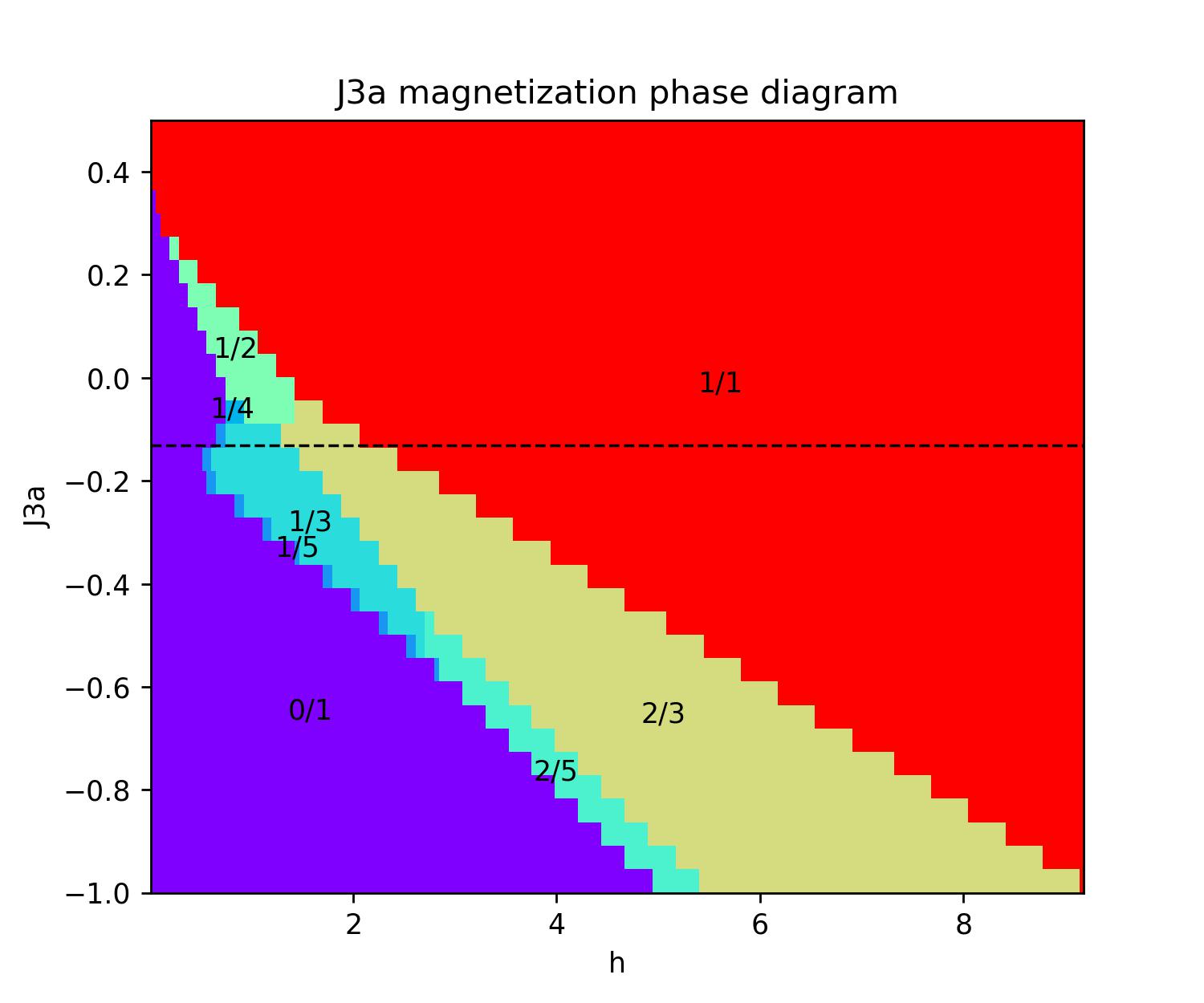}
\includegraphics[width=0.195\linewidth]{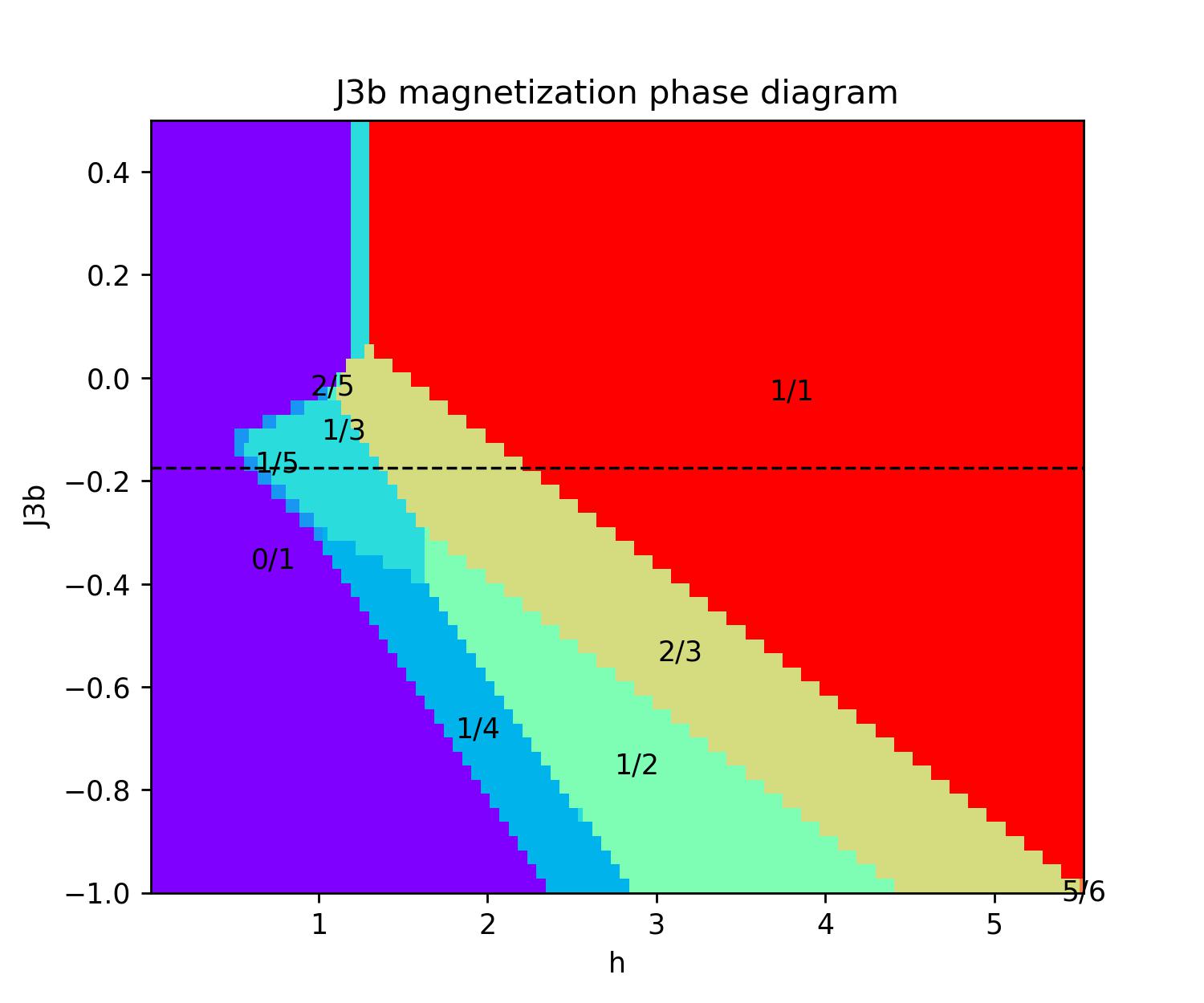}
\includegraphics[width=0.195\linewidth]{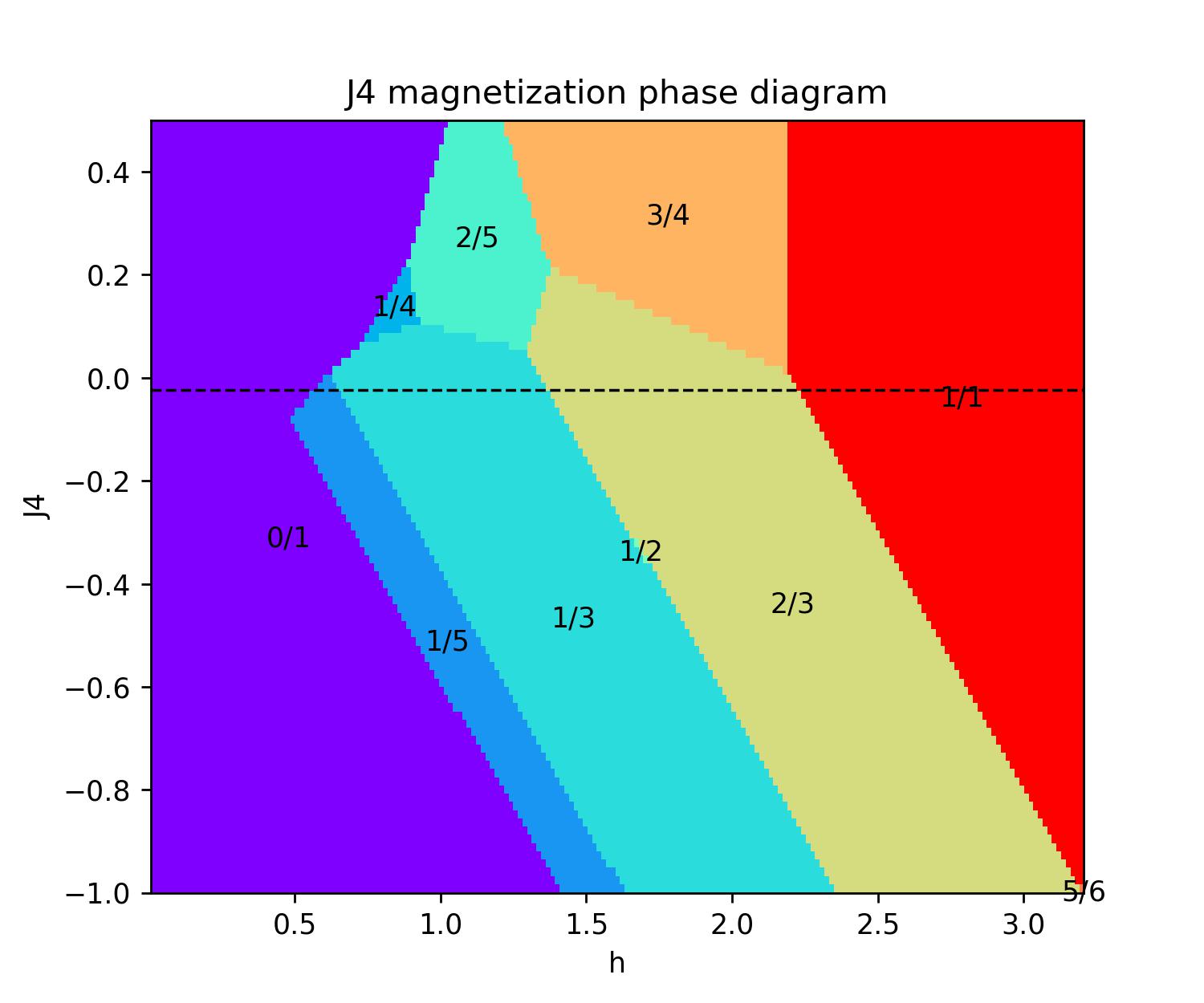}
\includegraphics[width=0.195\linewidth]{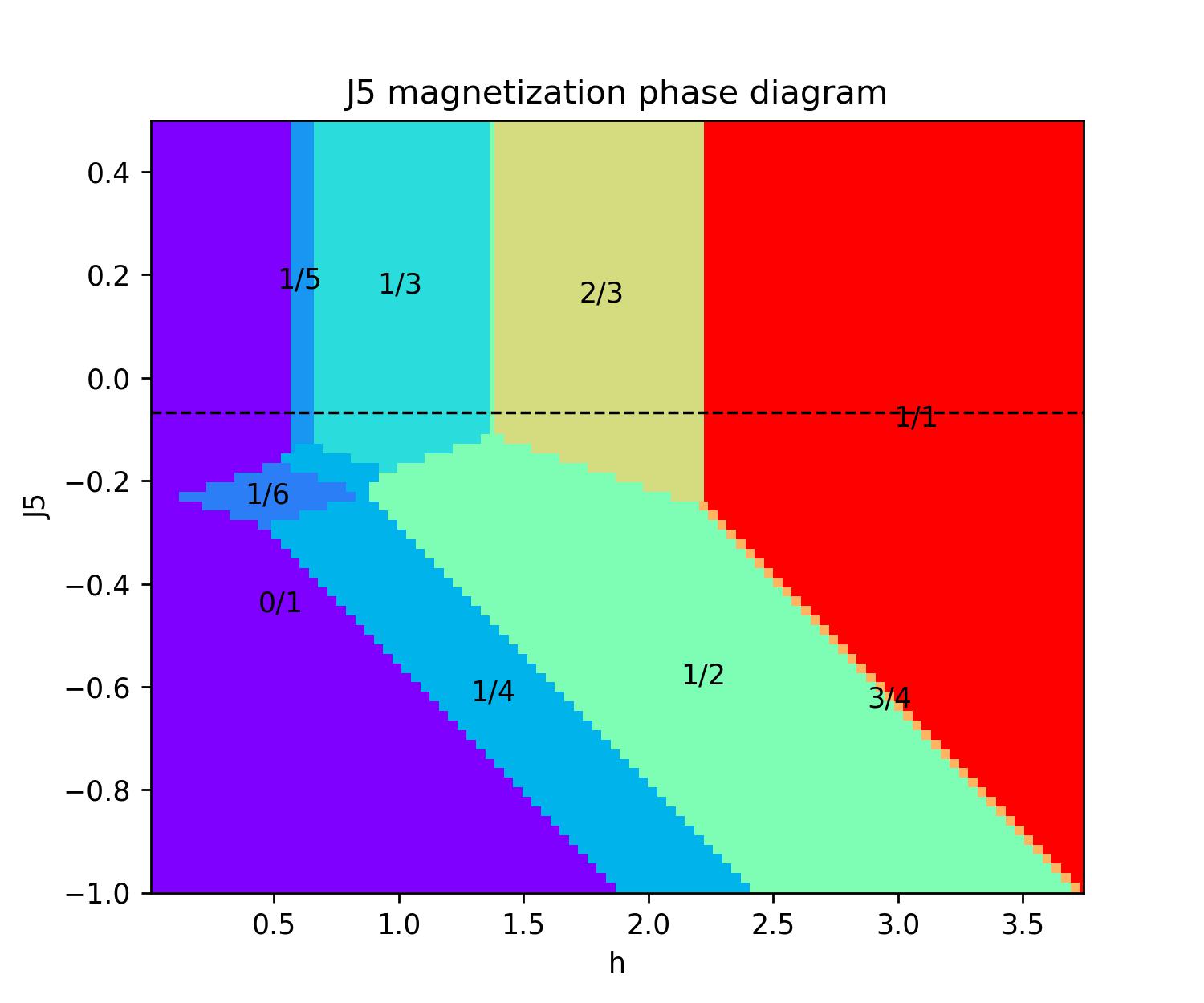}

\centering
\caption{Average magnetization phase diagrams for bond parameters $J_{2}, J_{3a}, J_{3b}, J_{4}, J_{5}$ for a magnetic field applied in the X (top) and Y (bottom) directions. Dotted line indicating the theorized value for $J_\alpha$}
\label{fig:PH_diag_frac}
\end{figure*}

\begin{figure*}[!h]
\includegraphics[width=0.195\linewidth]{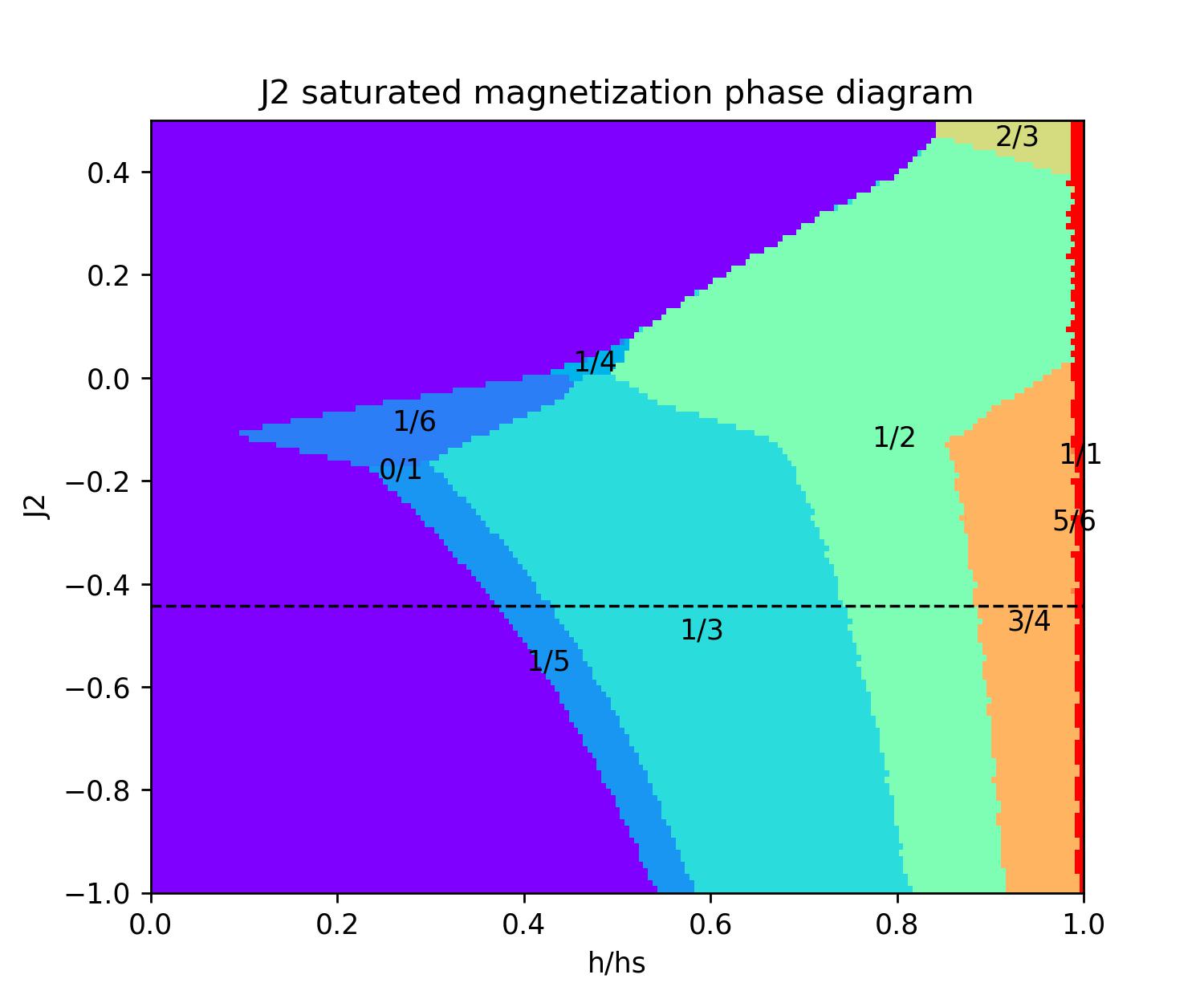}
\includegraphics[width=0.195\linewidth]{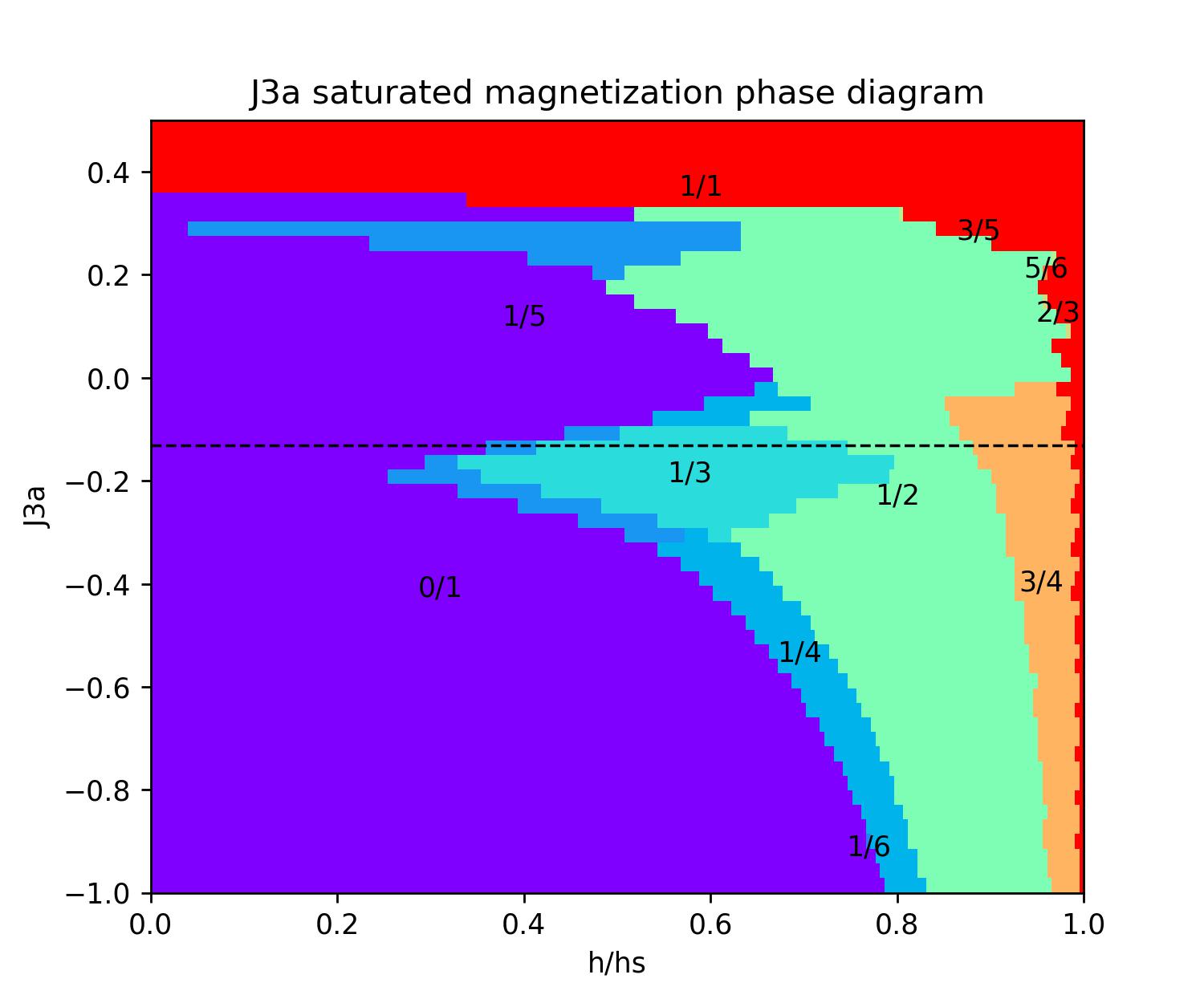}
\includegraphics[width=0.195\linewidth]{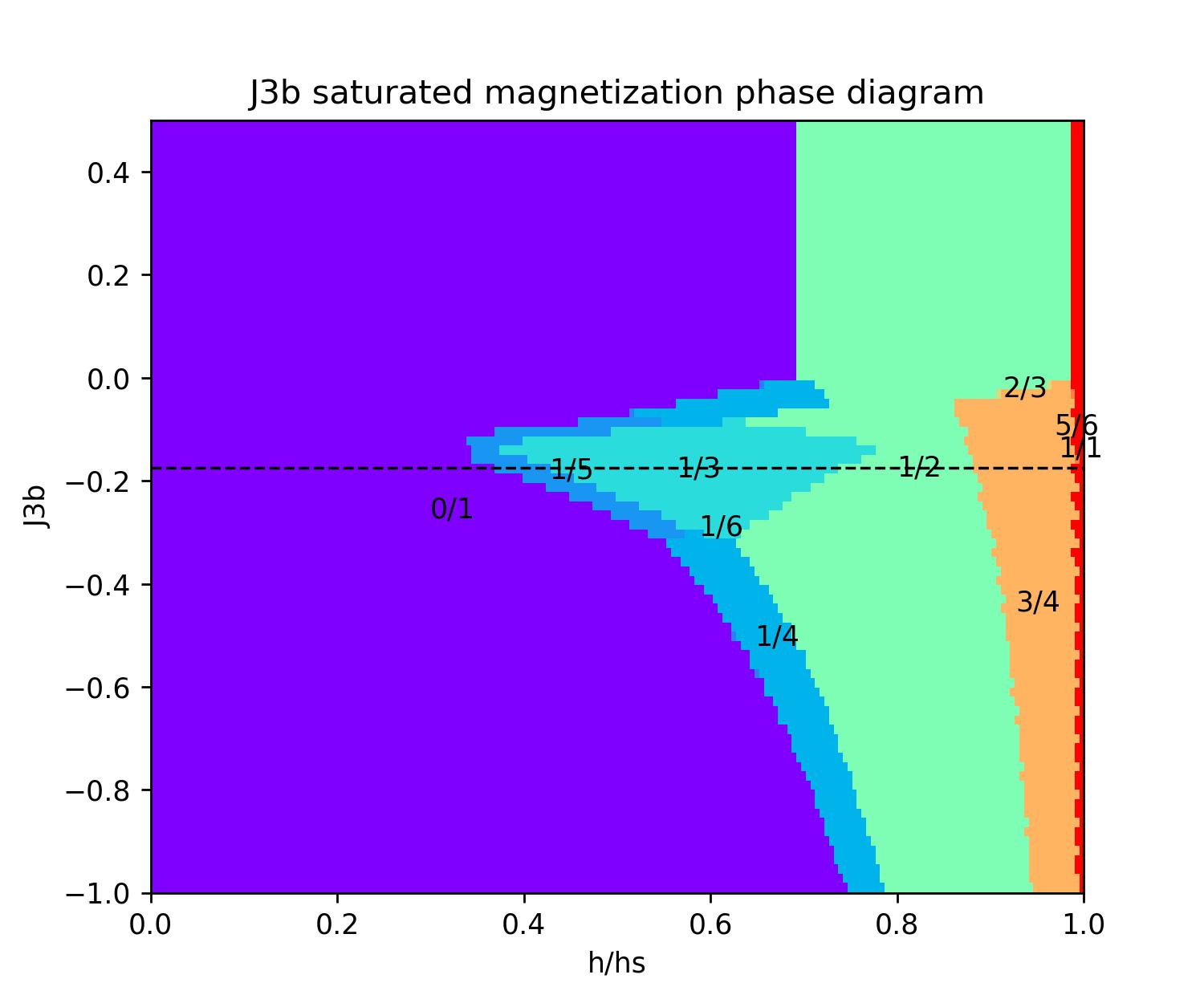}
\includegraphics[width=0.195\linewidth]{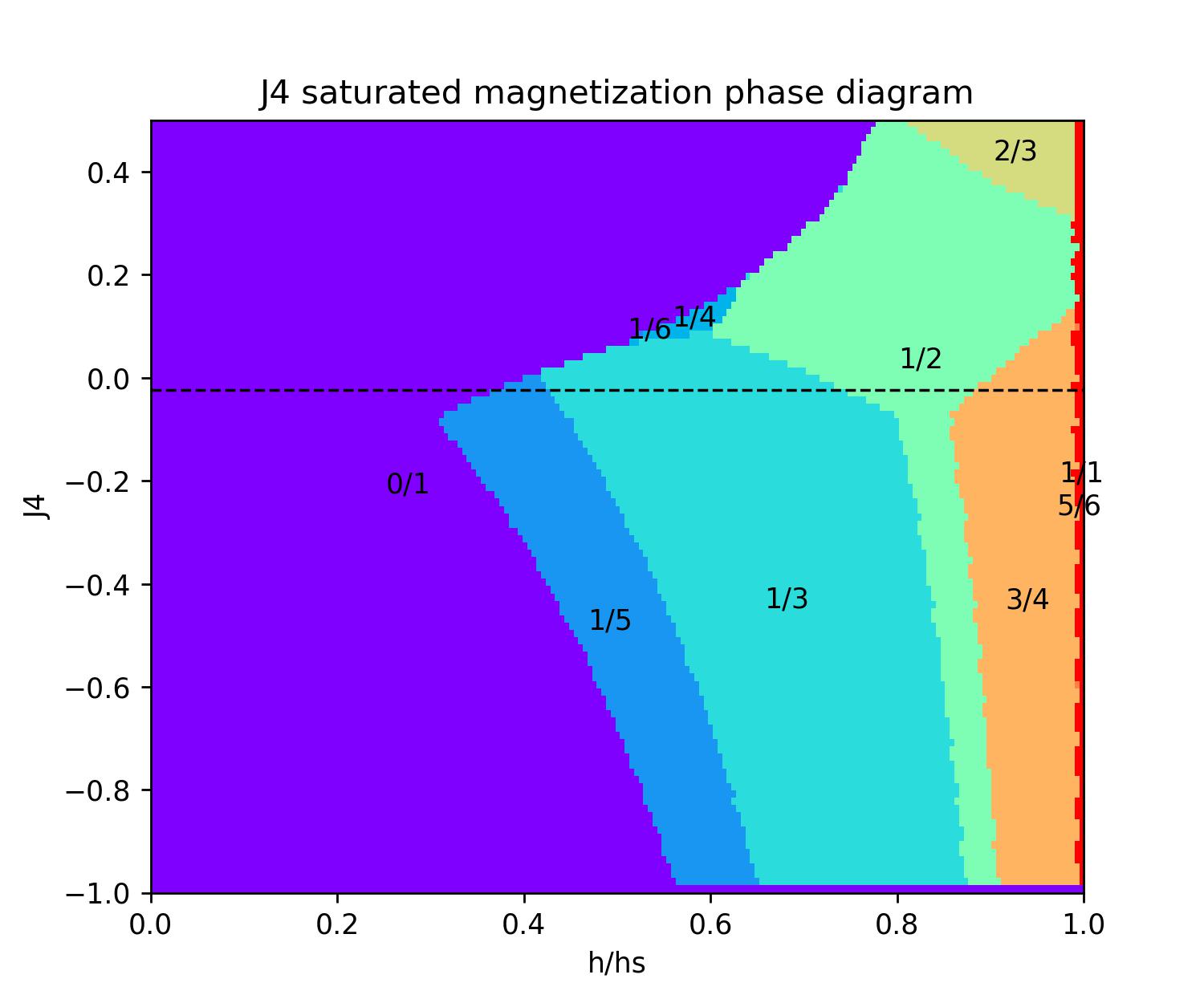}
\includegraphics[width=0.195\linewidth]{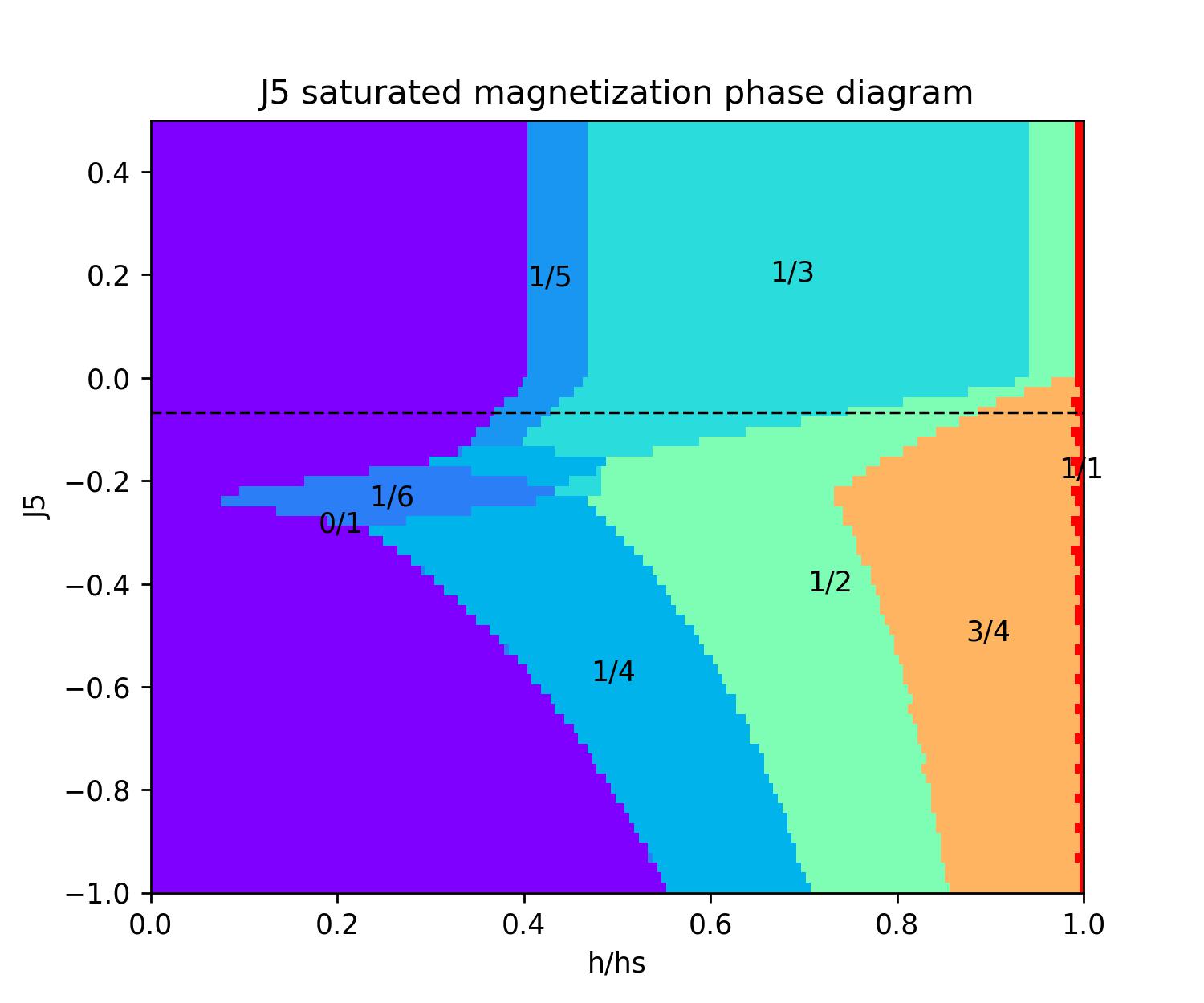}
\includegraphics[width=0.195\linewidth]{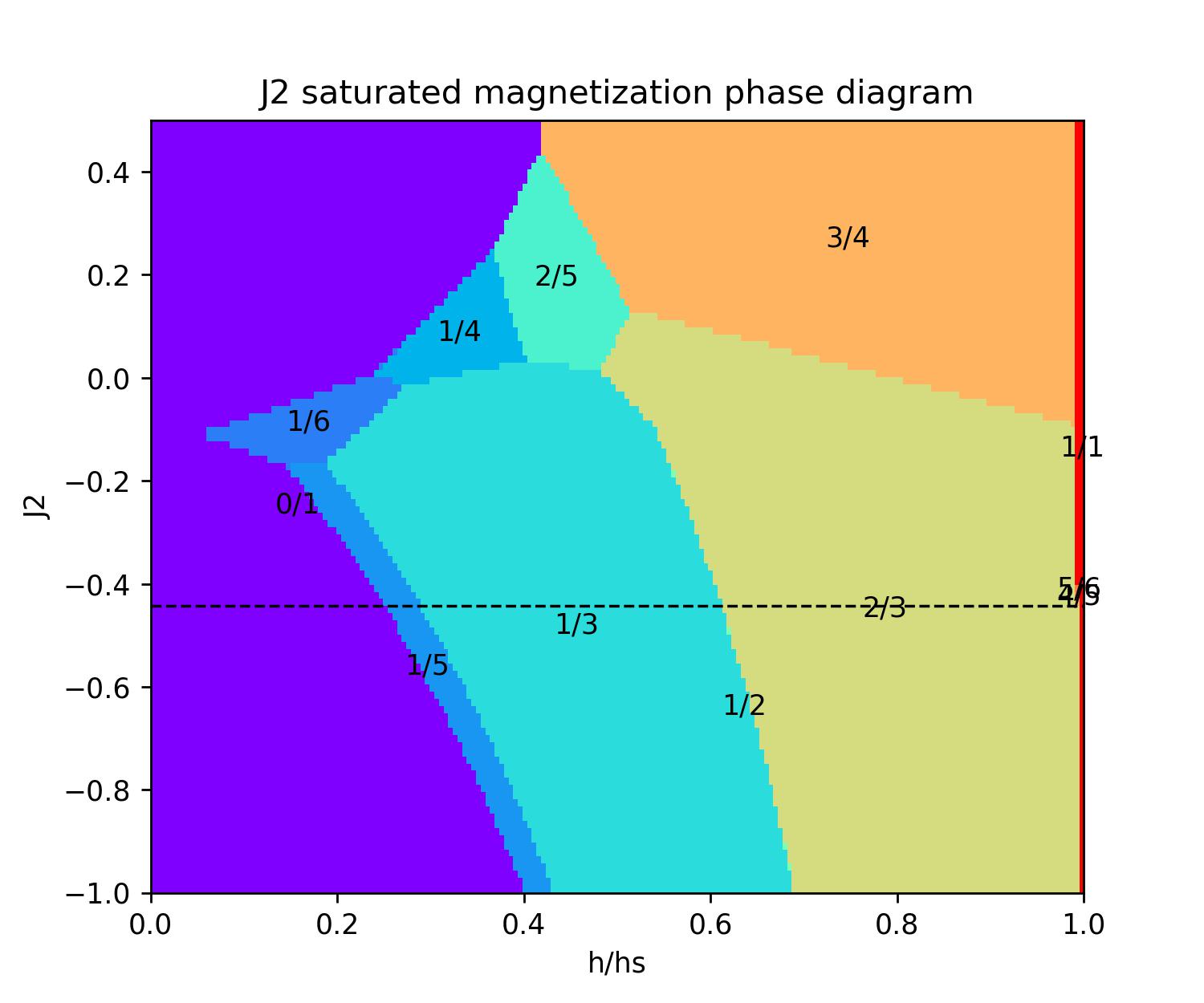}
\includegraphics[width=0.195\linewidth]{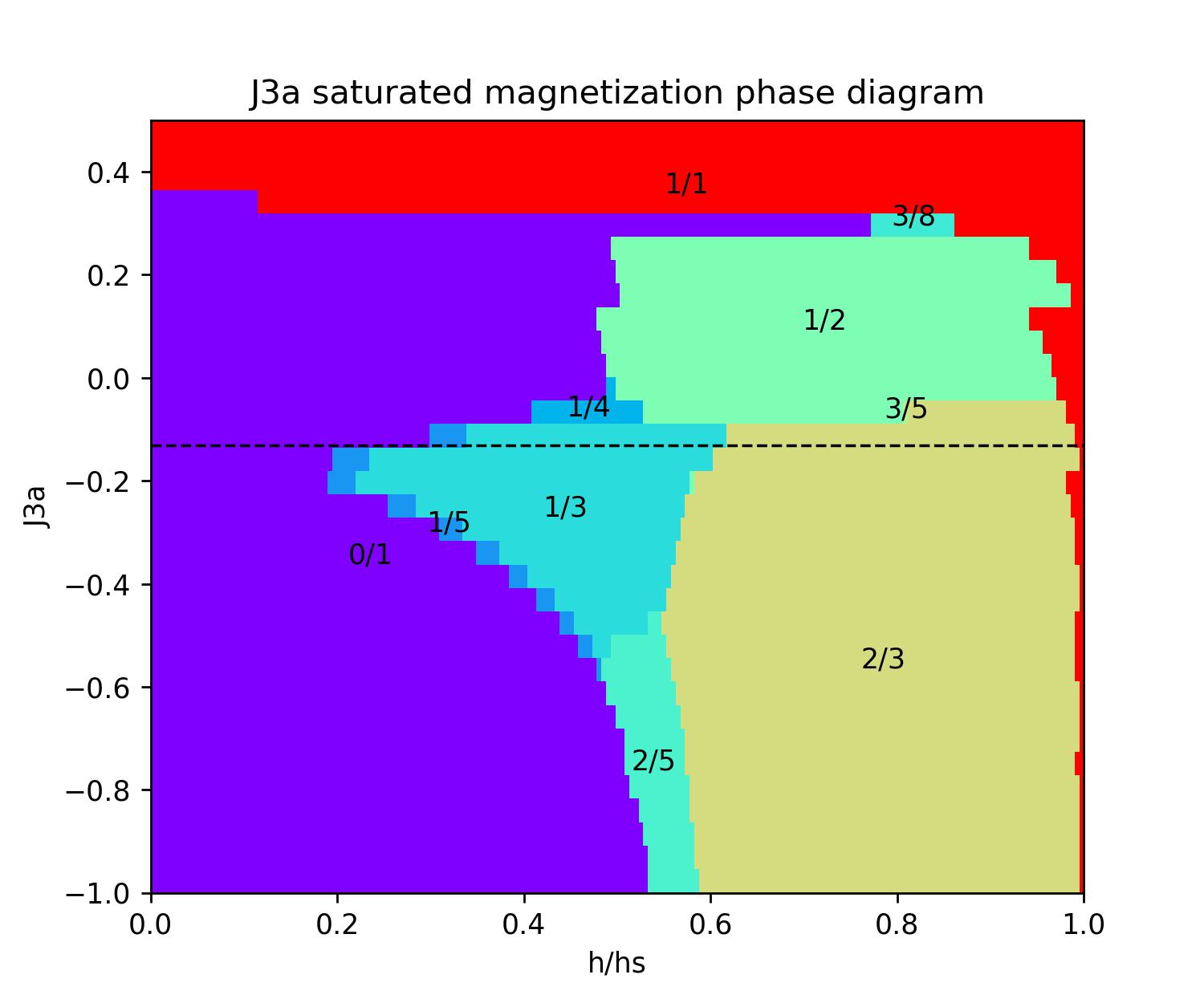}
\includegraphics[width=0.195\linewidth]{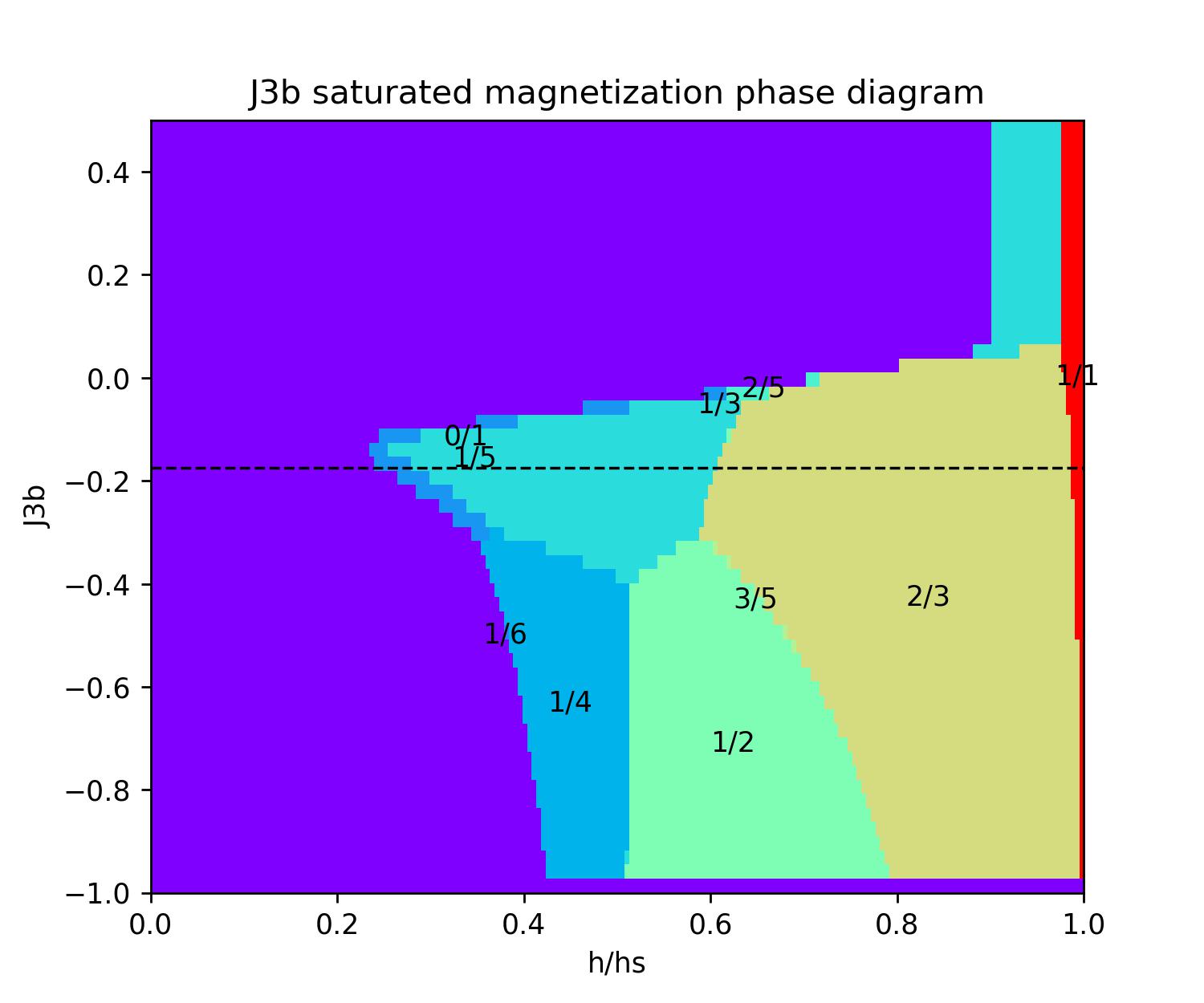}
\includegraphics[width=0.195\linewidth]{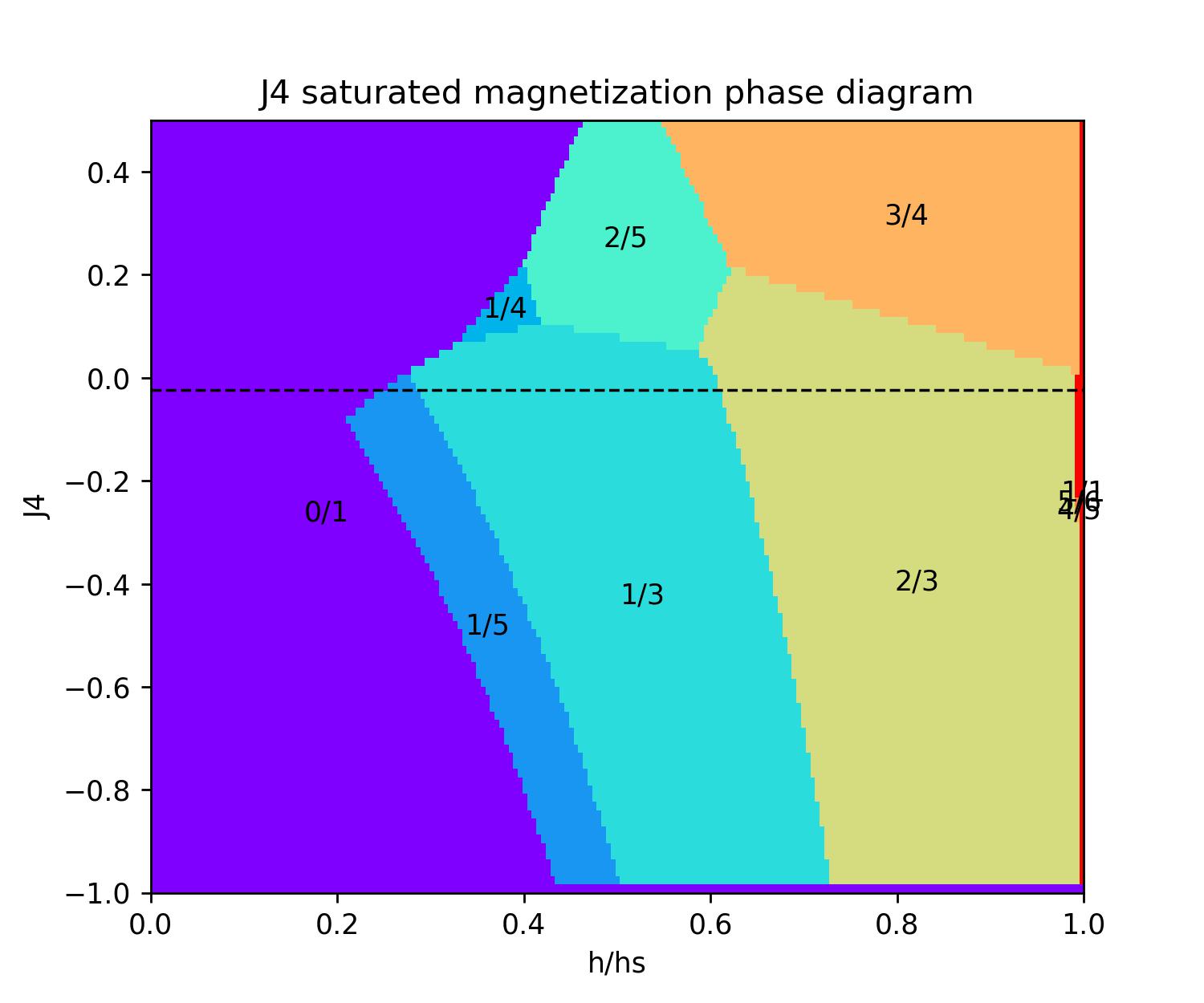}
\includegraphics[width=0.195\linewidth]{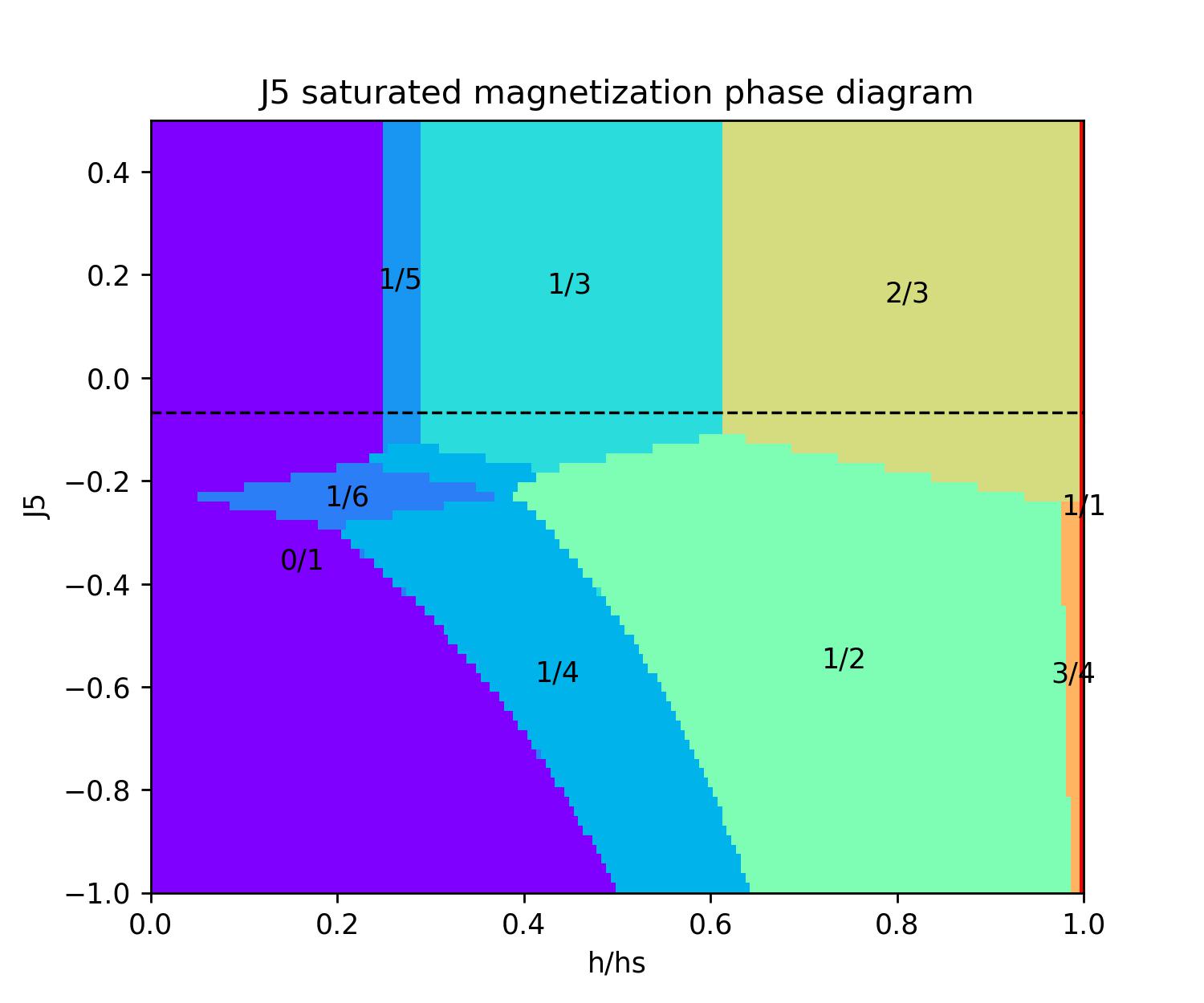}
\centering
\caption{Average magnetization phase diagrams for bond parameters $J_{2}, J_{3a}, J_{3b}, J_{4}, J_{5}$ for a magnetic field applied in the X (top) and Y (bottom) directions, normalized to saturation field. Dotted line indicating the theorized value for $J_\alpha$}
\label{fig:PH_diag_sat_frac}
\end{figure*}

\clearpage
\subsection{State Phase Diagrams}

\begin{figure*}[!h]
\includegraphics[width=0.195\linewidth]{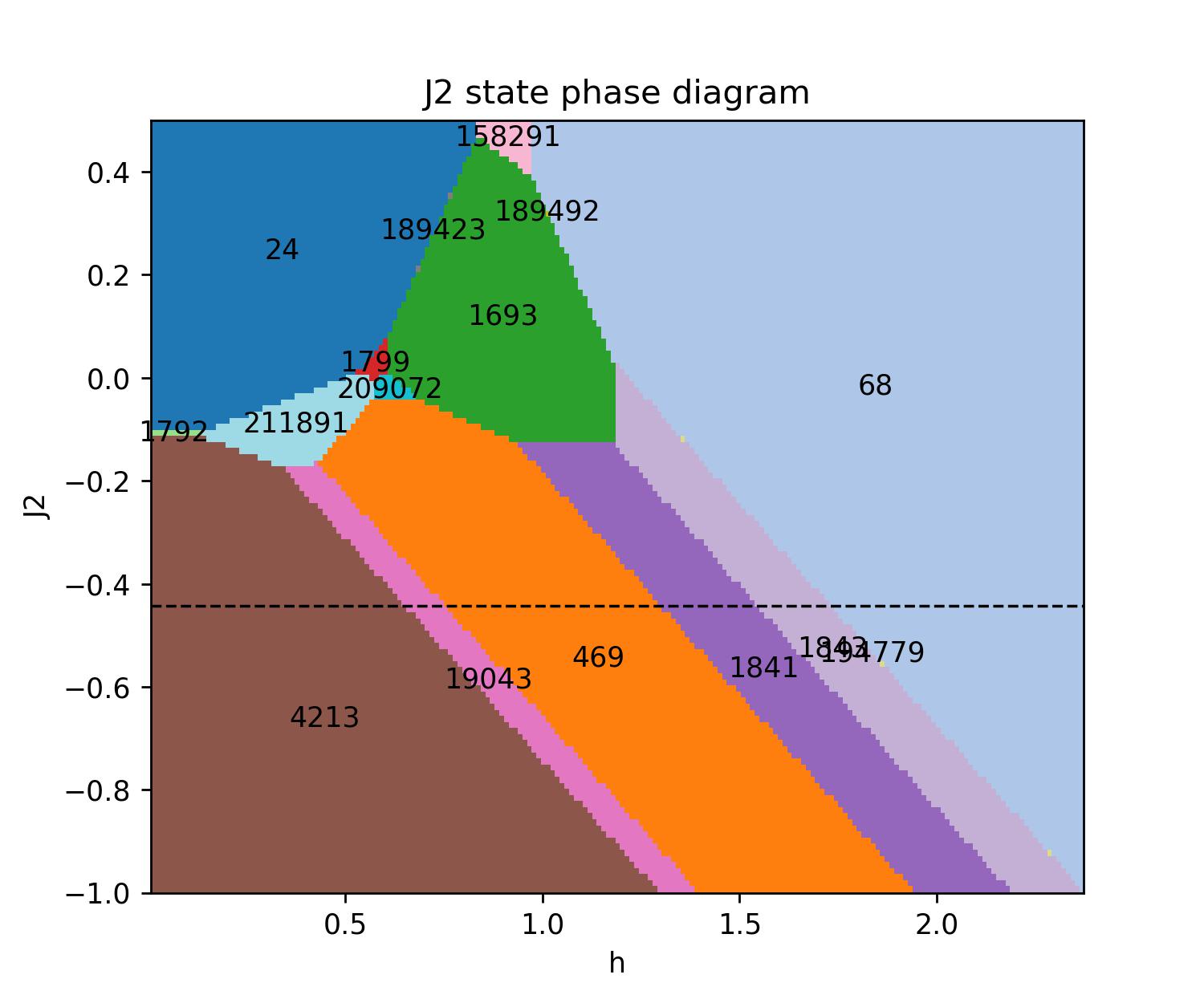}
\includegraphics[width=0.195\linewidth]{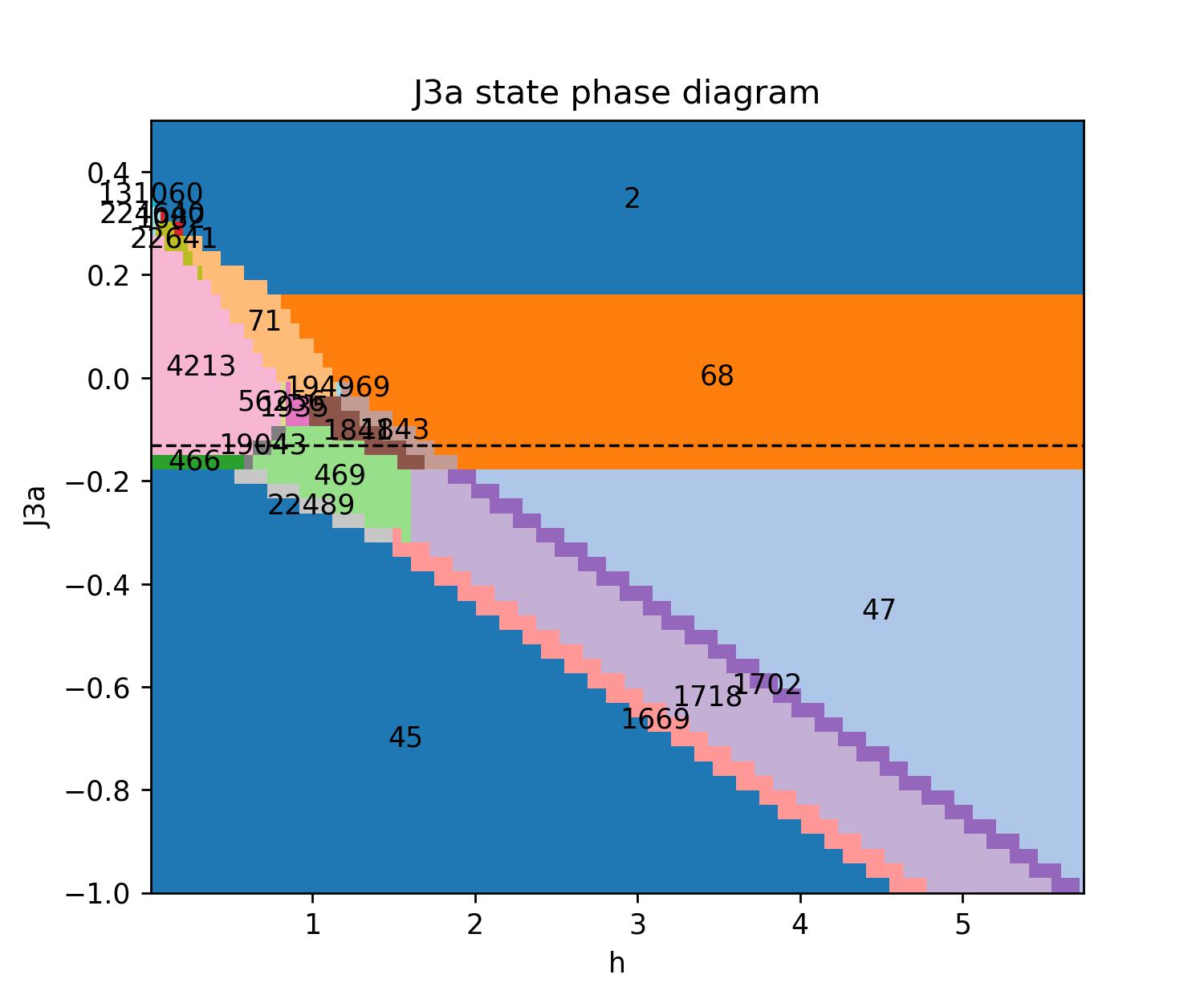}
\includegraphics[width=0.195\linewidth]{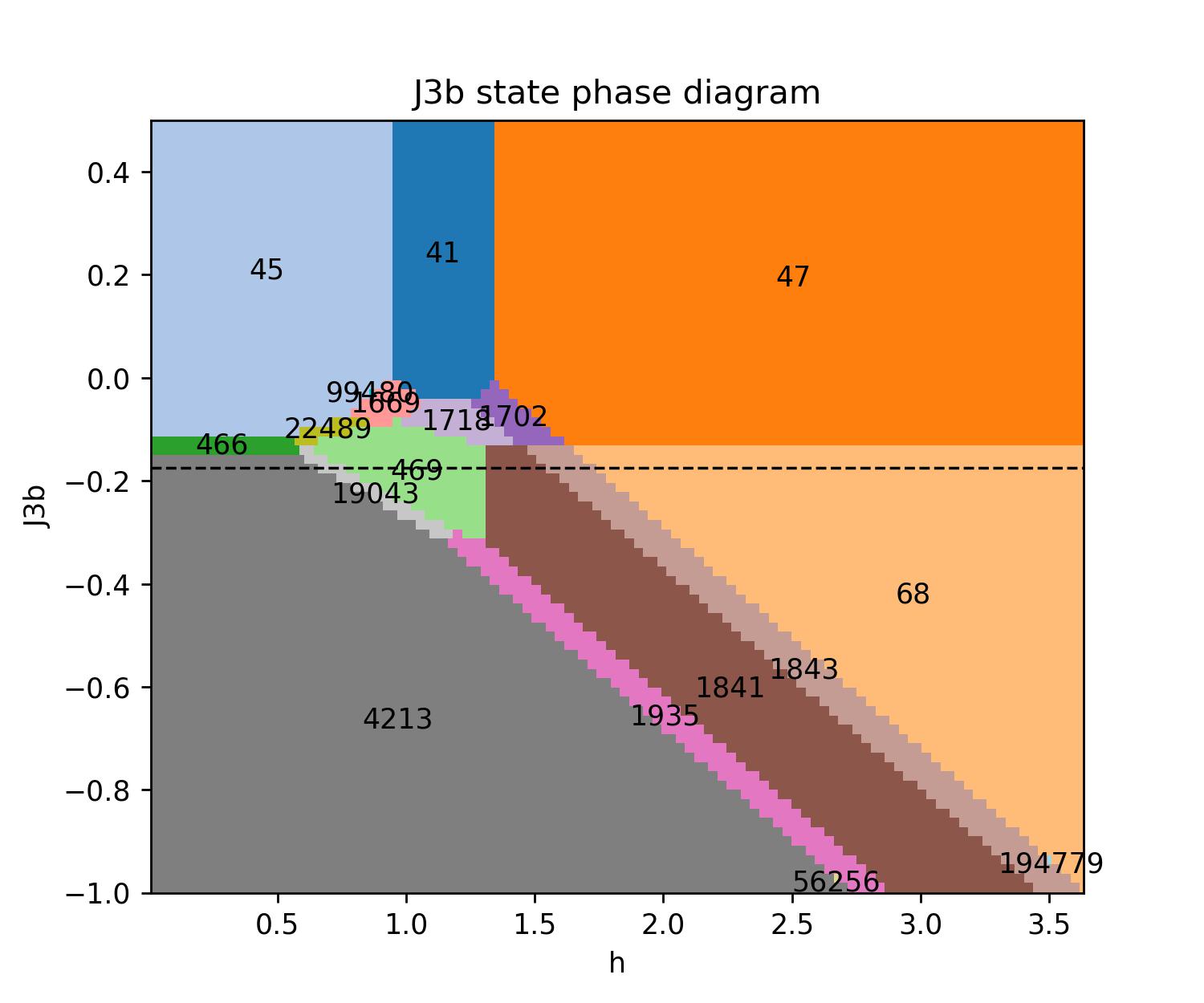}
\includegraphics[width=0.195\linewidth]{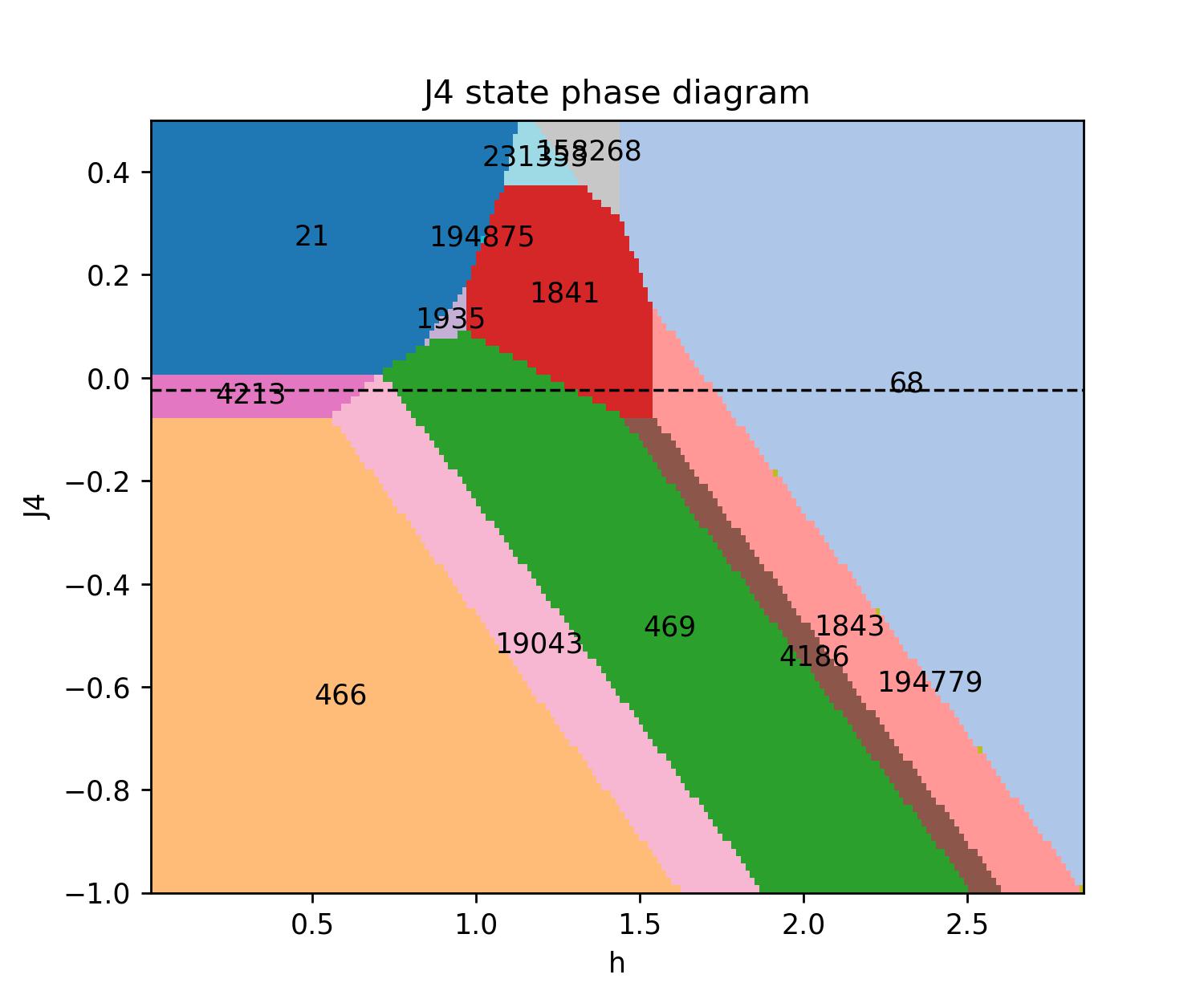}
\includegraphics[width=0.195\linewidth]{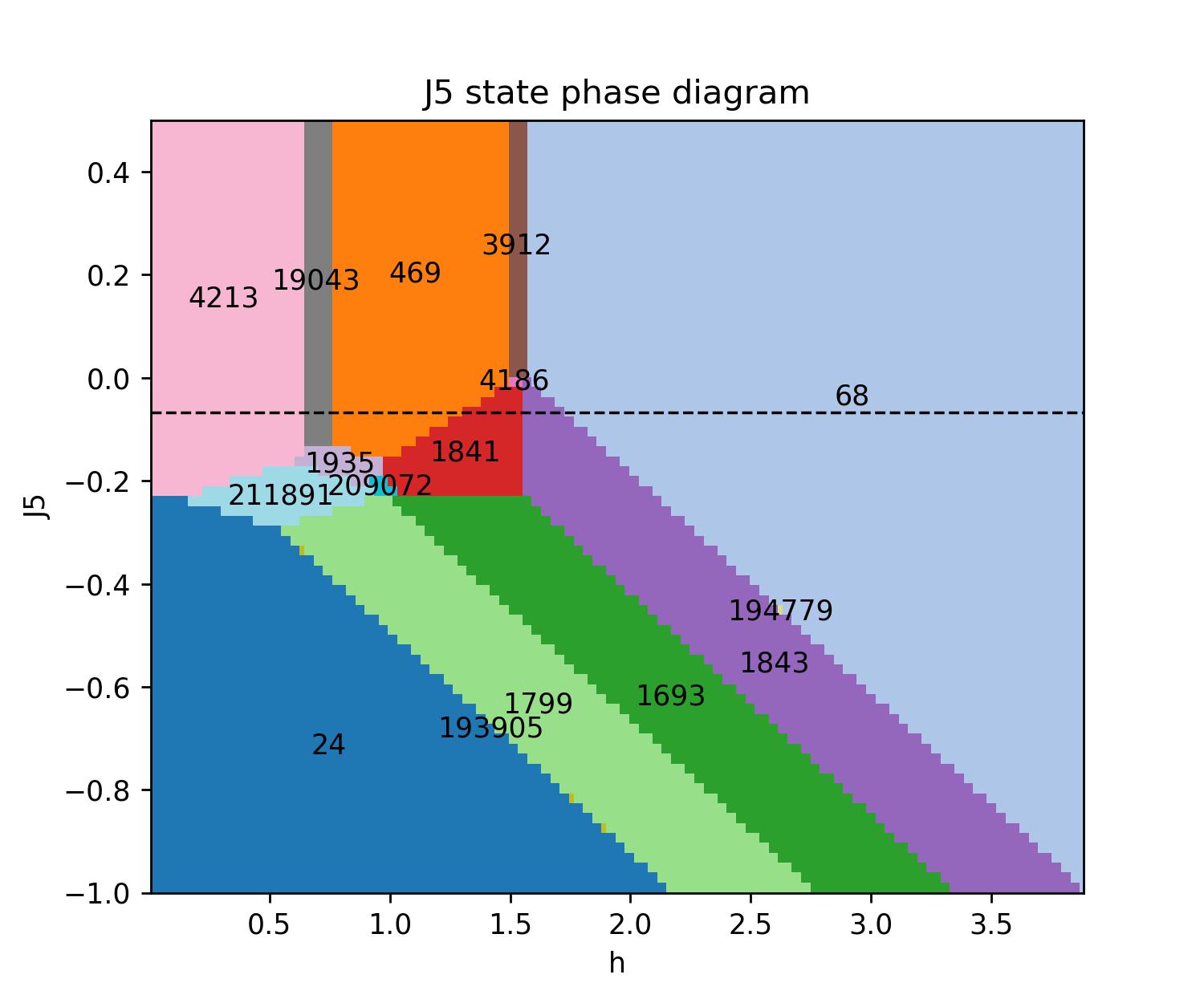}
\includegraphics[width=0.195\linewidth]{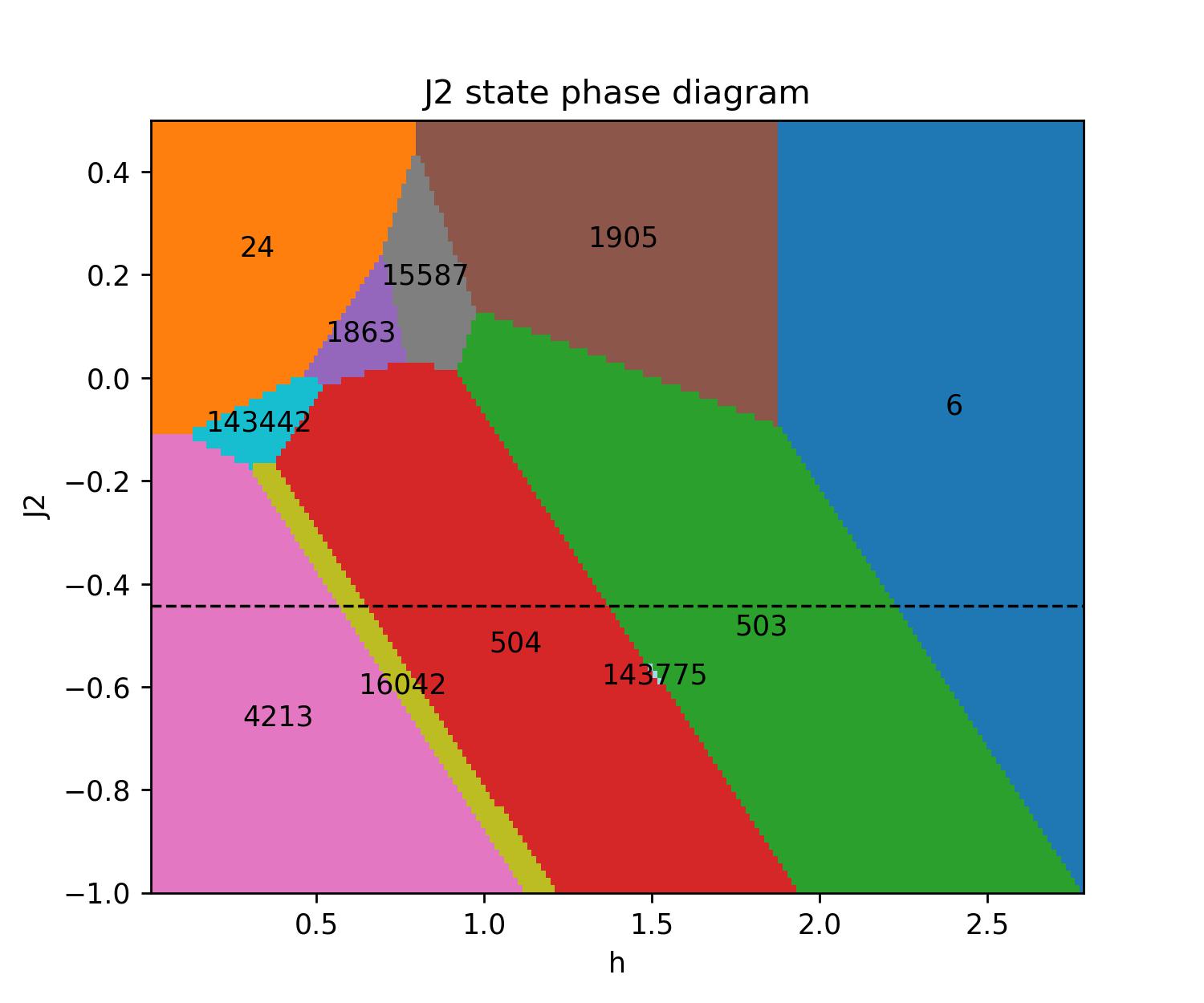}
\includegraphics[width=0.195\linewidth]{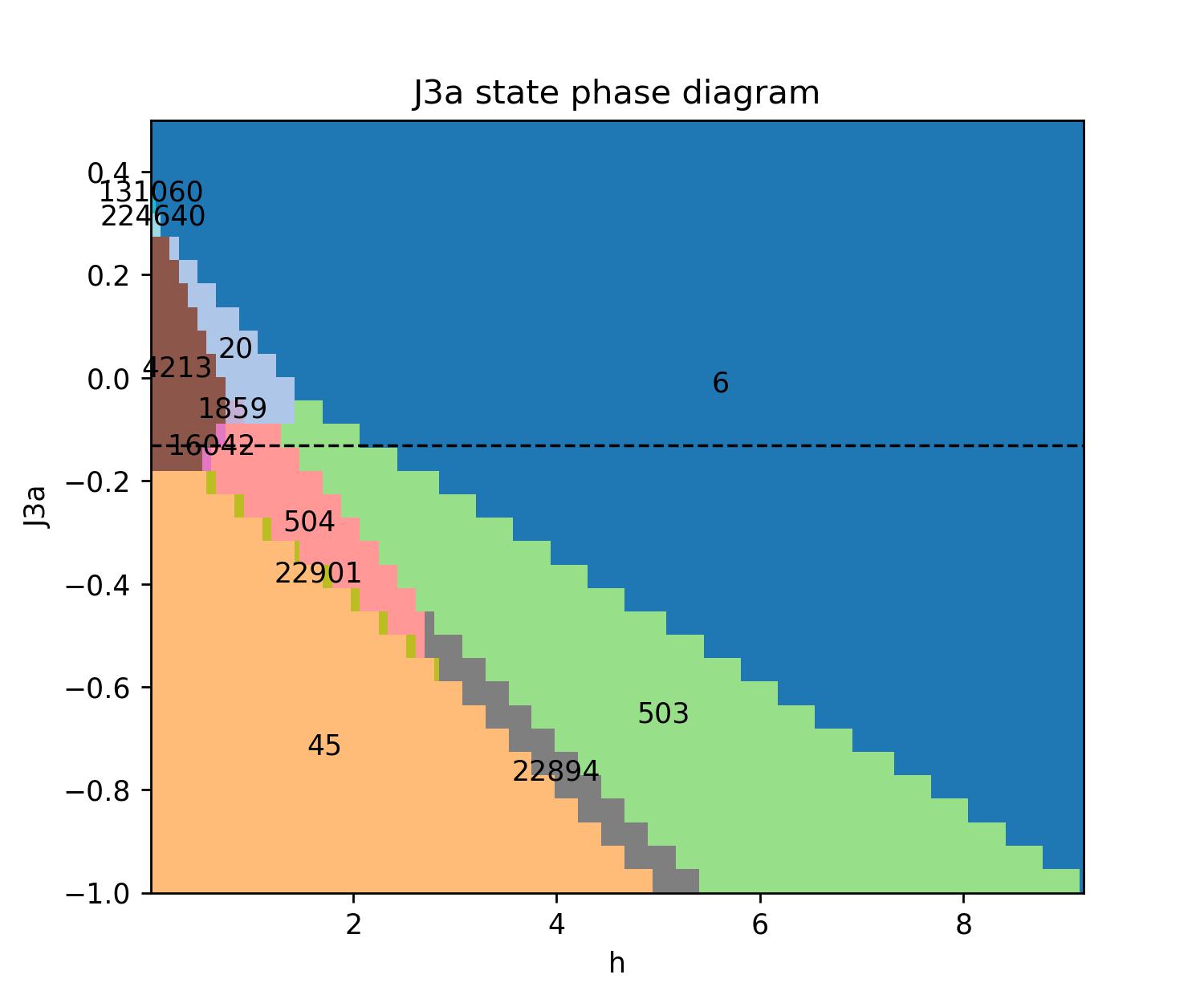}
\includegraphics[width=0.195\linewidth]{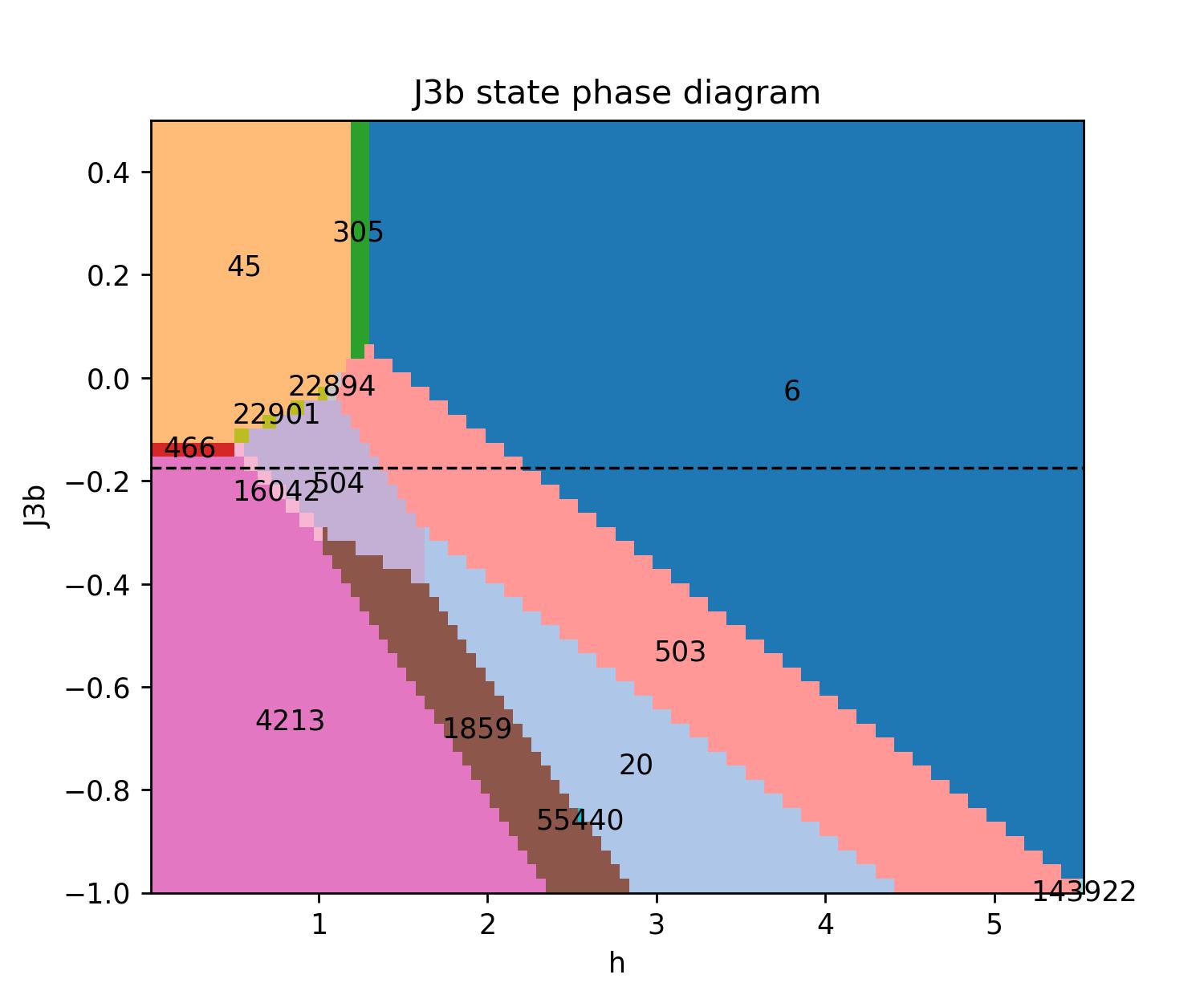}
\includegraphics[width=0.195\linewidth]{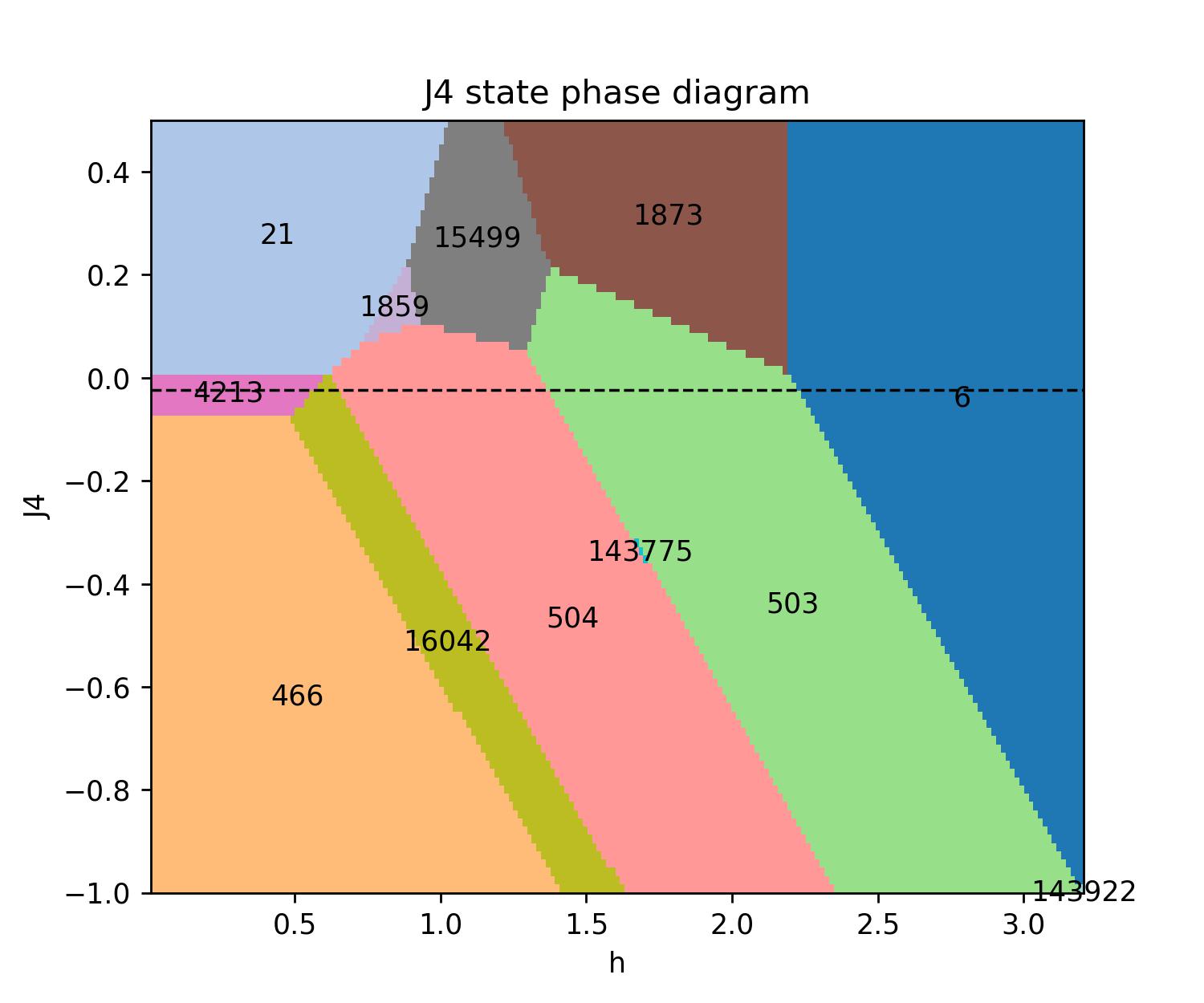}
\includegraphics[width=0.195\linewidth]{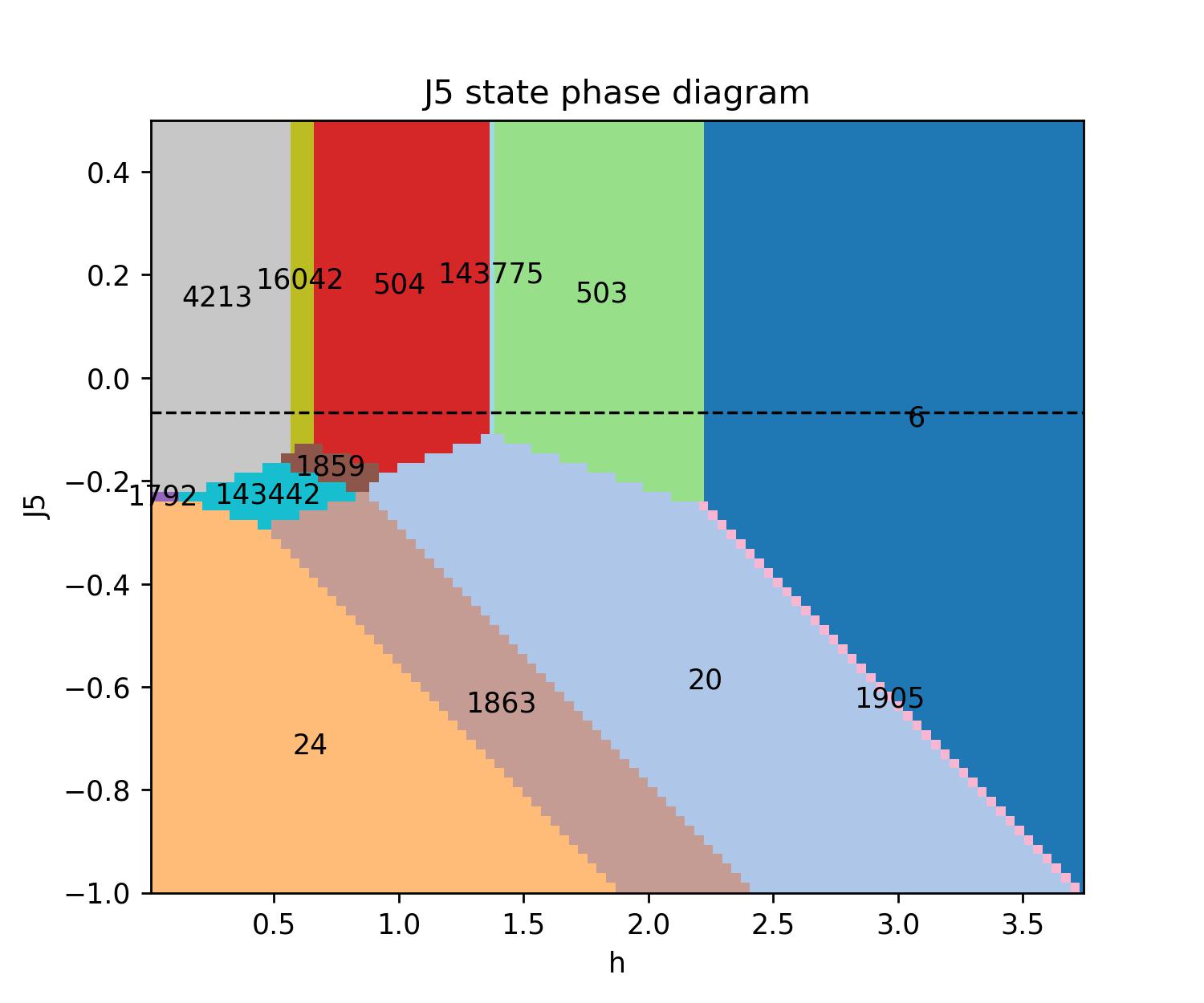}
\centering
\caption{Magnetic state phase diagrams for bond parameters $J_{2}, J_{3a}, J_{3b}, J_{4}, J_{5}$ for a magnetic field applied in the X (top) and Y (bottom) directions. Dotted line indicating the theorized value for $J_\alpha$}
\label{fig:PH_diag_state}
\end{figure*}

\begin{figure*}[!h]
\includegraphics[width=0.195\linewidth]{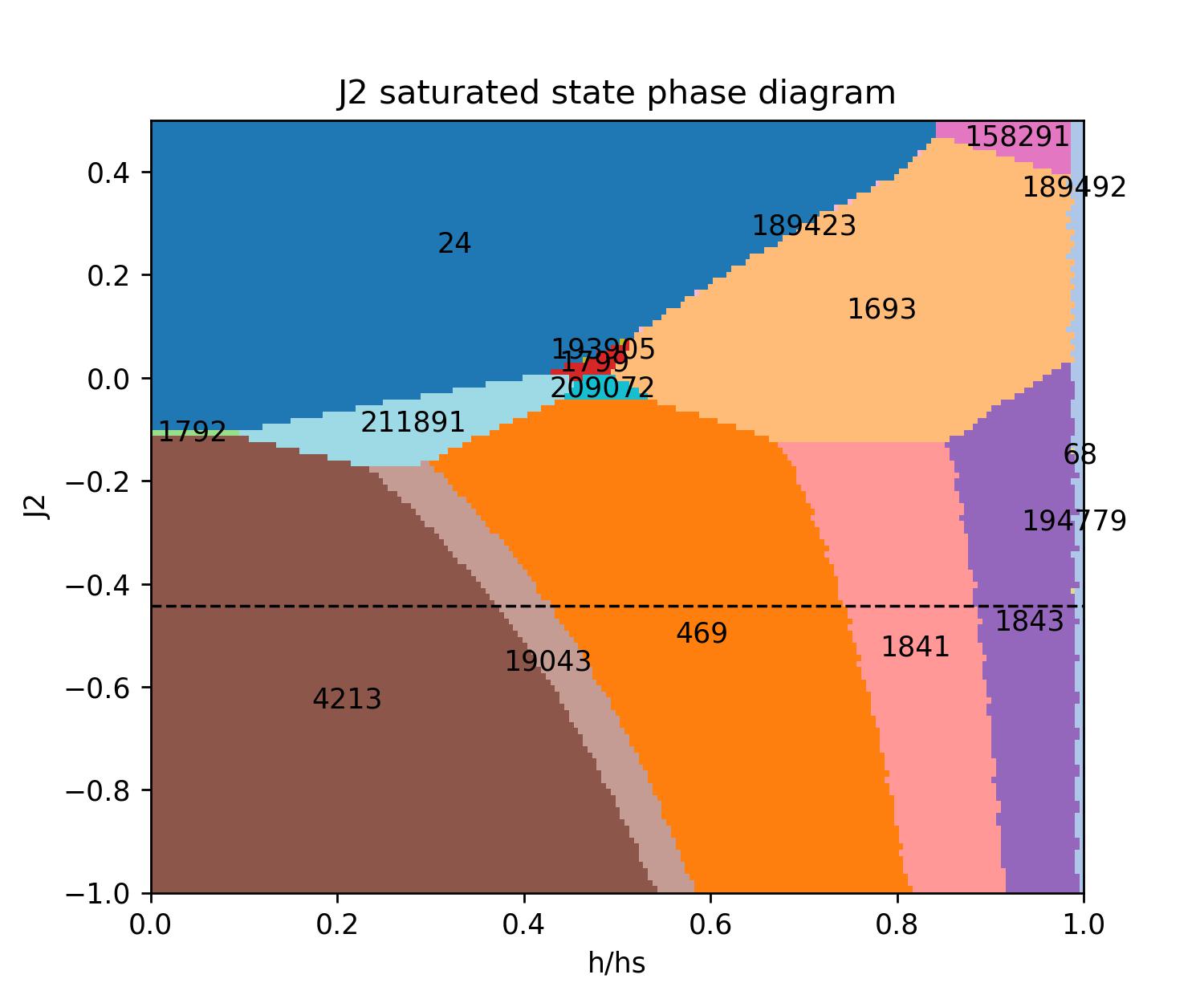}
\includegraphics[width=0.195\linewidth]{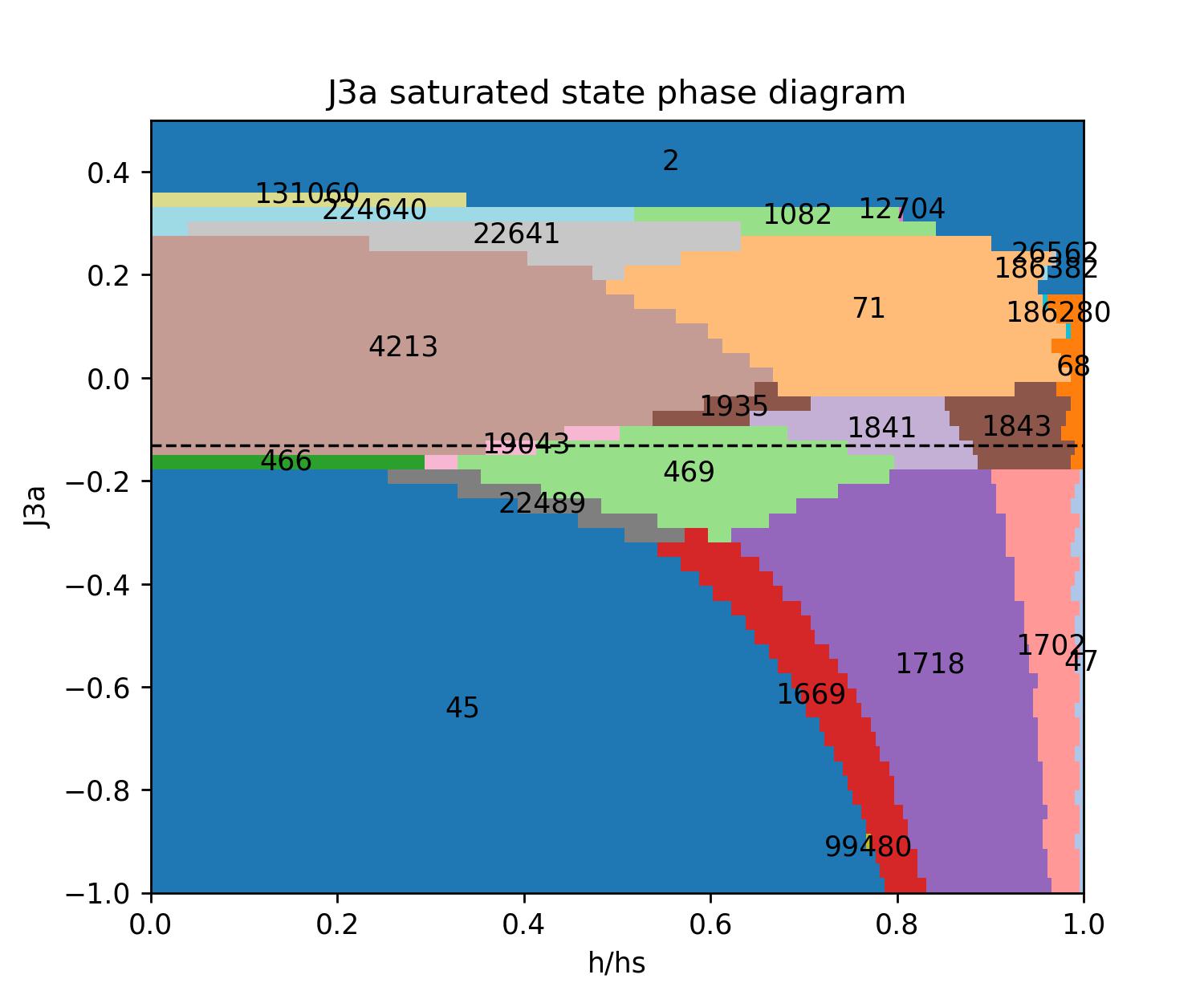}
\includegraphics[width=0.195\linewidth]{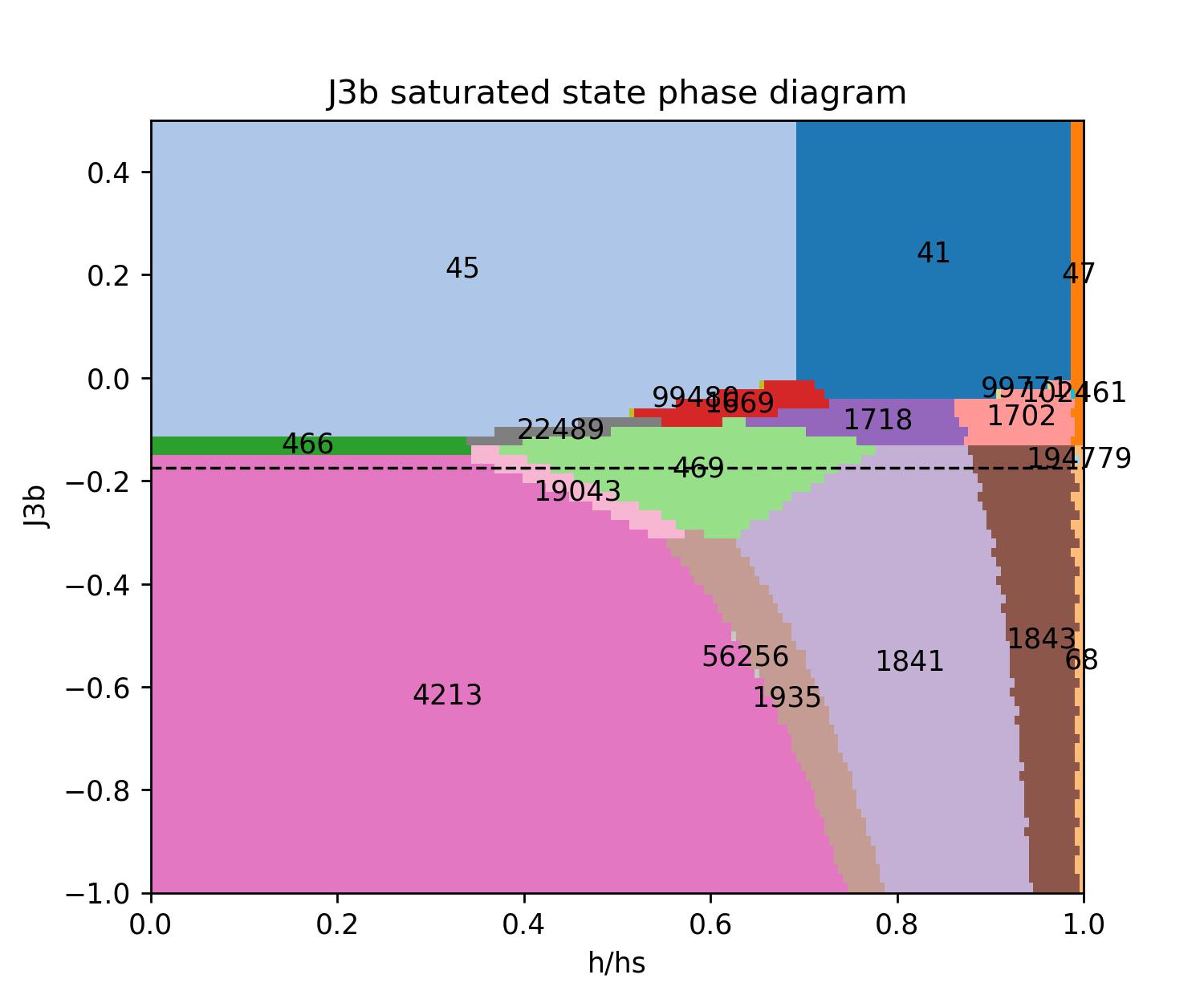}
\includegraphics[width=0.195\linewidth]{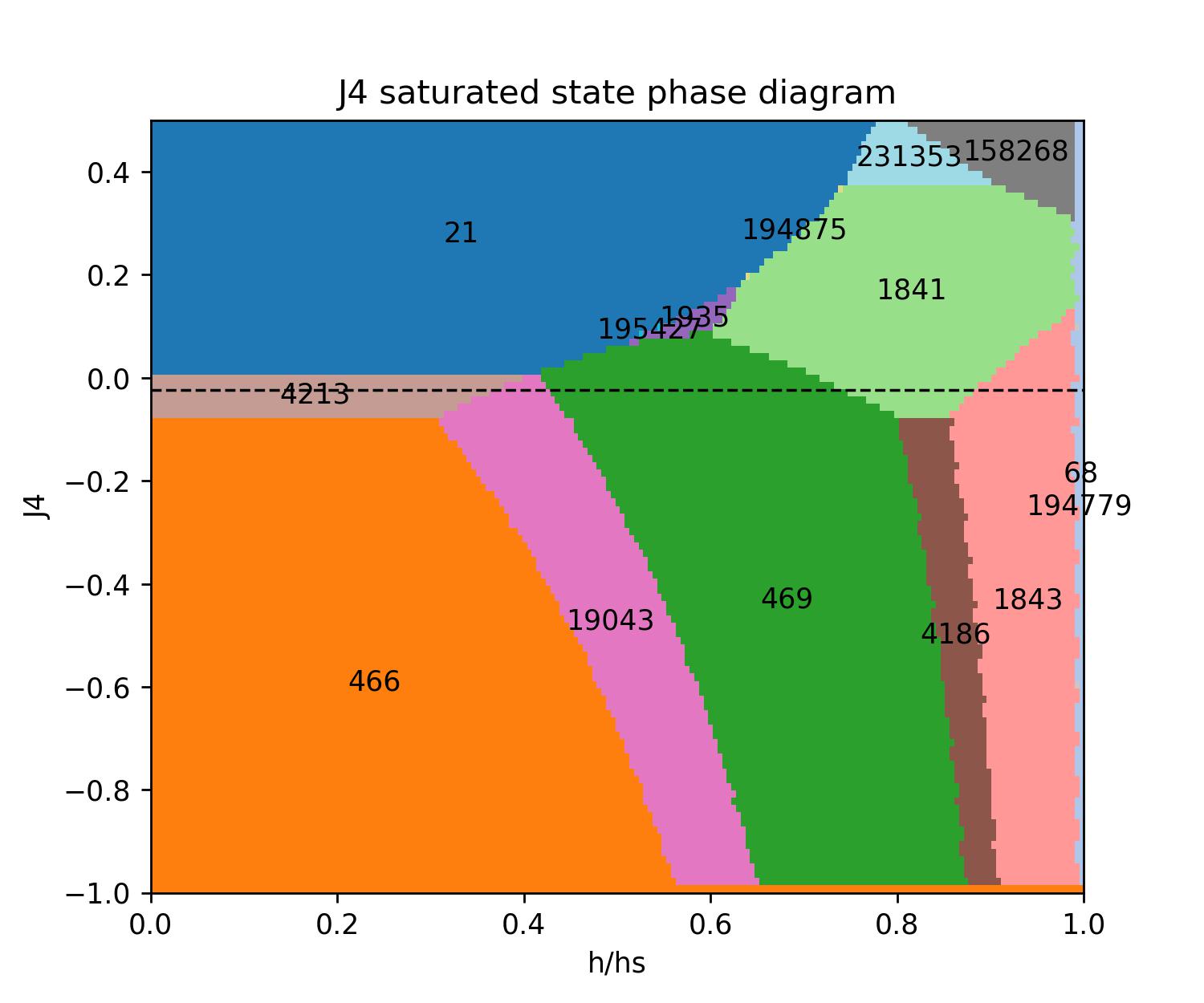}
\includegraphics[width=0.195\linewidth]{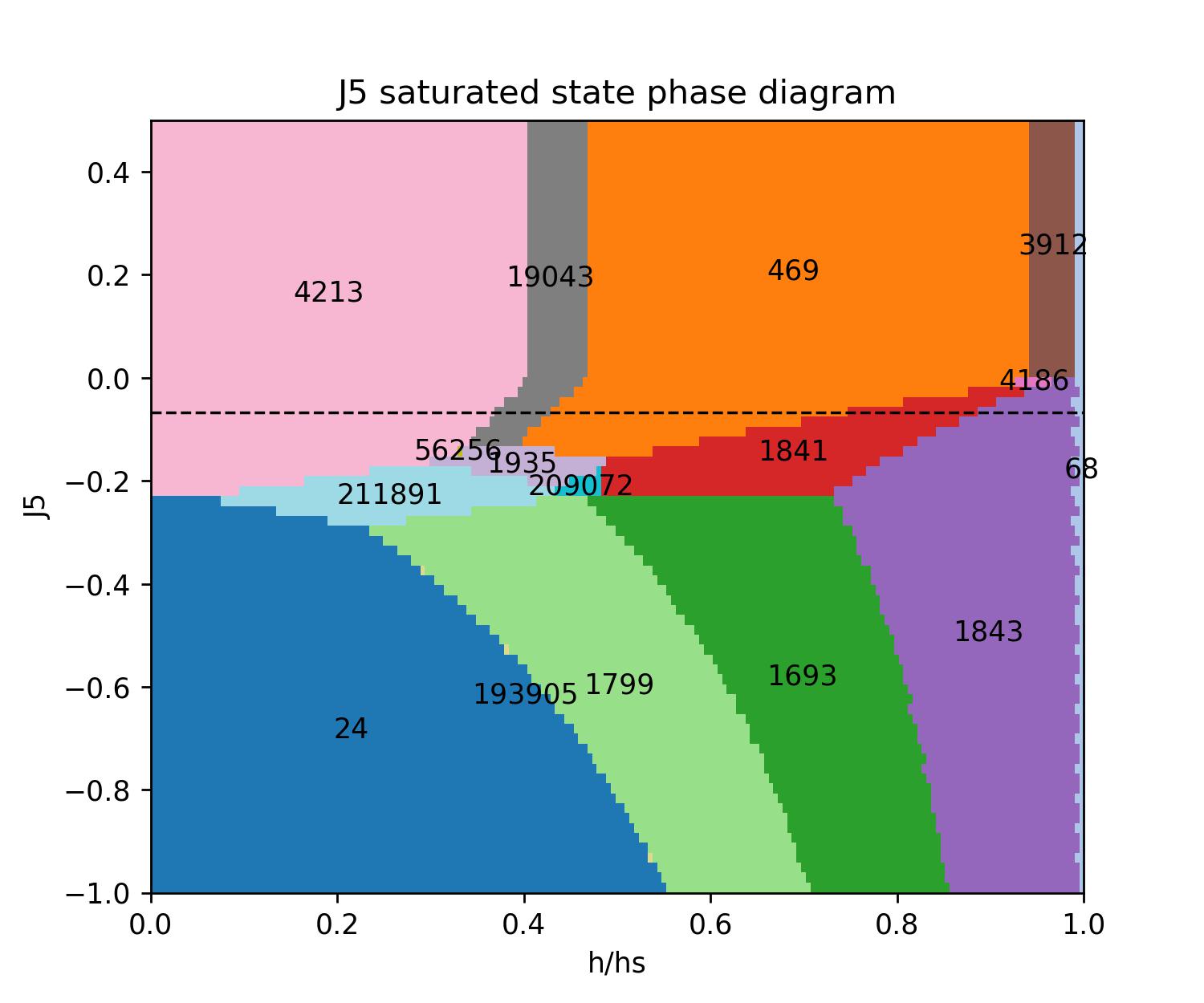}
\includegraphics[width=0.195\linewidth]{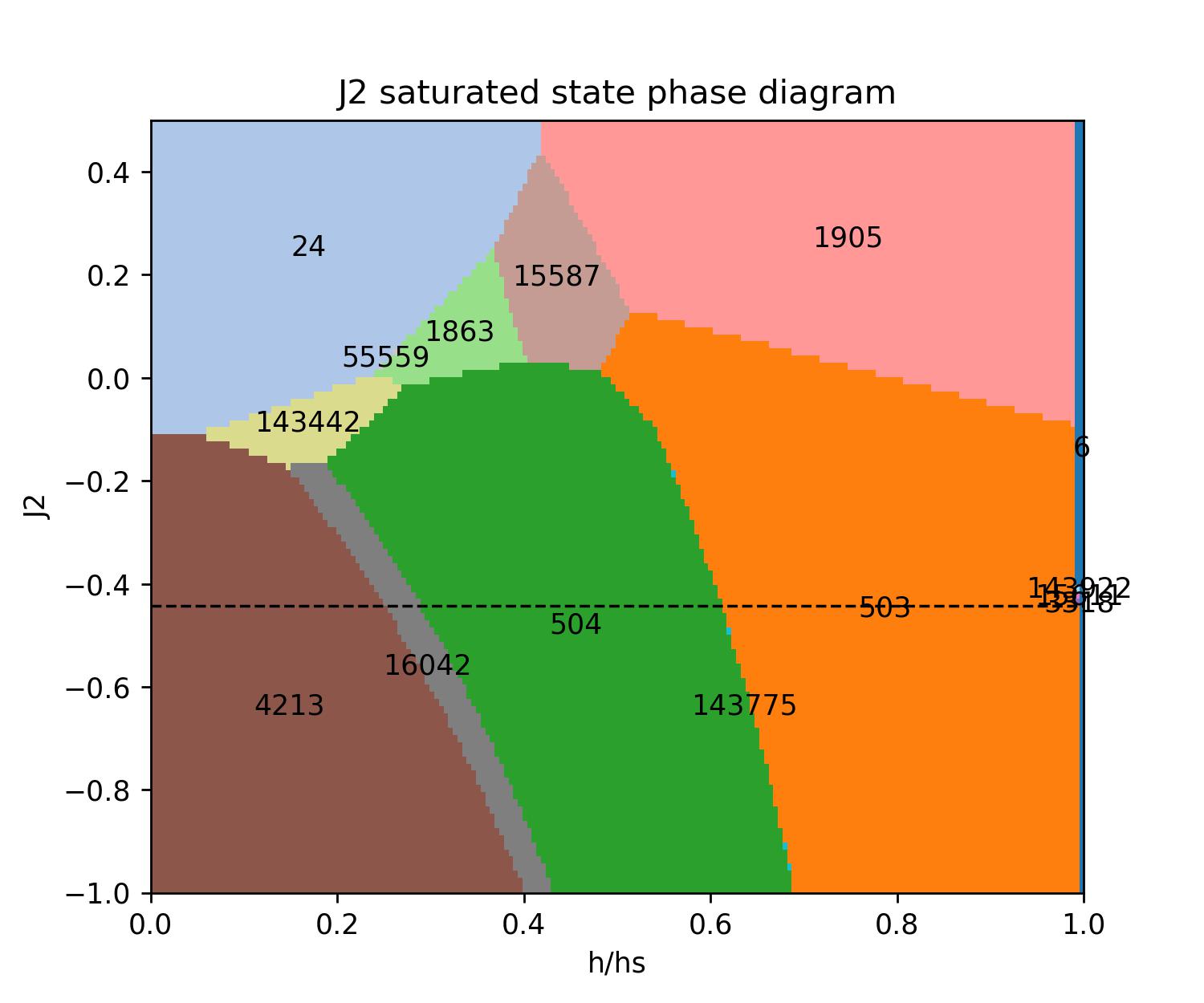}
\includegraphics[width=0.195\linewidth]{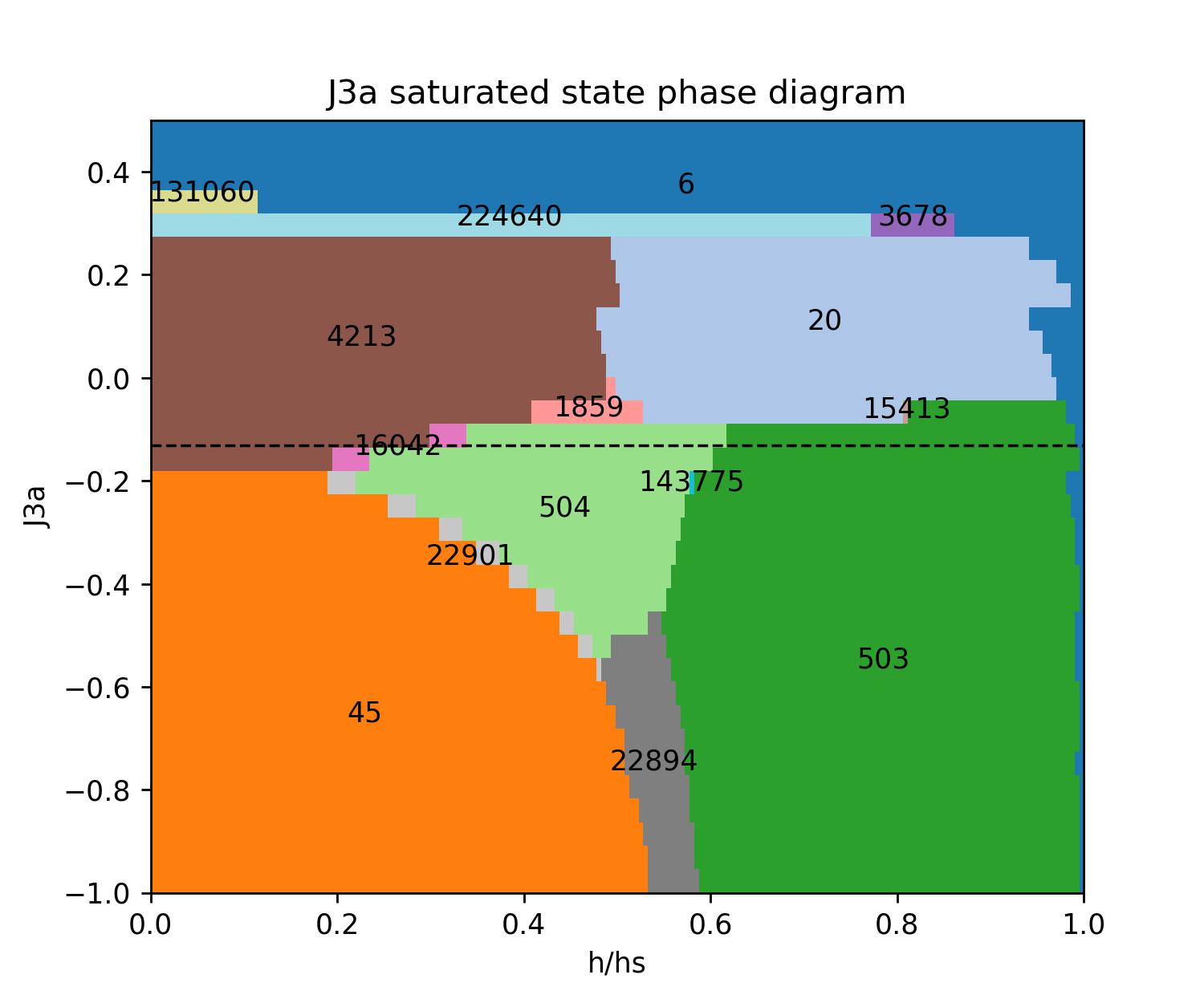}
\includegraphics[width=0.195\linewidth]{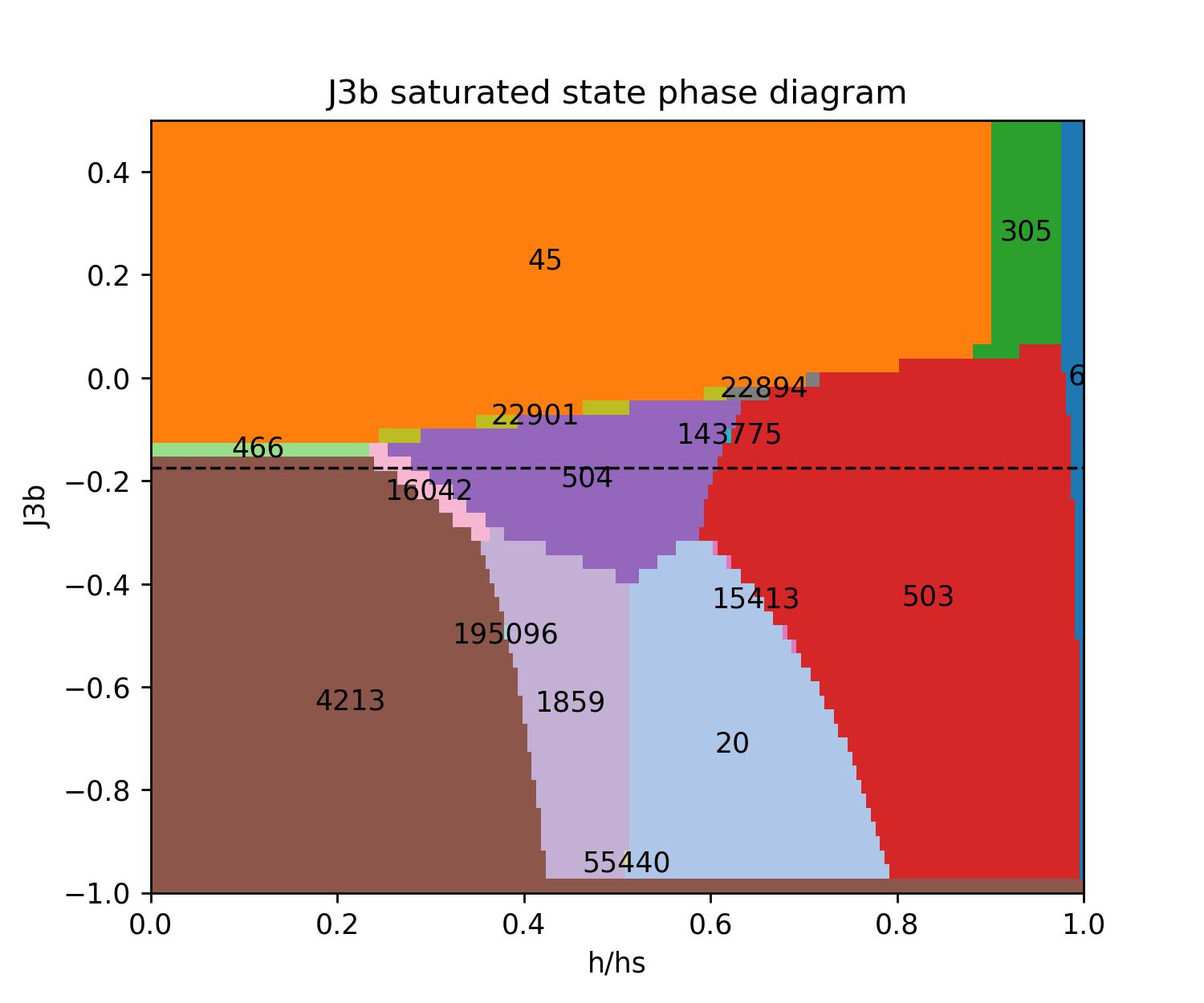}
\includegraphics[width=0.195\linewidth]{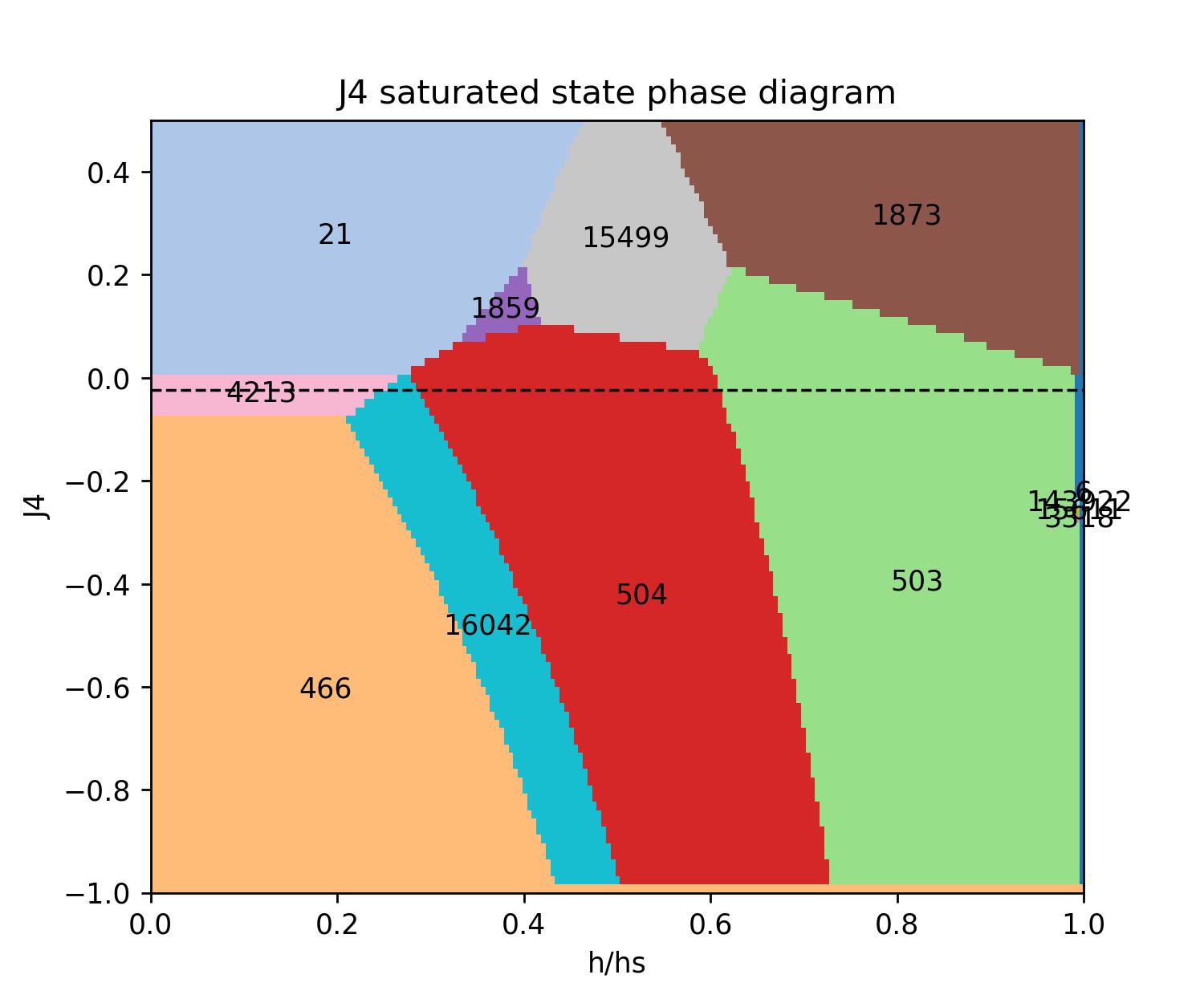}
\includegraphics[width=0.195\linewidth]{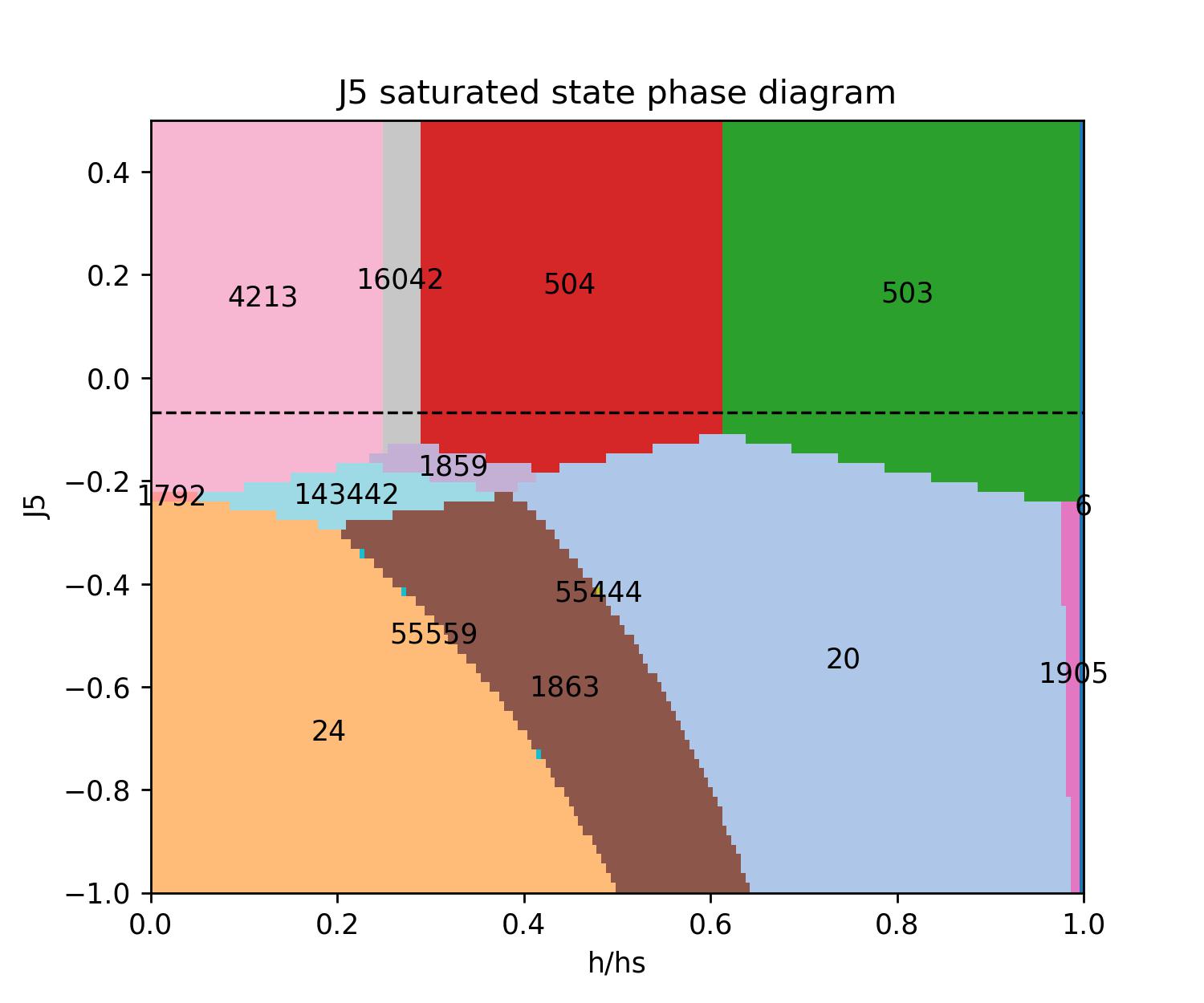}
\centering
\caption{Magnetic state phase diagrams for bond parameters $J_{2}, J_{3a}, J_{3b}, J_{4}, J_{5}$ for a magnetic field applied in the X (top) and Y (bottom) directions, normalized to saturation field. Dotted line indicating the theorized value for $J_\alpha$}
\label{fig:PH_diag_sat_state}
\end{figure*}

%\subsection{Further neighbor interactions}

%For completeness, we also generate and inspect phase diagrams for longer range interactions beyond the calculated parameters (up to the seventh nearest neighbor, $J_7$). We found that slight adjustments to these further neighbor interactions could still potentially contribute to lifting degeneracy in several of the found states due to the high frustration of the system. This can be seen directly in Fig \ref{fig:EX} where several potential magnetic unit cells are calculated to be extremely close in energy, and also in Fig \ref{fig:phase_J7x}.

%where the inclusion of even a small $|J_7|$ interaction $\left( J_7 < 0.1 |J_1| \right)$ could potentially give rise to new states.

%\begin{figure}[!h]
%\includegraphics[scale=0.5]{Figures_supplemental/J7_mag_phase_diag}
%\centering
%\caption{$J_{7}$ Magnetic phase diagram for a magnetic field applied in the X direction. Colors represent average magnetization in the X direction.}
%\label{fig:phase_J7x}
%\end{figure}

\clearpage
\section{Energy vs applied field}

Average energy per Ho atom can be plotted for each Magnetic Unit Cell considered. Below we can see the close competition between all 33 acceptable Magnetic Unit Cells with volume less than or equal to 6 times the Unit Cell volume. As described earlier, only the minimum energy state will be observed, meaning only the state corresponding to the bottom most line in each plot.

\begin{figure}[h!]
\centering
\includegraphics[scale=0.75]{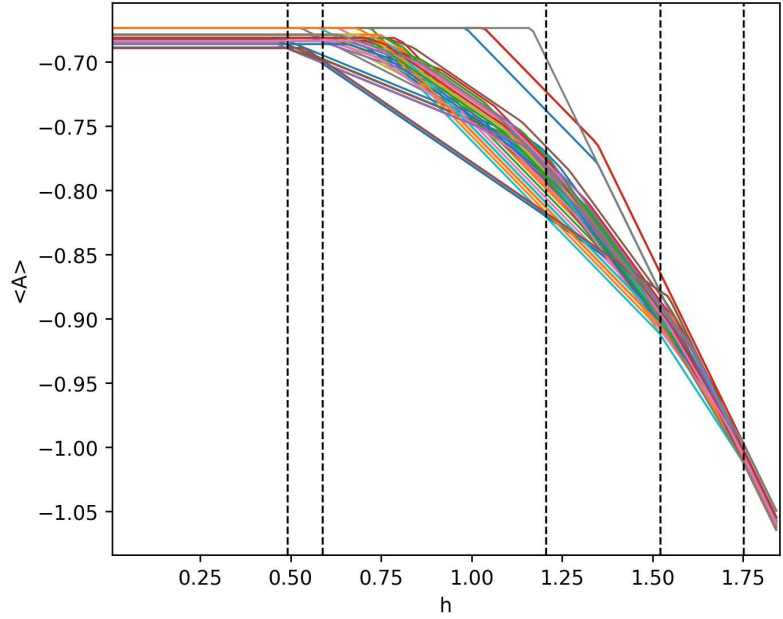}
\caption{Minimum average energy per Ho atom ($\left< A \right>$) of different Magnetic Unit Cell guesses as external magnetic field ($h||x$) is applied. (Vertical dotted lines indicating phase transitions)}
\label{fig:EX}
\end{figure}

\begin{figure}[h!]
\centering
\includegraphics[scale=0.75]{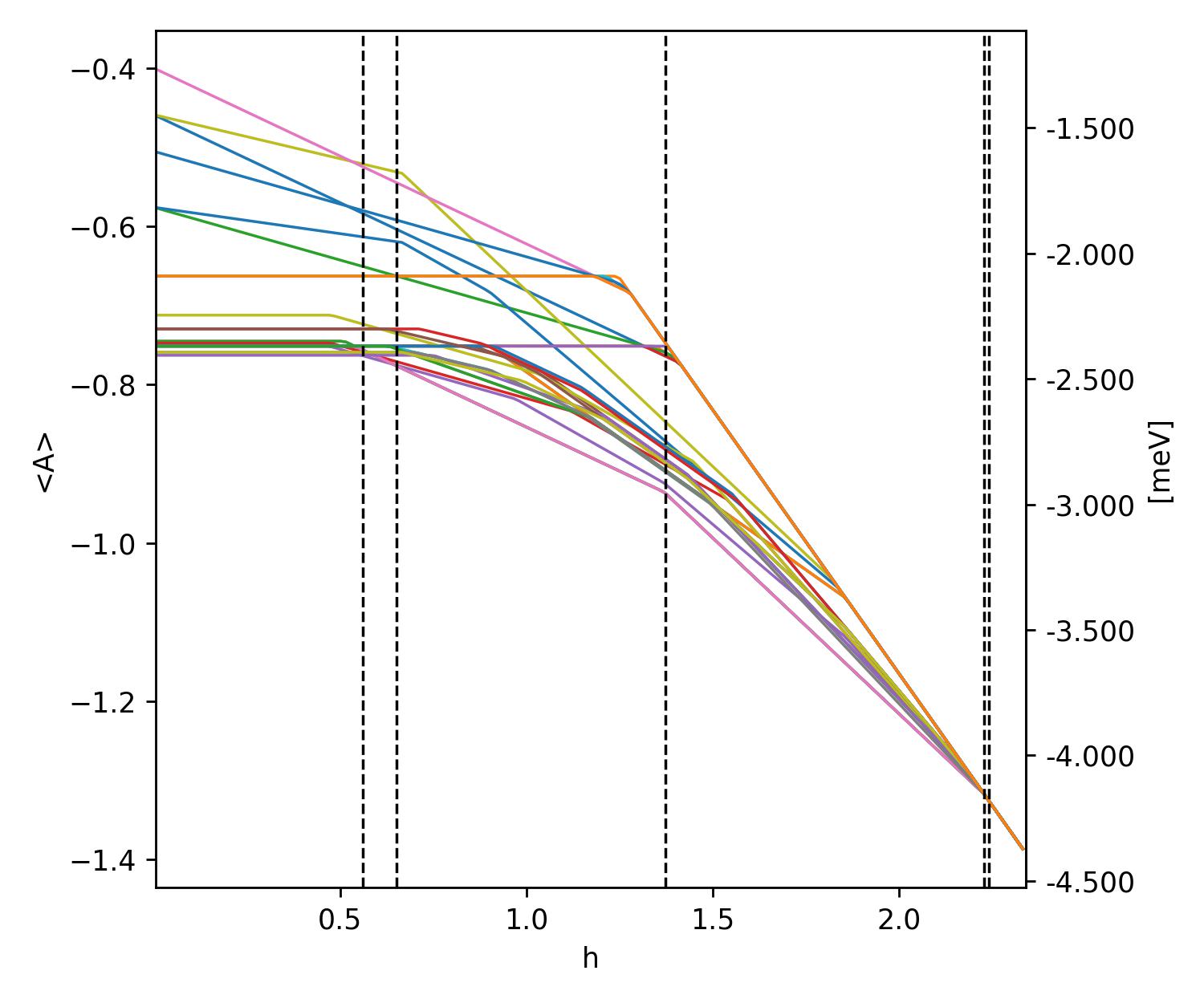}
\caption{Minimum average energy per Ho atom ($\left< A \right>$) of different Magnetic Unit Cell guesses as external magnetic field ($h||y$) is applied. (Vertical dotted lines indicating phase transitions)}
\label{fig:EY}
\end{figure}